\documentclass[
	12pt, 
	fleqn, 
	letterpaper, 
]{LegrandOrangeBook}

\usepackage[square,sort,comma,numbers]{natbib}
\usepackage{graphicx}
\usepackage{subcaption} 
\usepackage{aas_macros}
\usepackage{wrapfig}
\usepackage{xcolor}
\usepackage{tcolorbox}
\usepackage{multirow}
\usepackage[table,xcdraw]{xcolor}
\usepackage{longtable}

\hypersetup{
    colorlinks,
    linkcolor=orange,
    urlcolor=blue,
    anchorcolor=red,
    citecolor=red,
	pdftitle={Title}, 
	pdfauthor={Author}, 
	pdfsubject={Subject}, 
	pdfkeywords={Keyword1, Keyword2, ...}, 
	pdfcreator={LaTeX}, 
}

\definecolor{ocre}{RGB}{243, 102, 25} 

\chapterimage{AS17-134-20457HR_crop.jpg} 
\chapterspaceabove{6.5cm} 
\chapterspacebelow{6.75cm} 

\begin{document}


\titlepage 
{\includegraphics[width={\paperwidth}]{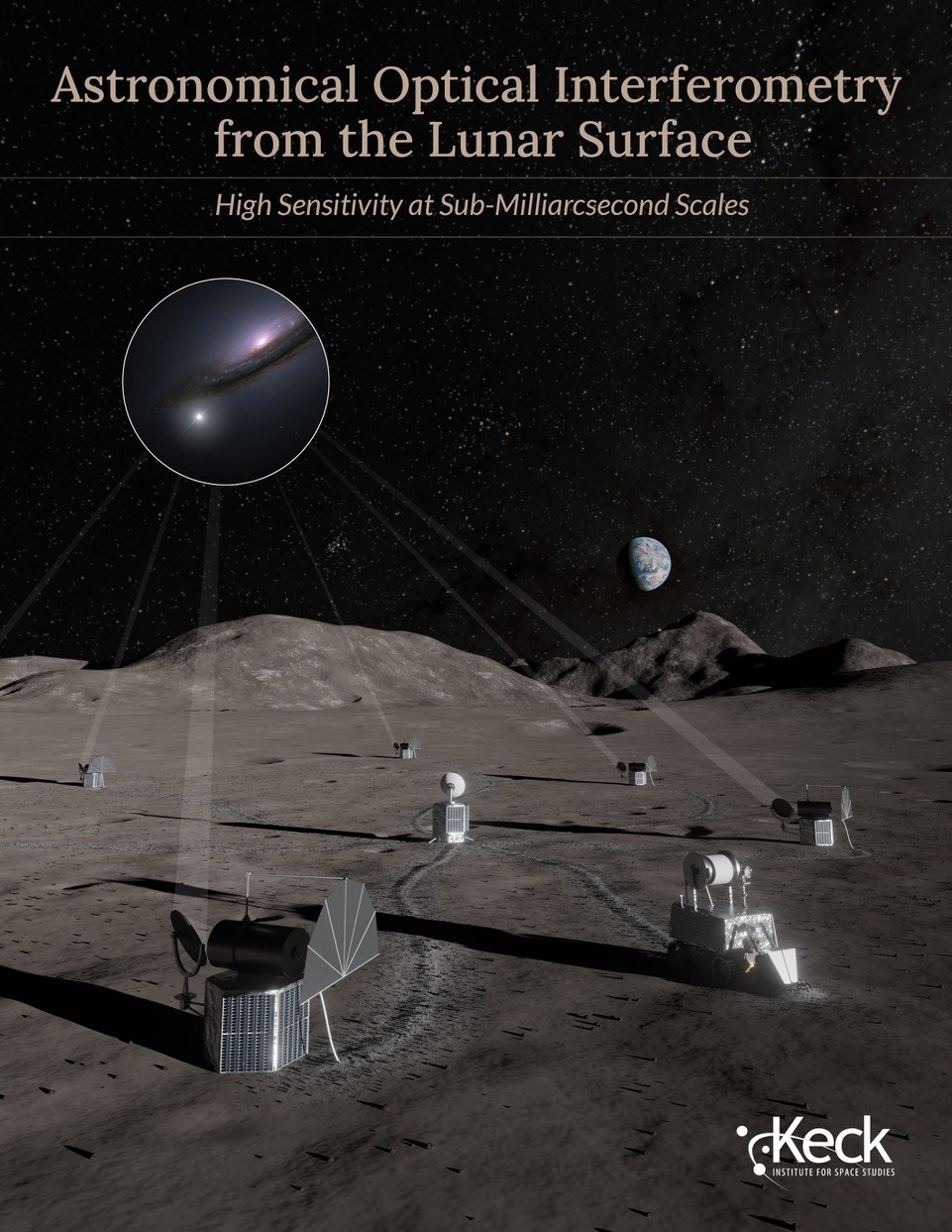}}



\thispagestyle{empty} 

~\vfill 

\noindent \textsc{Astronomical Optical Interferometry from the Lunar Surface}\\

\noindent Study Workshop: November 18-22, 2024\\

\noindent Study Leads:

Dr. Gerard T. van Belle (Lowell Observatory), \href{mailto: gerard@lowell.edu}{gerard@lowell.edu}

Dr. Stuart B. Shaklan (Jet Propulsion Laboratory, California Institute of Technology), \href{mailto: stuart.b.shaklan@jpl.nasa.gov}{stuart.b.shaklan@jpl.nasa.gov}

Dr. Shri R. Kulkarni (California Institute of Technology), \href{mailto: srk@astro.caltech.edu}{srk@astro.caltech.edu}\\

\noindent \textsc{Published by  W. M. Keck Institute for Space Studies (KISS)}

\noindent \textsc{\href{https://kiss.caltech.edu/workshops/lunar_interferometry/lunar_interferometry.html}{Workshop website}}\\ 

\noindent Study Report prepared for the W. M. Keck Institute for Space Studies (KISS)\\

\noindent The research was carried out in part at the Jet Propulsion Laboratory, California Institute of Technology, under a contract with the National Aeronautics and Space Administration (80NM0018D0004).\\

\noindent Pre-Decisional Information---For Planning and Discussion Purposes Only\\

\noindent The cost information contained in this document is of a budgetary and planning nature and is intended for informational purposes only. It does not constitute a commitment on the part of JPL and/or Caltech.\\

\noindent DOI: \href{https://authors.library.caltech.edu/records/z655j-jqm38}{10.26206/z655j-jqm38}\\

\noindent Recommended citation (long form):
Gerard van Belle, Tabetha Boyajian, Michelle Creech-Eakman, John Elliott, Kimberly Ennico-Smith, Dan Hillsberry, Kevin Hubbard, Takahiro Ito, Shri Kulkarni, Connor Langford, Laura Lee, David Leisawitz, Eric Mamajek, May Martin, Taro Matsuo, Dimitri Mawet, John Monnier, Jon Morse, Dave Mozurkewich, Paul Niles, Mark Panning, Lori Pigue, Aniket Sanghi, Gail Schaefer, Jeremy Scott, Stuart Shaklan, Locke Spencer, Aaron Tohuvavohu, Peter Tuthill, Karel Valenta, Jordan Wachs. 2025. "Astronomical Optical Interferometry from the Lunar Surface." G.T. van Belle, S.B. Shaklan, S.R. Kulkarni (Eds.) Report prepared for the W. M. Keck Institute for Space Studies (KISS), California Institute of Technology.\\

\noindent Recommended citation (short form):
G.T. van Belle, S.B. Shaklan, S.R. Kulkarni (Eds.). 2025. "Astronomical Optical Interferometry from the Lunar Surface." Report prepared for the W. M. Keck Institute for Space Studies (KISS), California Institute of Technology.\\

\noindent Acknowledgments

\noindent Director: Prof. Bethany Ehlmann

\noindent Executive Director: Harriet Brettle

\noindent Editing and Formatting: Gerard van Belle

\noindent Artwork Supervision: Lori Pigue

\noindent Cover Image: Artwork by Robert Hurt and Keith Miller under CC BY-NC 2.0 (Modified)  \\

\noindent Section Header Images: Contents and Sections 1, 3, 6: images from NASA/Apollo; Sections 2, 7, artwork by Gerard van Belle; Section 4, artwork by Britt Griswold; Section 5, SpaceX via NASA TV; Section 8, artwork by Robert Hurt and Keith Miller (excerpt from cover image)\\


\noindent \textit{First printing, August 2025} 

\noindent Copyright \copyright\ 2025 Caltech/Keck Institute for Space Studies. All rights reserved. 

\newpage

\noindent \textsc{Workshop Participants}

Left to right (back row): Dan Hillsberry (Argo Space), Laura Lee (Northern Arizona University), Stuart Shaklan (NASA JPL), Dave Mozurkewich (Seabrook Engineering), Taro Matsuo (Nagoya University), John Elliott (NASA JPL), Kimberly Ennico-Smith (NASA Ames), May Martin (ESA), Mark Panning (NASA JPL), Jeremy Scott (University of Arizona), Aaron Tohuvavohu (Caltech), John Monnier (University of Michigan), Peter Tuthill (University of Sydney), Eric Mamajek (NASA JPL), David Leisawitz (NASA GSFC), Locke Spencer (University Lethbridge), Michelle Creech-Eakman (New Mexico Tech), Connor Langford (University of Sydney), Jordan Wachs (MIT/SpaceRake), Tabetha Boyajian (Louisiana State University), Paul Niles (NASA JSC), Lori Pigue (U. S. Geological Survey, Astrogeology Science Center), Gail Schaefer (GSU CHARA), Jon Morse (AstronetX)

Left to right (kneeling): Kevin Hubbard (Honeybee Robotics), Takahiro Ito (Institute of Space and Astronautical Science, JAXA), Aniket Sanghi (Caltech), Karel Valenta (University of Sydney), Gerard van Belle (Lowell Observatory)

Not pictured: Shri Kulkarni (Caltech), Dimitri Mawet (Caltech)

\begin{figure}
    \centering
    \includegraphics[width=0.95\linewidth]{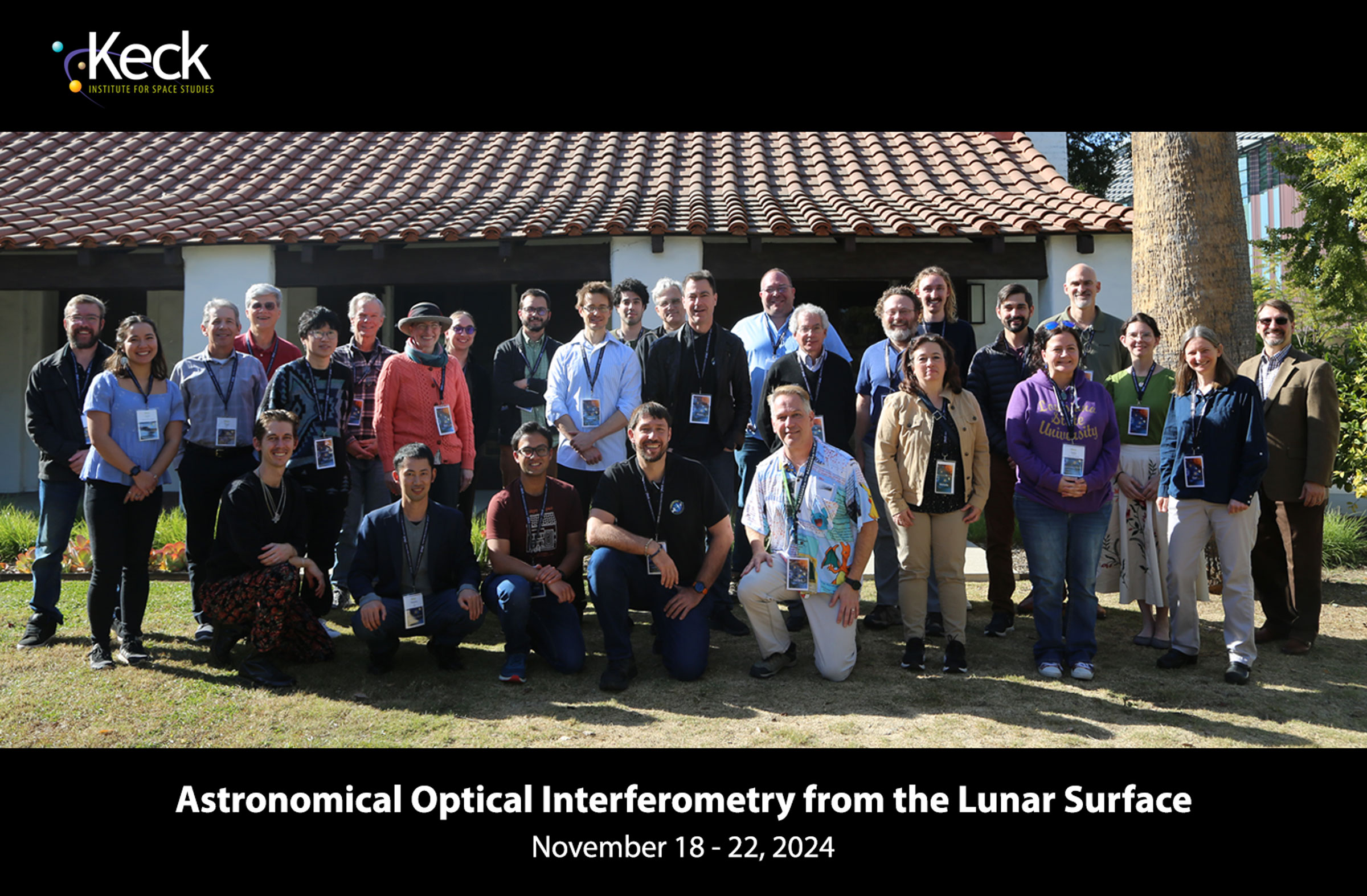} 
    \label{fig-group_photo}
\end{figure}

\newpage

\noindent \textsc{Acknowledgements}

The study "Astronomical Optical Interferometry from the Lunar Surface" was made possible by the W. M. Keck Institute for Space Studies, and by the Jet Propulsion Laboratory, California Institute of Technology, under contract with the National Aeronautics and Space Administration. Any use of trade, firm, or product names is for descriptive purposes only and does not imply endorsement by the U.S. Government.

The study team gratefully acknowledges the outstanding support of Harriet Brettle, Executive Director of the Keck Institute for Space Studies, as well as her dedicated staff, who made the study experience invigorating and enormously productive. Many thanks are also due to Charles Lawrence and the KISS Steering Committee for seeing the potential of our study concept and selecting it.

All of the participants dedicated their time, enthusiasm, and contributions to the workshop and report, for which we are deeply grateful. The workshop was memorable, enjoyable, and has laid the foundations for pioneering collaborations between the optical interferometry and lunar surface communities.  These collaborations would not have transpired without the spark of creativity, fanned into fire, by this KISS workshop.



\pagestyle{empty} 

\tableofcontents 



\pagestyle{fancy} 

\cleardoublepage 


\part{The Opportunity}


\chapterimage{AS16-114-18433HR_crop.jpg} 
\chapterspaceabove{8.25cm} 
\chapterspacebelow{7.75cm} 


\chapter{Introduction}

\section{Executive summary}

\begin{figure}
\begin{subfigure}[h]{0.48\linewidth}
\includegraphics[width=\linewidth]{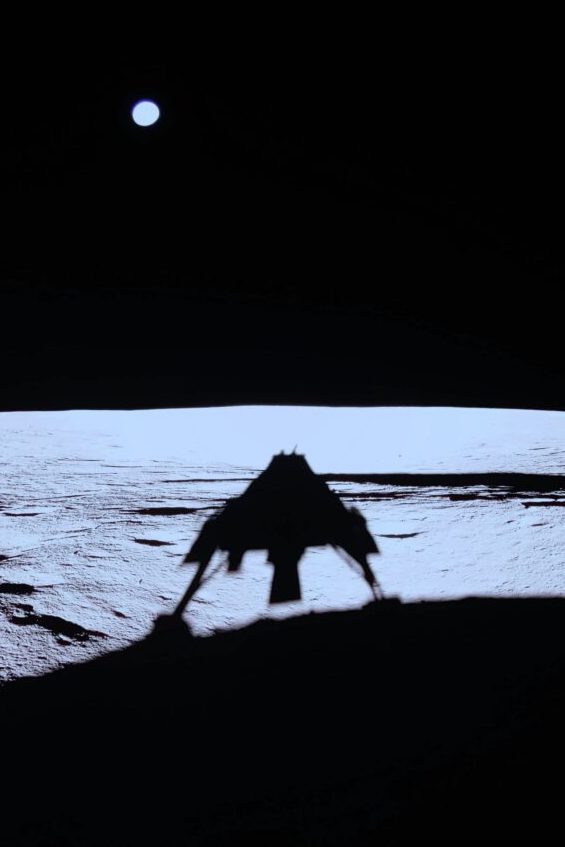}
\caption{Surface image from Firefly Aerospace's Blue Ghost Mission 1, which landed successfully on the Moon on 2 March 2025, at Mare Crisium with a dozen Commercial Lunar Payload Services (CLPS) science payloads on board \citep{firefly_mission_conclusion}. (Image credit: Firefly Aerospace)}\label{fig-firefly}
\end{subfigure}
\hfill
\begin{subfigure}[h]{0.48\linewidth}
\includegraphics[width=\linewidth]{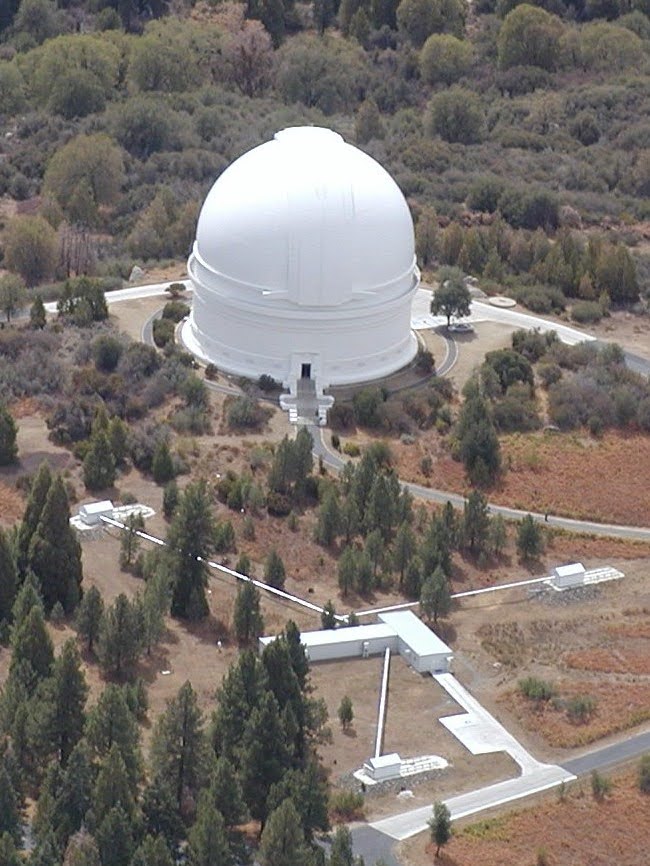}
\caption{The Palomar Testbed Interferometer (lower Y-shaped facility), which pioneered advanced optical interferometry techniques and operations from 1996 to 2008. (Image credit: Gerard T. van Belle)}\label{fig-PTI}
\end{subfigure}%
\caption{Motivations for the Keck Institute for Space Studies Study Program: rapid advances in lunar surface access (left) with missions such as those from NASA CLPS, along with multiple decades of demonstrated operations of terrestrial optical interferometry, such as the Palomar Testbed Interferometer\index{missions!Palomar Testbed Interferometer} (right).}
\label{fig-motivations_CLPS_interferometry}
\end{figure}

The lunar surface is a compelling location for large, distributed optical facilities, with significant advantages over orbital facilities for high spatial resolution astrophysics. The serious development of mission concepts is timely because of the confluence of multiple compelling factors. Lunar access technology is maturing rapidly, in the form of both US-based crewed and uncrewed landers, as well as international efforts (Figure \ref{fig-firefly}). Associated with this has been a definitive maturation of astronomical optical interferometry technologies at Earth-based facilities over the past three decades, enabling exquisitely sharp views on the universe previously unattainable, though limited at present by the Earth’s atmosphere (Figure \ref{fig-PTI}).  Importantly, the increasing knowledge and experience base about lunar surface operations indicates it is not just suitable, but highly attractive for lunar telescope arrays.

\begin{itemize}
    \item \textit{Unprecedented Imaging Potential}: Combining mature terrestrial optical interferometry with emerging lunar surface technologies could enable optical imaging with far greater resolution and sensitivity than current space or ground-based systems.

    \item \textit{Leveraging Existing NASA Funding}: NASA Astrophysics and Planetary Science programs could fund lunar interferometry missions through existing competitive processes, evaluated alongside orbital missions.

    \item \textit{Small-Scale Demonstration Opportunity}: A near-term, small mission---such as an Astrophysics Pioneers onboard a Commercial Lunar Payload System (CLPS) lander---could demonstrate the feasibility and value of lunar-based interferometry.

    \item \textit{Medium-Class Mission for Advanced Techniques}: A competitively selected medium-class mission (e.g., via Small Explorer or Medium Explorer missions) could enable precision interferometric methods like astrometry and nulling, supporting goals such as exoplanet reconnaissance.

    \item \textit{Large-Scale Mission for Breakthrough Science}: A Probe- or Flagship-class mission on the Moon could deliver unprecedented sub-milliarcsecond imaging across UV to MIR wavelengths, leveraging future lunar infrastructure for transformative astrophysics.
\end{itemize}

\section{Motivation}


This Keck Institute for Space Studies (KISS) Study Program was conducted to explore and establish the feasibility of mission concepts that can be realistically developed in the near term, within existing funding lines. The revolutionary opportunities of milli- to micro-arcsecond resolution science in the ultraviolet, visible, near-infrared, and mid-infrared are documented in this study report.  Additionally, this workshop report punctures myths about the lunar surface as a platform for astronomy.

By bringing together experts in the necessary yet disparate disciplines, we were able to highlight the specific opportunities afforded by lunar siting of interferometric telescopes. This convergence of expertise achieved the interdisciplinary focus required for demonstrating the realistic, immediate achievability of pioneering facilities. The study report highlights advances in the understanding of, and technology for, the lunar environment. This includes surface access, dust and thermal management, power and communication systems, and other issues. Specifically, the significant advantages for interferometer baseline management and simplification of input stations on the lunar surface are demonstrated. Within this framework, the already demonstrated capabilities of Earth-based interferometric facilities can be realized on the lunar surface, amplifying those capabilities for significant gains against the goals of the 2020 Astrophysics Decadal Survey (referred to as Astro2020 in this document).

The principal objective of this workshop was to assess the potential for lunar astronomical interferometry in the context of current flight opportunities and mission funding lines. A sober, wide-ranging assessment of the advantages and disadvantages of future lunar observatories was an important focus of this workshop.

Some notional "Big Questions" considered by this workshop included (with "Big Answers" in \S \ref{sec-Big-Questions-and-Answers}):
\begin{itemize}
    \item  What are key milestones on the way towards an interferometric lunar observatory?
    \item  What has changed in the last 5 to 10 years to make this a possibility?  What forthcoming developments will further enable this?
    \item  What can be done within the scope of each of the NASA Astrophysics funding lines---Pioneers, Small Explorer (SMEX), Medium Explorer (MIDEX), Probe, Flagship?
    \item  Are robotic or crewed missions best for the implementation of these ideas?
    \item  Are there implications that significantly impact the 2020 Astrophysics Decadal Survey, or could impact the next one?
    \item  What are the greatest challenges for, or misunderstandings about, astronomy from the lunar surface?
    \item  How do the cost, risk profiles, and science return of interferometric lunar observatories compare to orbital facilities?
\end{itemize}
A key outcome contained in this workshop report is to collect and document our findings in a comprehensive report for leaders and decision makers in the field.  Our intention is that the widest range of possible mission opportunities be available to competitive proposals, for uniquely addressing scientific questions of interest to the astrophysics community.

\subsection{Stakeholders}

During the KISS Workshop, the prospective stakeholders were kept in mind, which served to guide our discussions.  High on that list are, of course, astrophysicists and planetary scientists who would stand to benefit from these capabilities.  This includes the broader community, but also the specific interest groups such as professional societies (e.g., American Astronomical Society/AAS, Society of Photographic Instrumentation Engineers/SPIE, International Astronomical Union/IAU) and advisory groups (e.g., Exoplanet Exoploration Program Analysis Group/ExoPAG, Cosmic Origins Program Analysis Group/COPAG, Lunar Exploration Analysis Group/LEAG)  There is a significant intersection of those groups with organizations that would fund, design, and execute prospective missions.  This includes national and international space agencies (e.g., National Aeronautics and Space Administration/NASA, Japan Aerospace Exploration Agency/JAXA, European Space Agency/ESA, Canadian Space Agency/CSA), and potential industry partners that could implement these facilities, as prime contractors, supporting subcontractors, or even commercially contracted service providers.  Within space agencies are the discrete field centers (e.g., Jet Propulsion Laboratory/JPL, Goddard Space Flight Center/GSFC) that may play leading or supporting roles.  Along with potential government funding comes the possible involvement of related agencies (e.g., in the United States, the Office of Management and Budget/OMB and Office of Science and Technology Policy/OSTP).  Overarching all of these organizations with their specific areas of focus is the general public; spaceflight activities are a source of interest and inspiration for broad segments of the population, and lunar activities in particular carry that spark even further.















\chapterimage{science_drivers_crop.jpg} 
\chapterspaceabove{4.75cm} 
\chapterspacebelow{7.25cm} 

\chapter{Science Drivers}\index{Science Drivers}


Science cases for lunar optical interferometry are uniquely motivated in two ways: first, a lunar optical interferometer will deliver high spatial resolution relative to single-aperture space-based optical systems; and second, a lunar optical interferometer will enable significant strides in sensitivity relative to terrestrial optical interferometry.  The combination of high spatial resolution and high sensitivity carve out unique areas of discovery space, for which the following sections provide only a brief, incomplete exploration of the potential results.

\section{Ultra-high-resolution imaging}\index{Science Drivers!Imaging}\label{sec-ultra-high-res-imaging-science}
\begin{wrapfigure}{r}{8.75cm}
\vspace{-0.25cm}
\begin{minipage}[h]{1\linewidth}
\begin{tcolorbox}[colback=gray!5,colframe=green!40!black,title=Expanding the vision of astronomy]
Increased angular resolution has always led to new discoveries; improving our vision has let us see new and unexpected phenomena.
\end{tcolorbox}
\end{minipage}
\end{wrapfigure}
Advances in angular resolution have always led to advances in astrophysical knowledge.  While anticipated science cases are outlined below, it is important to note that the most striking examples are from completely unexpected findings.  From Galileo training the first astronomical telescope skyward and finding the rings of Saturn, to images of the "heart" on Pluto being discovered by New Horizons, angular resolution advances have always led to revolutionary discoveries.  The extreme resolution of optical interferometry has revealed to us: the oblate shape and gravity darkening of the star Altair \citep{vanBelle2001ApJ...559.1155V,Monnier2007Sci...317..342M}; that the brown dwarf Gliese 229B comprises two smaller "a and "b" components \citep{Xuan2024Natur.634.1070X}; and the non-Keplerian motion of stars around the black hole at the center of the Milky Way as they slingshot through distorted space-time \citep{GRAVITY2018A&A...618L..10G}\index{Instruments!GRAVITY}.
Imaging at milli- to micro-arcsecond scales, at new levels of sensitivity afforded by the lunar locale, will result in further paradigm-upending discoveries.

\begin{figure}
\includegraphics[width=1\textwidth, angle=0]{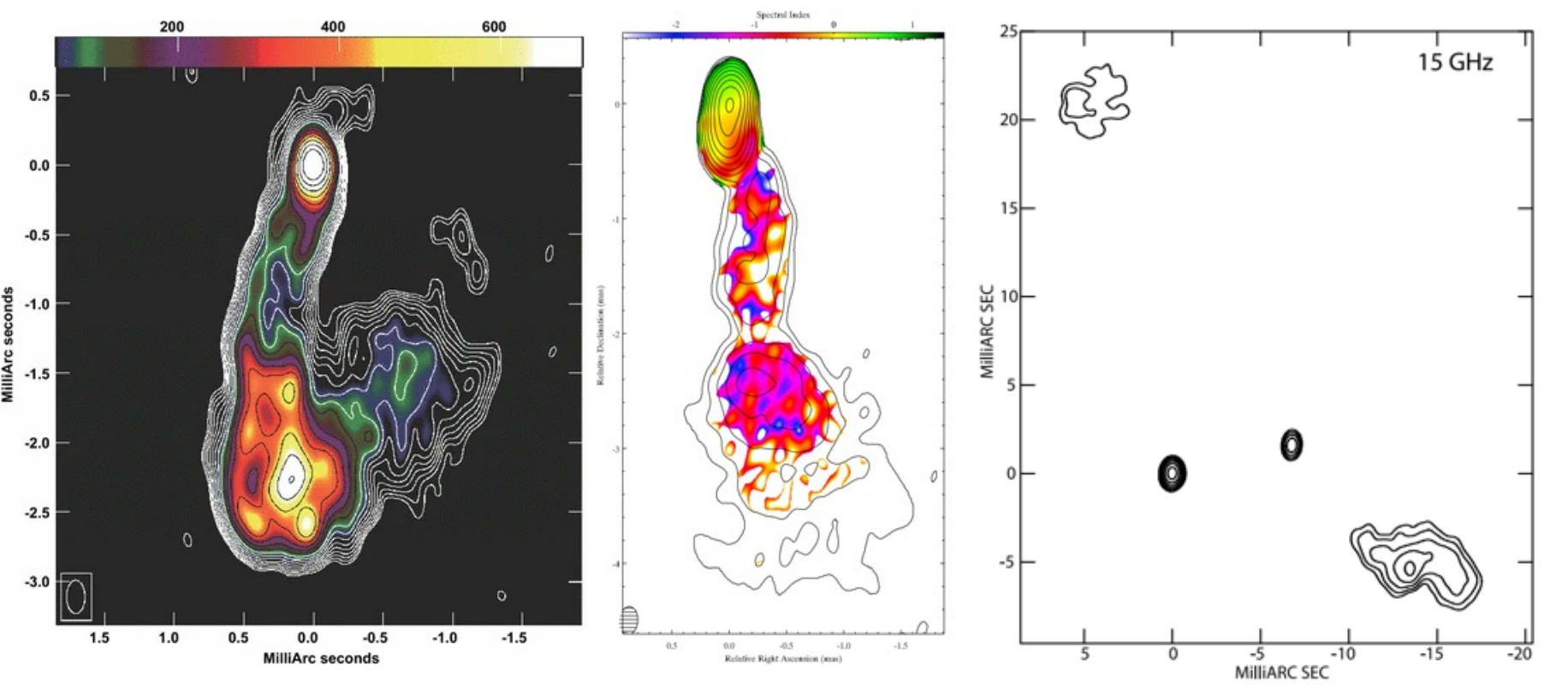}
\caption{ Radio VLBI observations of the inner regions of the radio jets in 3C~84 \cite{Nagai2014}, BL~Lacerta \cite{Gomez2016}, and the binary AGN candidate 0402+379 \cite{Rodriguez2006} (left to right).
The 3C~84 and BL~Lac jet structures, and the binary separation, are at milliarcsecond angular scales corresponding to $\sim$parsec linear scales.  The target brightnesses ($m_V \sim 12.5-17$) put them out of reach of terrestrial optical interferometry.
}
\label{fig:agn_examples}
\end{figure}

\textit{Active Galactic Nuclei (AGNs).}  The ability of supermassive black holes (SMBH)\index{Science Drivers!supermassive black holes} to launch powerful relativistic jets that can far exceed the sizes of their host galaxies, and with profound effects on those galaxies' evolution, has been an astrophysical mystery for decades.
By far, the dominant tool for studying the nuclear regions of the jet has been radio very long baseline interferometry (VLBI; Figure \ref{fig:agn_examples}), due to its ability to resolve on micro-arcsecond scales.
Studies have attempted to reconcile parsec-scale to kilo-parsec-scale misalignments in the radio structure using proxy methods for studying the optical structure; for example, \cite{Lambert2021} found that the optical centroids of quasars from Gaia coincide with downstream stationary radio features with high fractional polarizations in the jet, and that the optical emission on these scales arises from synchrotron emission in the jet.
\begin{figure}
\includegraphics[width=0.73\textwidth, angle=0]{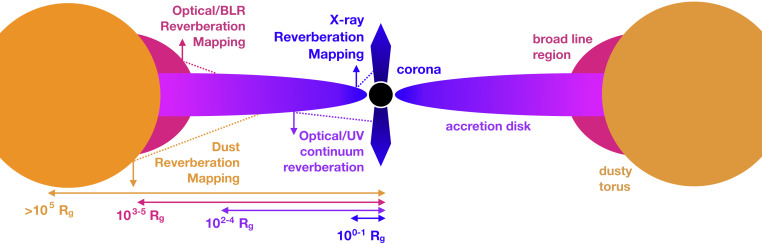}
\includegraphics[width=0.25\textwidth, angle=0]{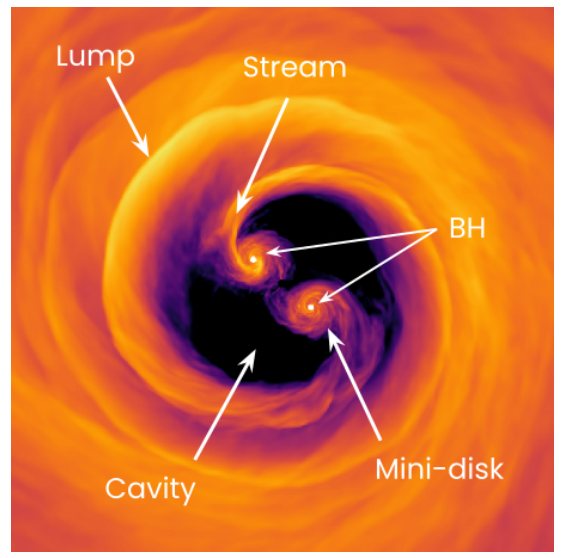}
\caption{Much of the expected interior structure of AGNs is inferred from much larger-scale imaging (Figure \ref{fig:agn_examples}); optical interferometry will be able to directly image these structures \citep{Cackett2021iSci...24j2557C,Gutierrez2024arXiv240514843G}.
}
\label{fig-agn_models}
\end{figure}
As SMBH in merging galaxies approach kilo-parsec to parsec-scale separations, when (or if) both black holes are activated by accretion and appear as a binary AGN is unclear \citep{Narayan1994ApJ...428L..13N,Blandford2019}. Until now, only radio interferometry was capable of measuring the binary fraction at the physical scales near 1 parsec, where the processes driving the merger are especially difficult theoretically. This is known as the "final parsec problem" \citep{Merritt2005LRR.....8....8M}.
Determining whether these SMBH radio binaries also appear in the optical as binaries would be the first step in answering important questions about the accretion process onto the binary pair at various separations.

\begin{figure}
    \begin{subfigure}[h]{0.53\linewidth}
        \includegraphics[width=0.95\linewidth]{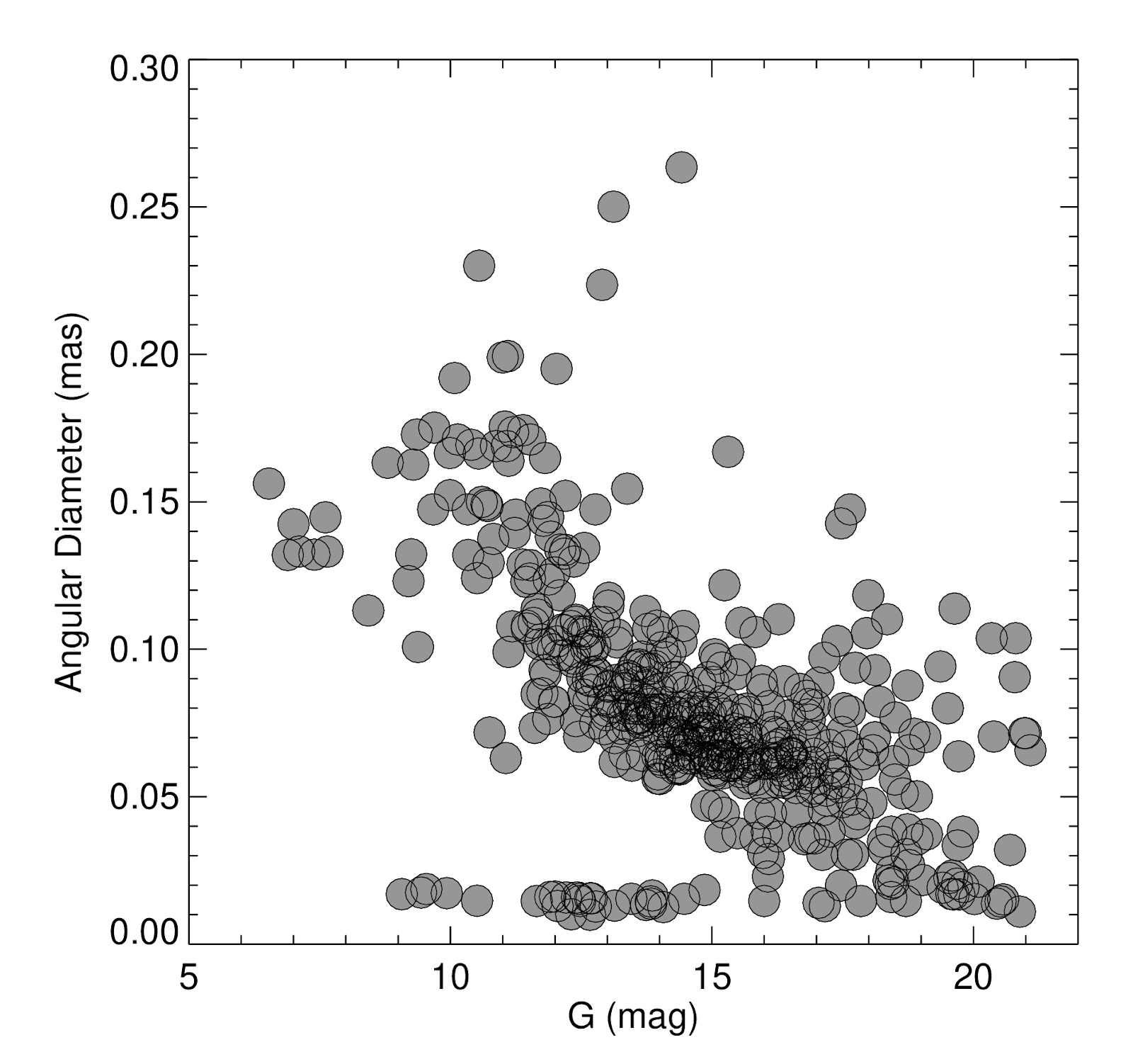}
        \caption{Estimated angular diameters of young stars in the Taurus star-forming region (140 pc). Membership selection from \cite{Krolikowski2021AJ....162..110K} with distances from \cite{Bailer-Jones2021AJ....161..147B} and sizes estimated from the evolutionary tracks from \cite{Baraffe2015A&A...577A..42B}.}\label{fig-YSO_diam_vs_Gmag}
    \end{subfigure}
\hfill
    \begin{subfigure}[h]{0.43\linewidth}
        \includegraphics[width=0.95\linewidth]{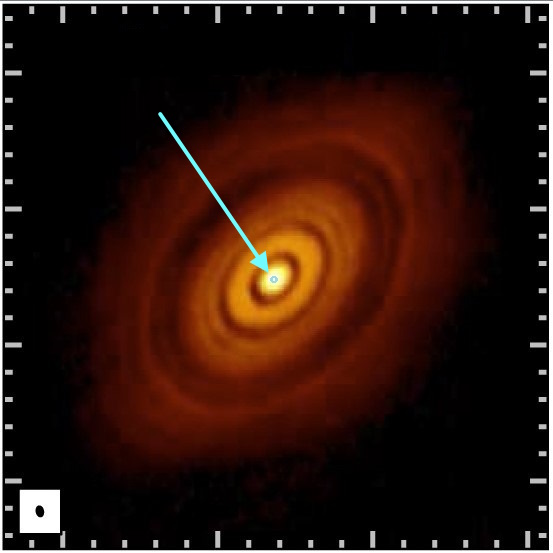}
        \caption{1.3~mm ALMA image of HL Tau \cite{ALMA2015ApJ...808L...3A}, with image scale $\sim$2" and linear scale of $\sim$200 AU edge-to-edge.  The $\sim$1 AU terrestrial planet region is inside the blue circle (highlighted by arrow) and requires significantly greater angular resolution.}\label{fig-HL_Tau}
    \end{subfigure}%
\caption{The angular resolution challenge of young stellar objects (YSOs).}
\end{figure}

\textit{Planet Formation.} The discovery of thousands of exoplanets\index{Science Drivers!exoplanets} in a multitude of architectures has challenged our theories of planet formation.  Given the ubiquity of exoplanets, planet formation must be a highly efficient process \citep{Burke2015ApJ...809....8B}, but theories that describe the formation and evolution of planets from protoplanetary disks around pre-main sequence stars have been poorly constrained because of a lack of sensitivity and resolution at the scales of planet formation (Figure \ref{fig-YSO_diam_vs_Gmag}). The 2020 Astrophysics Decadal Survey \citep{NASEM_Decadal_2021pdaa.book.....N} identified an understanding of the pathway to a habitable planet as one of the priority areas for the coming decade, and investigating how planets form and interact with their primordial disk and the pre-main sequence host star is a crucial element in understanding the morphology of our own solar system and the diversity of exoplanetary systems that have been discovered \citep{Raymond2022ASSL..466....3R}.  These systems are beginning to be explored with sub-millimeter observations (Figure \ref{fig-HL_Tau}), but missions such as MoonLITE\index{missions!MoonLITE} (\S \ref{sec-strawman-small}, \citep{vanBelle2024SPIE13092E..2NV}) can explore these YSOs in the visible, with higher resolution, for the first time.

\begin{figure}
    \centering
    \includegraphics[width=0.95\linewidth]{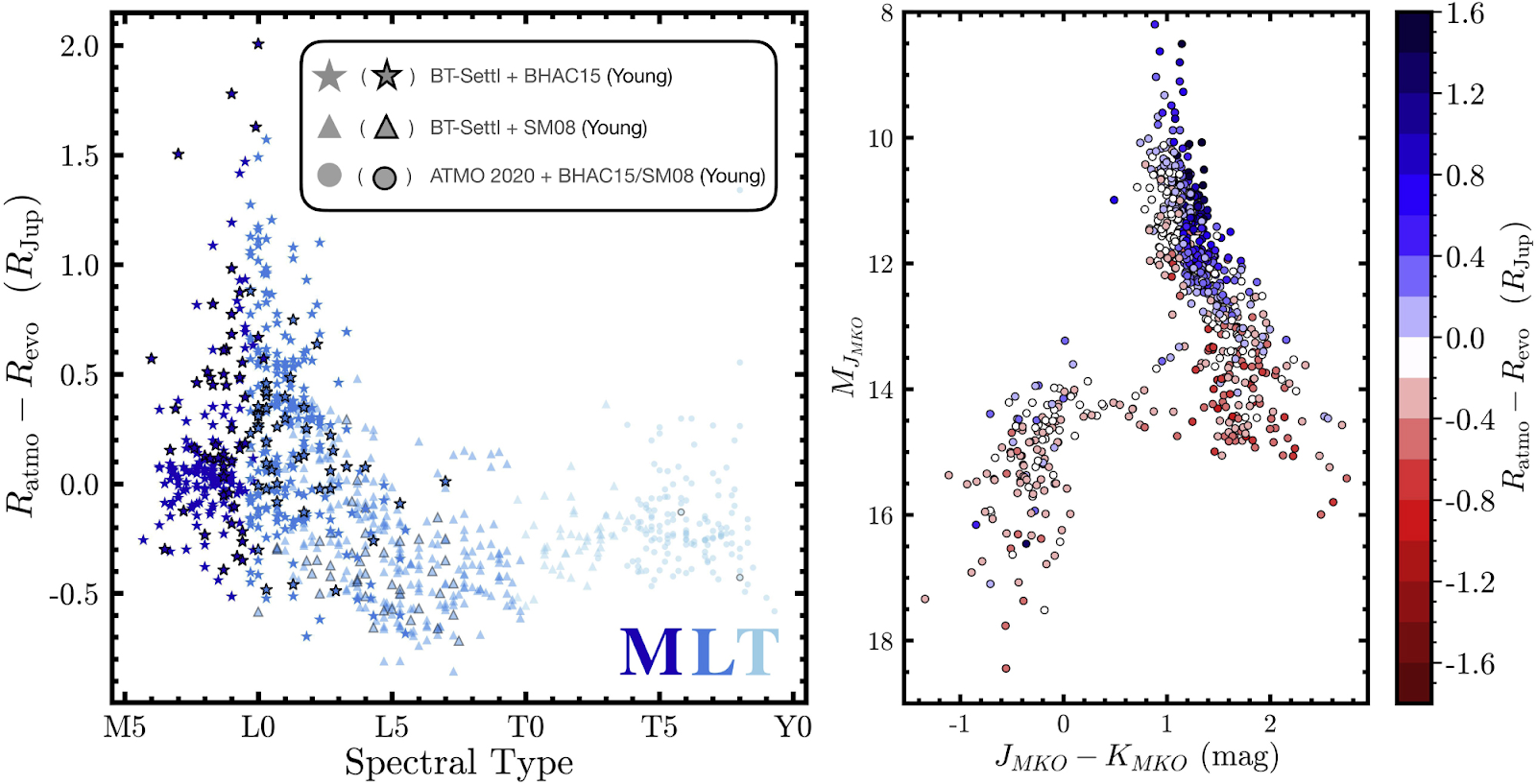}
    \caption{Difference between brown dwarf radii estimated with evolutionary and atmospheric models as a function of spectral type. Significant discrepancies exist at the M-to-L type transition boundary \cite{Sanghi2023ApJ...959...63S}.}
    \label{fig-brown_dwarfs}
\end{figure}

\begin{figure}
    \begin{subfigure}[h]{0.53\linewidth}
        \includegraphics[width=0.98\linewidth]{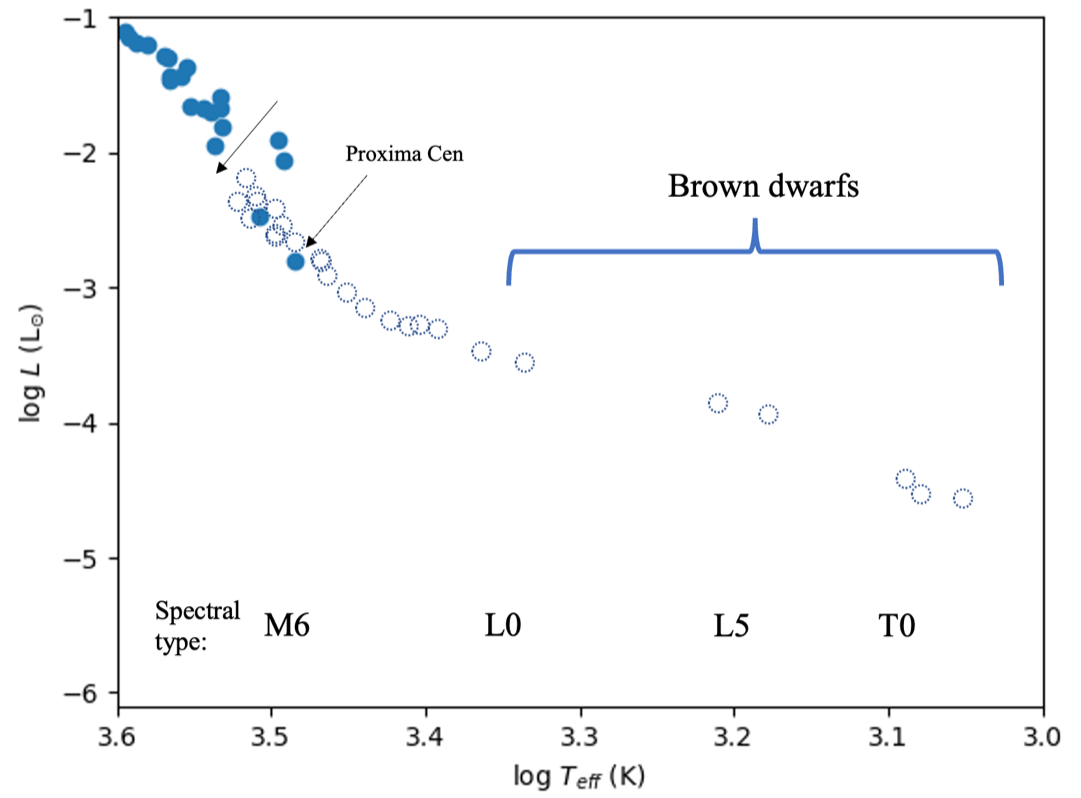}
        \caption{Known brown targets (dotted circles) for which a Pioneers-class mission such as Moon Lunar InTerferometry Express (MoonLITE) could directly measure angular sizes and empirically establish the effective temperature scale, extending that scale well beyond its current sensitivity-limited range (filled circles).} 
    \end{subfigure}
\hfill
    \begin{subfigure}[h]{0.43\linewidth}
        \includegraphics[width=0.98\linewidth]{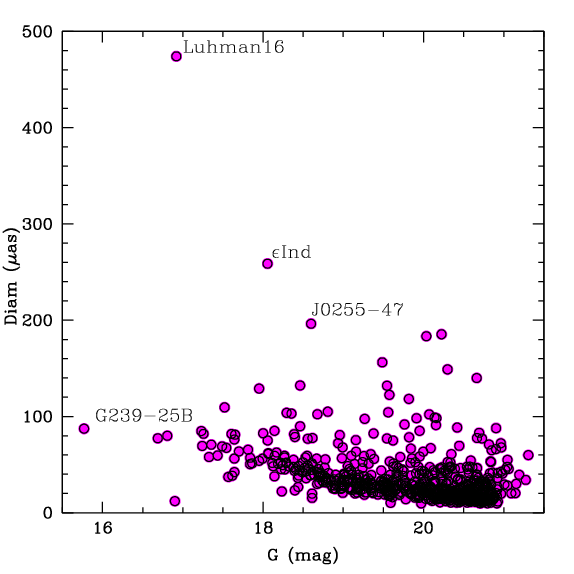}
        \caption{For a brown dwarf at 10 pc, this corresponds to an angular diameter of $\sim$0.1 mas ($\sim$100 $\mu$as). Predicted approximate angular diameters of nearby L/T brown dwarfs with Gaia Data Release 3 (DR3) G-band photometry and assuming radii of 1 $M_{Jup}$.}
    \end{subfigure} 
\caption{The combination of unprecedented angular resolution for a space-based optical system, with sensitivity unrivaled by terrestrial optical interferometers, will lead to empirical constraints into the nature of brown dwarfs.}
\label{fig-BD_diam_vs_Gmag}
\end{figure}

\textit{Surface imaging of brown dwarfs.}  Brown dwarfs\index{Science Drivers!brown dwarfs} show significant variability with indications that their top-level atmospheres have clouds, bands, and hotspots, rotating in and out of view \citep{Vos2023ApJ...944..138V,Zhou2022AJ....164..239Z}.  Suggested images of Luhman 16B  in comparison to Jupiter \citep{Apai2021ApJ...906...64A} are already examining the role of significant, persistent bands.  The nominal Jupiter-scale linear sizes of brown dwarfs mean that surface imaging at 0.1 to 1.0 mas should be possible for roughly a dozen nearby objects (Figures \ref{fig-brown_dwarfs}, \ref{fig-BD_diam_vs_Gmag}).  Fully resolved disk imaging can be preceded by simple size measurements (see \S \ref{sec-brown-dwarf-sizes}), which will already provide revealing discoveries.

\begin{wrapfigure}{R}{0.42\textwidth}
    \centering
    \includegraphics[width=0.41\textwidth]{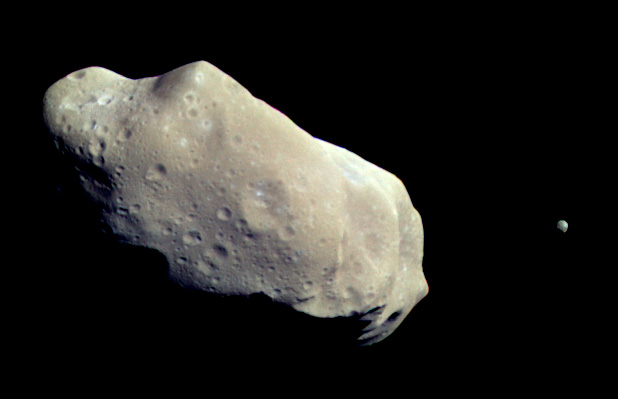}
    \caption{\label{fig-ida}The asteroid (243) Ida, and its Moon Dactyl (right)--which was serendipitously discovered from high-resolution imaging via a \textit{Galileo} close encounter.  (Image credit: NASA/JPL)}
\end{wrapfigure}

\textit{Physics of stellar explosions, and the extragalactic distance scale.}\index{Science Drivers!supernovae}  The ejecta of young supernovae undergo unconstrained expansion for a few hundred years after the supernova itself.  The initial expansion itself may have ejecta velocities of 5,000--10,000~km~s$^{-1}$ \citep{Mulligan2017MNRAS.467..778M}.  This expansion may be observed from the very initial stages, i.e., from when the shell is only 0.005 pc in diameter (assuming a distance of 0.6 Mpc) in galaxies in the Local Group \citep{Pun2002ApJ...572..906P}.  Using velocity information on the ejecta from spectroscopy, the distance to the supernova, and therefore to the host galaxy, is easily inferred.  This technique has already been applied with success to infer the distance to the Large Magellanic Cloud (LMC) (with SN 1987A) using high-resolution HST data \citep{Panagia1997AAS...191.1909P}, and to M81 (SN 1993J) using radio VLBI \citep{Bartel2004cosp...35.4112B}.  Radio VLBI, although successfully applied in the case of M81 \citep{Bietenholz2003ApJ...597..374B}, is not a guaranteed technique because a subset of supernovae do not become powerful radio emitters until much later in their dynamical expansion stage.  Shell diameter measurements out to the Virgo Cluster at 10 Mpc would be possible at 10-15 years after outburst.
Based on expected supernovae rates \citep{vandenbergh1991ARA&A..29..363V}, there should be an observable supernova within the Local Group once every 5 to 20 years, with a much greater frequency outside the Virgo Cluster.

\textit{Solar system exploration.}\index{Science Drivers!asteroids}  A milli- to micro-arcsecond interferometric imaging system could provide a capability to expansively and expeditiously explore small bodies in the solar system.  Such a system could provide "flyby without the flyby" (Figure \ref{fig-ida}) capability for investigating hundreds, if not thousands, of small bodies across the solar system, sidestepping the need for tasking individual spacecraft (with relatively small telescopes) for close encounters with these objects.  The surface morphology of small bodies can inform the impact histories \citep{cho2020}, regolith and surface process \citep{Harris2016ApJ...832..127H}, internal structure \citep{Busch2011Icar..212..649B}, geologic activity \citep{Jewitt2017AJ....153..223J}, composition and formation \citep{McCoy2025Natur.637.1072M}, and Yarkovsky-O'Keefe-Radzievskii-Paddack (YORP) effects \citep{Bottke2006AREPS..34..157B}.  Exploring asteroid binarity at high resolution can provide insights into asteroid formation and evolution \citep{Walsh_2015_article}, their internal structure and composition \citep{Barnouin2024NatCo..15.6202B}, and the gravitational dynamics and tidal evolution of these bodies \citep{Margot2015aste.book..355M}.

\subsection{Single baseline science}\index{Science Drivers!Single baseline science}\label{sec-single-baseline-science}

\begin{wrapfigure}{r}{8.25cm}
\vspace{-0.25cm}
\begin{minipage}[h]{1\linewidth}
\begin{tcolorbox}[colback=gray!5,colframe=green!40!black,title=Simple yet significant]
Even a two-aperture interferometer could reveal unique discoveries.
\end{tcolorbox}
\end{minipage}
\end{wrapfigure}
Single baseline interferometric observations are the simplest case of "imaging," available from a simple two-element interferometer.

The high precision and sensitivity of a lunar-based interferometer open up the parameter space inaccessible to terrestrial facilities.
 Factoring in the extended wavelength coverage available from the UV to the IR, the already-powerful terrestrial science case becomes even more compelling.



\subsubsection{Direct measurement of brown dwarf sizes}\index{Science Drivers!brown dwarfs}\label{sec-brown-dwarf-sizes}

Brown dwarfs are substellar objects more massive than gas giant planets (13 $M_{Jup}$; \cite{Spiegel2011ApJ...727...57S}) but less massive than stars (70 $M_{Jup}$; \cite{Dupuy2017ApJS..231...15D}). All brown dwarfs fuse deuterium \cite[e.g.,][]{Saumon1996ApJ...460..993S}, and those between 60 and 75 $M_{Jup}$ will also burn lithium \cite{Basri1998ASPC..134..394B}. Over 4000 brown dwarfs have been discovered as either free-floating objects or companions to hydrogen-fusing stars (as tracked in the UltracoolSheet \cite{Best2021AJ....161...42B, Best2024ApJ...967..115B} and Substellar and IMaged PLanet and Explorer Archive of Complex Objects (SIMPLE) databases \cite{Rodriguez2024ASPC..535...33R}).
Brown dwarfs cool throughout their lifetimes, progressing through a wide temperature range spanning the M, L, T, and Y spectral types \cite[e.g.,][]{Burrows1997ApJ...491..856B,Kirkpatrick2005ARA&A..43..195K,Cushing2011ApJ...743...50C}. This continuous cooling means that, for a given luminosity and effective temperature, and hence spectral type, the age and mass of a brown dwarf are not uniquely determined. Such an observational degeneracy makes it challenging to use colors, magnitudes, and spectral types to determine the physical properties of brown dwarfs, such as mass, age, and radius.

Typically, determining the fundamental properties relies on the use of models. These come in two flavors: atmospheric models and evolutionary models.  The bolometric luminosities derived by atmospheric model fits to brown dwarf spectral energy distributions (SEDs) can introduce systematic uncertainties, since such measurements depend on the fidelity of the fitting procedures adopted, the sampling of the model grid, and the differing input physics between various atmospheric models \cite[e.g.,][]{Stephens2009ApJ...702..154S,Zhang2021ApJ...921...95Z}. The other approach is to build and directly integrate a multiwavelength SED for brown dwarfs to empirically determine a bolometric luminosity \cite[e.g.,][]{Filippazzo2015ApJ...810..158F,Sanghi2023ApJ...959...63S} and combine it with an age estimate and evolutionary model to estimate masses and radii. This approach is limited by the uncertainty on the ages of brown dwarfs and is still model-dependent. Presently, the typical uncertainty on estimates of brown dwarf radii from this approach is $\sim$13\% (Figure \ref{fig-brown_dwarfs}); lunar optical interferometry will improve this by a factor of 10$\times$ (Figure \ref{fig-BD_diam_vs_Gmag}).

\subsubsection{Star and planet formation, and the evolution of young stars} \index{Science Drivers!star formation}\index{Science Drivers!planet formation}\index{Science Drivers!YSOs}


Young stars form in giant molecular clouds of gas and dust \citep{Chevance2023ASPC..534....1C}. As material accretes onto a dense protostellar core, gravitational contraction will cause the temperature and pressure in the core to increase until hydrogen burning is initiated. Over a few million years, the radius of a young star will decrease until the stellar radius stabilizes as it settles on the main sequence, where the outward pressure of nuclear fusion balances the inward pressure of gravitational collapse \citep{Hartmann1998ApJ...495..385H}. Direct measurement of the contraction of young stars will test models of stellar evolution at the youngest pre-main sequence ages \citep{Mathieu2007prpl.conf..411M}. The angular sizes of most young stars in nearby star-forming regions are just beyond reach of the current generation of terrestrial interferometers, both in terms of sensitivity and resolution.  As seen in Figure \ref{fig-YSO_diam_vs_Gmag}, extending interferometer sensitivity from the low to high double-digit magnitudes will greatly expand investigations of these objects into much more representative samples.


\subsubsection{Additional single-baseline science cases}\label{sec-additional-single-baseline-cases}

There is a wide variety of additional science cases for even the simple single-baseline size measurement case, as sensitivity increases for lunar- or space-based facilities over terrestrial facilities.  These cases are further extended when considering wavelength regimes that are problematic or outright inaccessible from the Earth's surface, from the ultraviolet to the mid-infrared, and across notable blind spots in between, such as the atmospheric water bands.  Such cases include:


\begin{itemize}
    \item \textit{White dwarf diameters.}  The diameter of the two nearest white dwarfs\index{Science Drivers!white dwarfs}, Sirius B and Procyon B, are expected to be $\sim$30$\mu$as, within reach of a $\sim$few hundred meter baseline operating in the UV.  Discrepancies in the mass of Sirius B as determined by the gravitational redshift method \citep{Joyce2018MNRAS.479.1612J}, astrometric methods \citep{Bond2017ApJ...840...70B}, and the theoretical mass-radius relation \citep{Fontaine2001PASP..113..409F} indicate that a direct radius measurement is required to resolve the disagreements and provide new insights on the physics of degenerate matter.
    \item \textit{Direct measure of Hubble Constant.} When combined with a redshift and a known physical scale, measuring the angular diameter of objects at cosmological distances allows direct inference of distances and the Hubble Constant, independent of other techniques that are currently in tension. In the case of AGN, the physical size of the accretion disk can be inferred via reverberation mapping \citep{Cackett2021iSci...24j2557C}, and the angular size via optical interferometry \cite{Dalal2024PhRvD.109l3029D}. For supernovae, the physical scale can be inferred via the photosphere velocities measured with spectroscopy, and the angular diameter evolution can be time resolved with optical interferometry \citep{Kim2025PhRvD.111h3047K}. Generically, single-baseline optical interferometry of faint objects enables access to bright targets at meaningful cosmological distances, and combining these measurements with other data and well-understood physics will enable precision distance measurements with independent and uncorrelated systematic uncertainties.\index{Science Drivers!Hubble constant}
    \item \textit{Measurement of the Planck scale.}  It has been hypothesized that the granular nature of spacetime foam at the Planck scale\index{Science Drivers!Planck scale} ($\sim10^{-35}$ meters) has a de-cohering effect on electromagnetic (EM) waves propagating across cosmological distances \cite{Lieu2003ApJ...585L..77L}.  While disputed in subsequent works \citep[e.g.,][and references therein]{Maziashvili2016PhRvD..94l4044M}, testing this theory can place significant limits on models for electromagnetic propagation through spacetime foam \cite{Ng2022Univ....8..382N,Carlip2022arXiv220914282C}.  The observational experiment is straightforward: observe point-like extragalactic sources such as quasi-stellar objects (QSOs) and investigate if there is a systematic reduction in point-source visibility with increasing redshift indicative of non-coherence caused by the fabric of space.

\end{itemize}

\section{Astrometry}\index{Science Drivers!astrometry}\index{Astrometry!Science drivers}\label{sec-astrometry}

\begin{figure}
    \centering
    \includegraphics[width=0.95\linewidth]{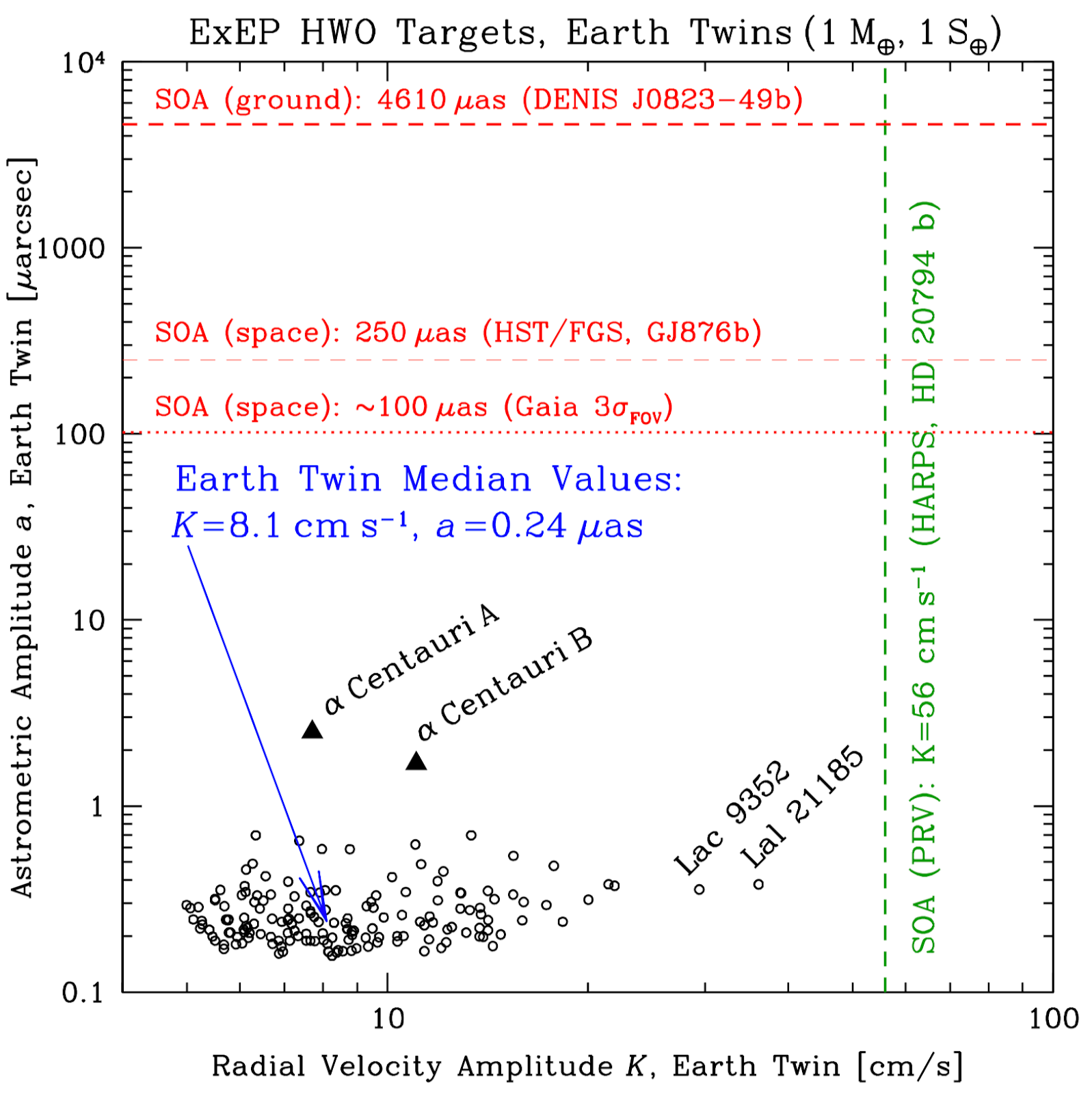}
    \caption{State-of-the-art (SOA) astrometric reflex motion signature versus radial velocity amplitude signature for possible Earth twins to be observed by HWO.  The current state of the art for RV detections is well above the $\sim$1-10 cm/s needed (green line), and could be fundamentally limited by stellar RV jitter noise.  Astrometric single-measurement precision at the 0.1$\mu$as level, enabled by $>$100~m lunar interferometric baselines, would guide HWO observations and provide masses for atmospheric retrievals from HWO spectra.} 
    \label{fig-astrometric_exoearths}
\end{figure}

Astrometric single-measurement accuracy at the 0.1-$\mu$as level would be a transformative capability in the search for true Earth-analog exoplanets.  Advances of the current state-of-the-art in radial velocity (RV) searches have stalled at the $\sim$50~cm/s level (Figure \ref{fig-astrometric_exoearths}), roughly an order of magnitude too large for detection of such exoplanets.  This impasse is due to stellar surface RV jitter noise, which stubbornly refuses to be characterized as a subtractable signal in RV data \citep{Blackman2020AJ....159..238B,Zhao2022AJ....163..171Z}.

Astrometric jitter for a Sun-like star, in contrast, is primarily expected to be due to starspots, has been modeled to be at the single-epoch level of $\leq \pm 0.5$ milli-$R_\odot$ = 0.25~$\mu$as for a sun-like star in the visible at 10~pc \citep[][and references therein]{Shapiro2021ApJ...908..223S}, and averages down to 0.05~$\mu$as over $\sim$80 days
\citep{Sowmya2021ApJ...919...94S}.
Sun-as-a-star observations \citep{Makarov2010ApJ...717.1202M,Lagrange2011A&A...528L...9L} indicate the astrometric jitter of our own host star is consistent with this modeling.  Errors at this level indicate no fundamental astrometric detection systematic error for exo-Earths (Figure \ref{fig-astrometric_exoearths}).
Rapidly rotating stars are expected to be a factor of $\sim$10$\times$ worse \citep{Sowmya2022ApJ...934..146S}, but these can be easily vetted via spectroscopy.



Advances in our ability to precisely measure the positions, motions, and parallaxes (and thereby distances) of celestial objects have traditionally led to major advances in astronomy and our understanding of the universe. The ESA Hipparcos \citep{Perryman1997A&A...323L..49P} and Gaia \citep{Gaia2016A&A...595A...1G} missions advanced our knowledge of distances and motions of the brightest hundreds of thousands and billions of stars, leaping from the milli-arcsecond accuracy regime to the hundreds to tens of micro-arcsecond regime, respectively.

\subsection{Masses and orbits to inform the Habitable Worlds Observatory (HWO)}\label{sec-masses-for-HWO}

%

\begin{wrapfigure}{r}{8.25cm}
\vspace{-0.25cm}
\begin{minipage}[h]{1\linewidth}
\begin{tcolorbox}[colback=gray!5,colframe=green!40!black,title=Reconnaissance for HWO]
HWO has a fundamental hurdle from the outset: Where to point?
\end{tcolorbox}
\end{minipage}
\end{wrapfigure}
While Gaia is beginning to enable the detections of some large, Jupiter-sized planets \citep{Sozzetti2023A&A...677L..15S,Sozzetti2024eas..conf.1626S}, a major, $>100\times$ leap in astrometric accuracy to the sub-micro-arcsecond regime is needed to enable the efficient detection of small, Earth-mass planets on temperate orbits around nearby Sun-like stars. 
The Astro2020 Decadal Survey \citep{NASEM_Decadal_2021pdaa.book.....N} had a key recommendation to not just find $\sim$25 habitable zone planets, but to search them for biosignatures.
The recommended NASA mission concept, the Habitable Worlds Observatory\index{missions!Habitable Worlds Observatory}, (HWO) \citep{Feinberg2024SPIE13092E..1NF} will take advantage of significant advancements in key technologies over the past decade spurred by investments recommended by the Astro2010 Decadal Survey (e.g., starlight suppression technologies, coronagraph designs, deformable mirrors, etc.) \citep{2010nwnh.book......} and advancements in astrobiology, planetary and exoplanetary science (e.g., planetary atmospheres, interiors, evolution, Earth science, exoplanet observations and analysis), to design a mission to search for and characterize potentially habitable worlds orbiting nearby stars.
Astrometric single-measurement precision at the 0.1$\mu$as level, enabled by $>$100~m lunar interferometric baselines, would guide HWO observations and provide masses for atmospheric retrievals from HWO spectra.

Although this is a key recommendation of the Astro2020 Decadal Survey, it is important to acknowledge we don't even know if these planets exist, or even currently have a plan in advance of HWO to find out where they are, and what their characteristics are.
While it seems increasingly reasonable to presuppose that such objects orbit nearby stars, preparatory science activities are urgently needed to inform a future exo-Earth survey with HWO by the time it launches, to enhance the science return of the flagship mission, and to enable the analysis of the imaging and spectroscopy of exoplanets (including potentially Earth-like planets). Using HWO to find HWO targets is a plausible but extravagantly inefficient and expensive prospect.  Early exoplanet science yield simulations \cite{Morgan2021JATIS...7b1220M} for space-based direct imaging concepts analogous to HWO find that prior knowledge of the existence and orbital properties of exoplanets can increase total exoplanet science yield by many tens of percent, and rapidly accelerate the timescale to achieve science yield for the mission.

\subsection{Other science enabled by micro-arcsecond astrometry}\index{Science Drivers!astrometry}

Additional astrophysics is expected from astrometry at the sub-microarcsecond level, as characterized in detail by the science expectations for the unflown Space Interferometry Mission \citep{Unwin2008PASP..120...38U}.  This includes a broad menu of stellar and galactic astrophysics, as well as the studies of quasars and AGNs.  At this level, planets can also be detected around young stars---quite challenging via radial velocity due to intrinsic noise of these stars \citep{Grandjean2020A&A...633A..44G}----which will provide insights into how planetary systems are born and the timescales on which they evolve.  Observing the cores of AGNs at these levels will characterize the dynamics of central accretion disks around SMBH\index{Science Drivers!supermassive black holes}, and their emergent relativistic jets; photo-center motion at the $\sim$10~$\mu$as level \citep{Unwin2008IAUS..248..288U} will potentially be seen at one-year timescales.

\section{Nulling}\index{Science Drivers!nulling}

Although indirect planet detection methods (such as transits and radial-velocity observations) have revolutionized our understanding of the ubiquity and diversity of exoplanets, the promise of directly imaging exoplanets, particularly those in the habitable zone (the region around a star where liquid water and possibly life can exist) has until now been out of reach due to the truly daunting observational challenge entailed. The overwhelming glare of the host star against the faint mote of planetary light imposes a punishing contrast requirement, while simultaneously, the high spatial resolutions needed are degraded by the Earth’s turbulent atmosphere. Overcoming these obstacles to usher in the era of direct imaging of exoplanets will reveal a wealth of information beyond that achievable with indirect methods, such as unambiguous orbital parameters (and hence planet masses). However, the most enticing scientific motivation for nulling is arguably the most sought-after in all of exoplanet research: the ability to isolate planetary light for study to recover full spectra, revealing the properties and chemistry of atmospheres, surfaces, and enabling the search for active biomarkers \citep{Monnier2019BAAS...51c.514M}.

Nulling interferometry from terrestrial facilities has emphasized longer wavelengths into the infrared, which has advantages both for the technology and the science. The seeing is more benign, allowing better phase tracking, while the exoplanetary science is more easily accomplished at long wavelengths where the contrast to the host star is somewhat less daunting. Although nulling can be readily accomplished with bulk-optic beam combiners (e.g., the Keck Interferometer Nuller\index{Instruments!KI Nuller} \citep{Serabyn2012ApJ...748...55S}, or the Large Binocular Telescope Interferometer (LBTI) nuller \citep{Hinz2014SPIE.9146E..0TH}), there has been growing interest in photonic beam combination for this science. While photonics offer many advantages by leveraging the compact, stable technology platform, the primary driver is the ability to deliver clean single-mode filtered channels that facilitate ideal starlight cancelation and therefore deep nulls. Non-nulled "bright" channels that provide an ideal metrology signal to facilitate stabilization of the fringes (see, for example, discussion of the Guided-Light Interferometric Nulling Technology (GLINT) instrument \citep{Norris2020MNRAS.491.4180N})\index{Instruments!GLINT}. Photonic nulling instruments have been successfully deployed at Palomar \citep{Serabyn2019MNRAS.489.1291S} and Subaru \citep{Martinod2021NatCo..12.2465M} telescopes.

In the lunar context, nulling technology and many of these design choices can be reassessed to present new opportunities unshackled by the limitations of the Earth's atmosphere. New transparency windows at all wavelengths are available from the Moon; observation at short wavelengths, potentially the blue/visible or even ultraviolet (UV), comes within scope, enabled by phase-stable conditions that particularly empower nulling architectures to tune for maximum cancelation of unwanted starlight. Mid-wavelength infrared windows are also potentially opened by low thermal noise backgrounds and transparency windows
unobtainable from the Earth's surface.













\section{Mature operations and current results}\label{sec-mature-operations}
\begin{wrapfigure}{r}{8.75cm}
\vspace{-0.25cm}
\begin{minipage}[h]{1\linewidth}
\begin{tcolorbox}[colback=gray!5,colframe=green!40!black,title=Optical interferometry's impact]
Over 1,000 refereed science articles have been published that utilize optical interferometry data, providing unique insights at the milli-arcsecond scale.
\end{tcolorbox}
\end{minipage}
\end{wrapfigure}
Operational terrestrial optical interferometry facilities, including Very Large Telescope Interferometer (VLTI)\index{missions!VLTI}, Center for High Angular Resolution Astronomy (CHARA)\index{missions!CHARA}, and Navy Precision Optical Interferometer (NPOI)\index{missions!NPOI}, soon to be joined by Magdalena Ridge Observatory Interferometer (MROI), have been on-sky in a routine sense since the 1990s.  The extensive bibliography of science
from the CHARA Array\footnote{\url{https://www.chara.gsu.edu/astronomers/journal-articles}} and from the VLTI\footnote{\url{https://www.eso.org/sci/facilities/paranal/telescopes/vlti/science.html}} shows the unique impact that milli- to micro-arcsecond scale spatially resolved observations has on astrophysics.
\pagebreak

The 10 most-cited optical interferometry papers since 2004 include (from most to least cited):
\begin{itemize}
    \item "A Survey of Stellar Families: Multiplicity of Solar-type Stars", Raghavan et al. 2010 \citep{raghavan2010}\index{Science Drivers!stellar multiplicity}
    \item "A geometric distance measurement to the Galactic center black hole with 0.3\% uncertainty", GRAVITY Collaboration et al. 2019 \citep{Gravity2019A&A...625L..10G}\index{Instruments!GRAVITY}
    \item "Stellar Diameters and Temperatures. II. Main-sequence K- and M-stars", Boyajian et al. 2012 \citep{Boyajian2012ApJ...757..112B}\index{Science Drivers!stellar diameters}
    \item "Southern Massive Stars at High Angular Resolution: Observational Campaign and Companion Detection", Sana et al. 2014 \citep{Sana2014ApJS..215...15S}\index{Science Drivers!massive stars}
    \item "Detection of orbital motions near the last stable circular orbit of the massive black hole SgrA*", GRAVITY Collaboration et al. 2018 \citep{GRAVITY2018A&A...618L..10G}
    \item "A diversity of dusty AGN tori. Data release for the VLTI/MIDI AGN Large Program and first results for 23 galaxies", Burtscher et al. 2013 \citep{Burtscher2013A&A...558A.149B}
    \item "Stellar Diameters and Temperatures. III. Main-sequence A, F, G, and K Stars: Additional High-precision Measurements and Empirical Relations", Boyajian et al. 2013 \citep{Boyajian2013ApJ...771...40B}
    \item "Mid-infrared sizes of circumstellar disks around Herbig Ae/Be stars measured with MIDI on the VLTI", Leinert et al. 2004 \citep{Leinert2004A&A...423..537L}\index{Science Drivers!YSOs}
    \item "Fundamental Properties of Stars Using Asteroseismology from Kepler and CoRoT and Interferometry from the CHARA Array", Huber et al. 2012 \citep{Huber2012ApJ...760...32H}\index{Science Drivers!asteroseismology}\index{missions!CHARA}
    \item "Stellar Diameters and Temperatures. I. Main-sequence A, F, and G Stars", Boyajian et al., 2012 \citep{Boyajian2012ApJ...746..101B}
\end{itemize}
This sampling of articles represents the "gold standard" of science that has come from the current generation of optical interferometers.  Even within the context of limited-sensitivity terrestrial observing, the collected science of these articles has had an outsized impact in shaping our current understanding of exoplanets, stars, black holes, and beyond.  In addition to demonstrating the readiness of optical interferometry technology for a lunar platform, these results portend an entirely new generation of results to come from high-resolution observing empowered by the high sensitivity from the lunar surface.

\part{The Technology}

%

\chapterimage{Apollo-14-Elements-de-l-ALSEP-AS14-67-9376_crop.jpg} 
\chapterspaceabove{6.75cm} 
\chapterspacebelow{7.25cm} 

\chapter{The Lunar Environment and Interferometry}




\section{Geologic context}
\begin{wrapfigure}{r}{8.25cm}
\vspace{-0.25cm}
\begin{minipage}[h]{1\linewidth}
\begin{tcolorbox}[colback=gray!5,colframe=green!40!black,title=Challenging but useful]
The lunar surface environment presents unique opportunities, with challenges that are surmountable.
\end{tcolorbox}
\end{minipage}
\end{wrapfigure}
The unique lunar geologic environment is an important consideration for any surface mission. Regolith properties, topography, surface roughness, boulder density, regolith albedo, lighting conditions, seismic environment, and areas of instability and landslip will factor into the daily and long-term operations of an instrument, and can have regional implications for a distributed campaign. Detailed topographic, geologic, and morphologic mapping campaigns at a variety of regional and local scales are necessary for landing site characterization.  A dominant surface process on the Moon throughout its history has been meteorite impacts\index{Lunar Environment!meteorite impacts}, and the resulting regolith properties need to be factored into landing site studies. At large scale, impacts have broken bedrock into pieces that blanket the majority of the Moon’s surface, with sizes ranging from $>$ 1~km to $<$ 1~nm .

Discussions around the effects of lunar dust and the abrasivity of regolith have their own sections within this chapter, though an important consideration of a surface blanket of regolith is its inherent instability. Many regions have been compacted over time via seismic vibrations, but regolith deposits on slopes are at risk of dry flows (landslips). The Apollo 17 mission landed next to one of these dry flows that was deposited on the floor of the Taurus-Littrow Valley from one of the surrounding mountains. Regions with dry flows or nearby steep slopes should be avoided. The upper 1-5~cm of the regolith is also unstable and compacts when pressure is applied, meaning deployed instruments will also initially settle by a few centimeters.

The Moon is generally more mountainous and topographically varied on the farside and smoother on the nearside, at regional scales \citep{Zhu2019JGRE..124.2117Z,Head2024GeoRL..5110510H}. The lunar South Pole is topographically rough, but contains sizable flatter regions poleward of -75$^o$ line of latitude. While it may be advantageous to deploy astronomical instruments at topographically high points, many of these are small and unstable or inaccessible, severely limiting the sites available for such deployment. The Lunar Reconnaissance Orbiter’s\index{missions!Lunar Reconnaissance Orbiter} (LRO) Lunar Orbiter Laser Altimeter (LOLA)\index{instruments!Lunar Orbiter Laser Altimeter} instrument provides topographical and slope maps that can be used to determine potential landing sites---these are available to view on the Lunar Reconnaissance Orbiter Camera\index{instruments!Lunar Reconnaissance Orbiter Camera} (LROC) Quickmap website\footnote{\url{https://quickmap.lroc.asu.edu/}} \citep{Smith2010SSRv..150..209S}.

On a similar note, boulder density\index{Lunar Environment!boulder density} and surface roughness are essential considerations when performing a landing site study. To avoid boulder-rich regions, data from the Diviner Lunar Radiometer Experiment  \cite[Diviner, aboard LRO][]{Paige2010SSRv..150..125P} can be used to derive the rock abundance at the surface and further constrain potential landing sites \cite{Bandfield2011JGRE..116.0H02B,Powell2023JGRE..12807532P}. Additionally, LRO Narrow-Angle Camera\index{instruments!Narrow-Angle Camera} (NAC) imaging can be used to identify and map boulders $<$1~m in diameter. Regolith properties can also be evaluated using H-Parameter analysis \cite{Hayne2017JGRE..122.2371H}, which gives an estimate of regolith density with depth. Again, these datasets are available on the LROC Quickmap resource, and datasets can be downloaded for performing Geographic Information System (GIS) queries and for tailoring analyses (e.g., changing boulder size in rock abundance maps) to refine the potential number of landing sites based on set mission requirements.

Line-of-sight to the Earth needs to be considered for communication requirements. The nearside will have permanent communications with Earth, as will many parts of the boundary regions between nearside and farside. The farside and parts of the lunar South Polar region will not have direct communication, though development of cis-lunar communications and additional lunar platforms will provide a reliable communications relay to the Earth. Additionally, the Earth will remain in the sky indefinitely in a specific location from a nearside site, which has to be taken into consideration for Earthshine, seeing, and sky viewing opportunities, as targets rotate by this location. This effect is reduced from a nearside-farside boundary location while maintaining direct communications with Earth. Finally, considering the other lunar science goals that can be addressed by visiting boundary/farside locations could provide other funding sources, collaborations, and more science objectives addressed per site.

Equatorial and polar regions have both advantages and disadvantages to consider as sites for conducting interferometry activities. Firstly, polar regions have limited sky views to a single hemisphere, though targeting both lunar North and South Poles with separate missions/landers will solve this issue. Polar regions also have permanently shadowed regions (PSRs)\index{Lunar Environment!permanently shadowed regions} and cold traps that could be used to the advantage of the mission (e.g., detector cooling). Equatorial regions have full sky views (minus Sun/Earth locations) but have day/night cycles with large temperature variations, on the order of 28 days with no cold traps or PSRs available for use. Equatorial regions are also easier to access from Earth with respect to fuel requirements and orbital trajectories. Finally, there are dust lifting and charging effects related to the day/night cycle that will have the greatest effect at the lunar equator. The lunar poles, with a low Sun angle, will experience far less extreme charge differences and dust levitation between day/night conditions.

\pagebreak
The leading hemisphere of the Moon experiences frequent impacts and, to some extent, enhanced maturation of the lunar regolith. Conversely, the trailing hemisphere has a 1.2--1.5$\times$ lower rate of impacts \citep{Gallant2009Icar..202..371G}, so this may be a consideration in landing site selection studies, particularly for larger flagship-level missions.

\section{Seismology} \label{sec-seismology}
\begin{figure}
    \centering
    \includegraphics[width=0.95\linewidth]{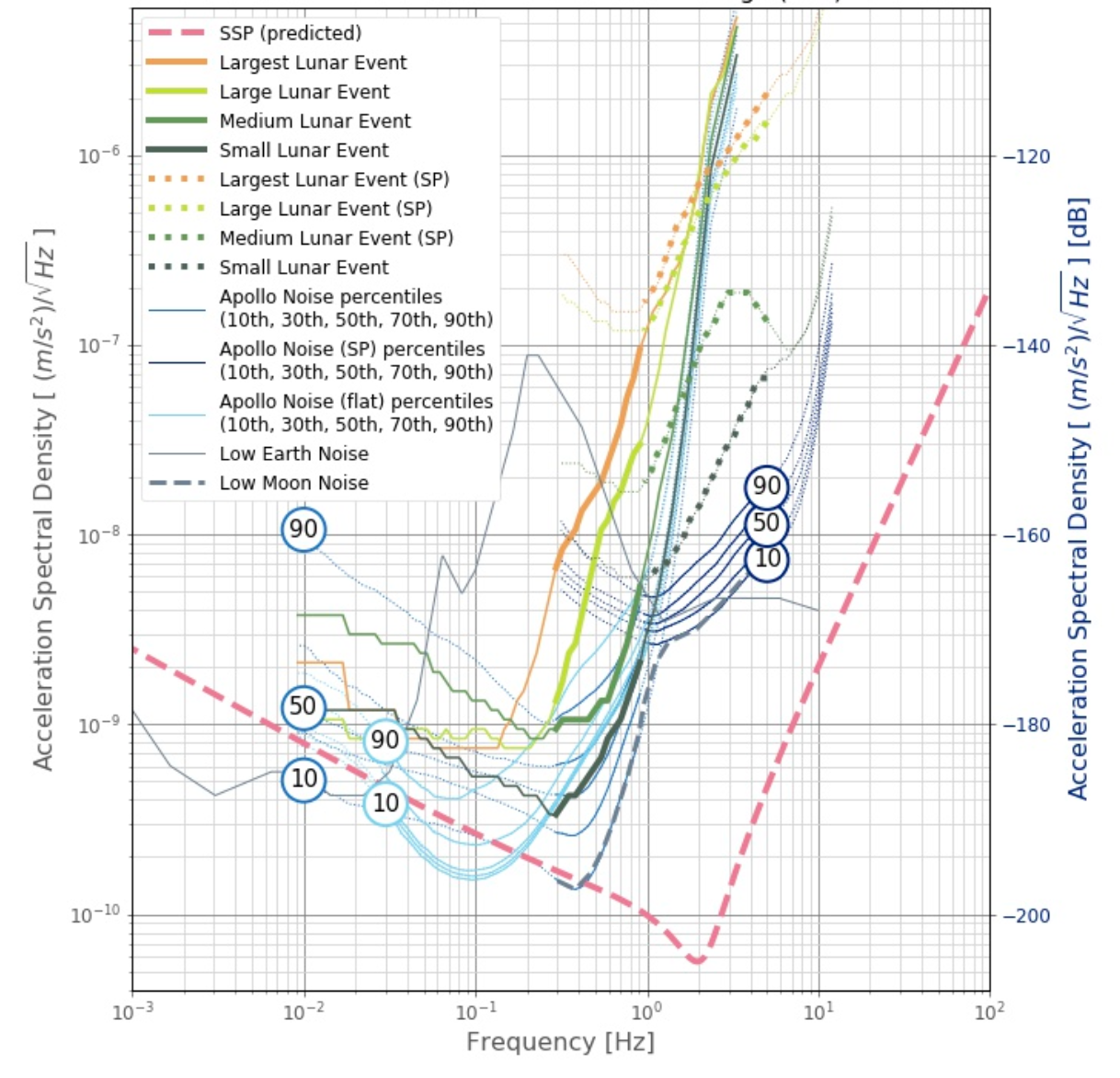}
    \caption{Amplitude spectral density of seismic signals based on Apollo data in units of acceleration spectral density, from \cite{Nunn2021PSJ.....2...36N}. Lines labeled with numbers are percentiles of noise levels recorded on the Apollo mid-period and short-period instruments. This represents an upper boundary of the lunar seismic noise environment, as this was likely limited by instrument self-noise rather than lunar noise. Solid and dotted lines represent spectra of individual Moonquakes and impacts that can exceed 100 nm/s$^2$ near 1 Hz (displacements of 2.5 nm).}
    \label{fig-SEIS1}
\end{figure}

\begin{figure}
    \centering
    \includegraphics[width=0.95\linewidth]{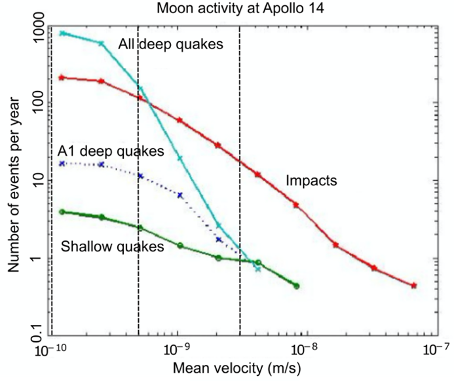}
    \caption{
    Events per year versus mean velocity (which is the same as displacements at 0.45~Hz in meters when divided by $\pi$) from Apollo 14 seismometer data.   The A1 cluster represents one of the most active deep moonquake source regions \citep{Nakamura1978LPSC....9.3589N}.  Impacts form the greatest risk, though impacts at the nm displacement level only occur at roughly biweekly intervals (right dashed line).  Moonquakes overtake impacts (left dashed line) at only the 0.1~nm displacement level. (Adapted from \cite{Lognonne2015trge.book...65L}).}
    \label{fig-SEIS2}
\end{figure}

\index{Lunar Environment!seismology}
\begin{wrapfigure}{r}{8.0cm}
\vspace{-0.25cm}
\begin{minipage}[h]{1\linewidth}
\begin{tcolorbox}[colback=gray!5,colframe=green!40!black,title=A stable surface]
Seismic disturbances on the Moon are generally below the level that would affect interferometric observations.
\end{tcolorbox}
\end{minipage}
\end{wrapfigure}
Various instruments from the Apollo Passive Seismic Experiments recorded seismic signals on the Moon between 1969 and when the instruments were turned off in 1977 \cite[e.g.,][]{Nunn2020SSRv..216...89N}. Figure \ref{fig-SEIS1} (from \cite{Nunn2021PSJ.....2...36N}) summarizes the noise levels recorded in those experiments. Note that the noise levels here are limited by Apollo instrument sensitivity. If we look only at the lowest noise level at the level of the Apollo mid-period sensor peaked mode near 0.45 Hz, we see it is at the instrument noise floor of $\sim$0.2~nm/s$^2$, which corresponds to $\sim$25~pm. Models of background noise driven by micrometeorite impacts \cite{Lognonne2009JGRE..11412003L} suggest actual lunar noise may be 2-3 orders of magnitude below Apollo instrument sensitivity. However, there are many events recorded in the Apollo seismic data that exceed these noise floors. \cite{Lorenz2018Icar..303..273L} looked at occurrence rates of motions exceeding various seismic thresholds, which showed the most frequent occurring motions above this background noise level are associated with "thermal Moonquakes," often interpreted as very local events due to thermal changes \cite{Duennebier1974JGR....79.4351D}, and these exhibited motions of the order of tens of nanometers occurring tens to thousands of times per year, although they occurred most frequently near dusk and dawn when temperatures were varying most rapidly. Other Moonquakes also showed similar or larger amplitudes \cite{Nakamura1981STIA...8311825N,Lognonne2015trge.book...65L}, but these occurred much less frequently on the order of a few times per year (Figure \ref{fig-SEIS2}).

The noise estimates in Figures \ref{fig-SEIS1} and \ref{fig-SEIS2} focus on motions recorded on the vertical components of the Apollo seismometers. In the Apollo data, however, the horizontal components generally show higher noise than the vertical components, with noise levels generally between $1-5 \times 10^{-9}$~m/s$^{2}$ near the peak sensitivity of the Apollo instruments at 0.45 Hz. Because the high seismic scattering of the Moon suggests motion should be roughly equi-partitioned between different components of motion, this is typically interpreted as indicating that these horizontal signals are driven by small tilt signals rather than translational ground motions. If all of this signal is interpreted as tilt, the acceleration signal can be treated as the projection of the gravity vector into the horizontal component of acceleration, $a_H(t) = -g_M \sin(\tau(t))$, where $a_H$ is the observed horizontal acceleration, $g_M$ is the lunar gravitational acceleration of $\sim$1.6~m/s$^{2}$, and $\tau$ is the tilt in radians. For small tilts, $\sin(\tau)$ is approximately $\tau$, so the tilt can be estimated as $\tau=-a_H/g_M$ and therefore we can estimate tilts of a few nanoradians (i.e., hundreds of $\mu$as). This suggests isolation systems may be important for high-precision astrometry, but perhaps not for imaging systems (depending on observational wavelength); monitoring of seismic signals will be important regardless.  \textbf{In general, the seismic environment of the Moon is not an obstacle for optical interferometry} \citep{Mendell1998sp98.conf..451M}.

\clearpage      

\section{Lunar regolith and dust environment}\label{sec-lunar-regolith-and-dust}
\begin{wrapfigure}{r}{8.75cm}
\vspace{-0.25cm}
\begin{minipage}[h]{1\linewidth}
\begin{tcolorbox}[colback=gray!5,colframe=green!40!black,title=Dirty but not debilitating]
Dust is not a fundamental barrier for lunar telescopes, as shown by years of operations for the Lunar Ultraviolet Telescope aboard the Chang'e 3 lander.
\end{tcolorbox}
\end{minipage}
\end{wrapfigure}
The average particle size\index{Lunar Environment!dust}\index{Lunar Environment!regolith} of the lunar regolith is around 60-100~$\mu$m, depending on maturity and the regolith’s bulk composition in any given location. The finer particles ($<$50~$\mu$m) have been demonstrated to be highly abrasive and corrosive during Apollo operations for suit seals, sample return box seals, respiration, equipment wear, surface charging, and thermal insulation/overheating of equipment. Roughly 20\% of the regolith, by weight, consists of micron and sub-micron sized particles \citep{Manka1973ASSL...37..347M,Colwell2005Icar..175..159C}.  The properties of the regolith include:
\begin{itemize}
    \item Highly angular particle shapes (formed via impacts with no water/air on the Moon to erode particles) \citep{Greene1975LPSC....6..517G,McKay1991lsug.book..285M}
    \item Electrostatically charged particles via friction and ionizing radiation exposure \citep{Mishra2019ApJ...884....5M}
    \item Chemically active particles, again due to ionizing radiation and the Moon’s dry environment
    \item Thermally insulating properties, due to regolith being composed of rock, mineral, and glass particles with low rates of thermal conductivity
    \item Levitation of dust particles at sunrise and sunset \citep{Berg1976LNP....48..233B}
    \item The upper few centimeters of regolith are highly reworked fine grained particles, exposed to solar weathering and space weathering and mixing processes known as impact gardening \citep{Jensen1982LPI....13..364J,Su2024ApJ...976L..30S}
\end{itemize}

\begin{wrapfigure}{R}{0.64\textwidth}
    \centering
    \includegraphics[width=0.63\textwidth]{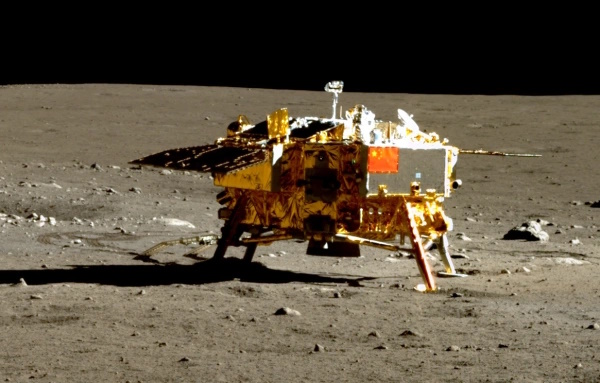}
    \caption{\label{fig-change3}The Chang'e 3 lander, which included the Lunar Ultraviolet Telescope (LUT), which operated for more than three years without problems from lunar dust. (Image credit: CNSA)}
\end{wrapfigure}

These properties, combined with the vacuum environment on the Moon’s surface, cause regolith particles to be attracted to equipment and make these surfaces difficult to clean, resulting in overheating and wear (particularly of moving parts). Technologies are being developed to effectively remove the regolith from surfaces that may be applicable to various interferometry designs; the recent Firefly Blue Ghost Mission 1\index{missions!Firefly Blue Ghost Mission 1} flew the Electrodynamic Dust Shield (EDS)\index{instruments!Electrodynamic Dust Shield} experiment, which lifted and removed lunar regolith using electrodynamic forces on the glass and thermal radiator surfaces \citep{firefly_mission_conclusion}. Likewise, more effective seals are being developed
for protecting vital equipment from dust exposure.

Regardless of the operational considerations that need to be taken into account by telescope systems because of lunar dust, the landing and multi-year operations of the Lunar Ultraviolet Telescope (LUT)\index{Instruments!LUT} aboard the Chang'e 3 lander\index{missions!Chang'e 3} (Figure \ref{fig-change3}) is proof that such a thing is possible.   To mitigate the impact of dust levitation during sunrise/sunset, the Chang'e~3 concept of operations (CONOPS) included closing a telescope entrance aperture shutter during these transition times.  \textbf{Optical to ultraviolet telescope operations are possible from the surface for telescopes atop the dusty regolith} \citep{Wang2015Ap&SS.360...10W}.




\subsection{Mitigation strategies}

Various mitigation strategies have been proposed to reduce dust lifting from the lunar surface and to reduce the effects of dust on instrumentation. Reducing the number of moving parts could reduce the effects of the abrasion of lunar regolith on instruments. Placing sensitive parts of instruments inside some kind of shielding, or by housing them in habitats, could protect moving parts from abrasive lunar particles. Active mitigations have been proposed and are in development for removing lunar dust from landed instruments. Electrodynamic removal of dust grains may alter the charged nature of the grains and their propensity to attach to electronics and instruments.

\section{Temperatures}\label{sec-lunar-temperatures}\index{Lunar Environment!temperatures}

\begin{figure}
    \centering
    \includegraphics[width=0.90\linewidth]{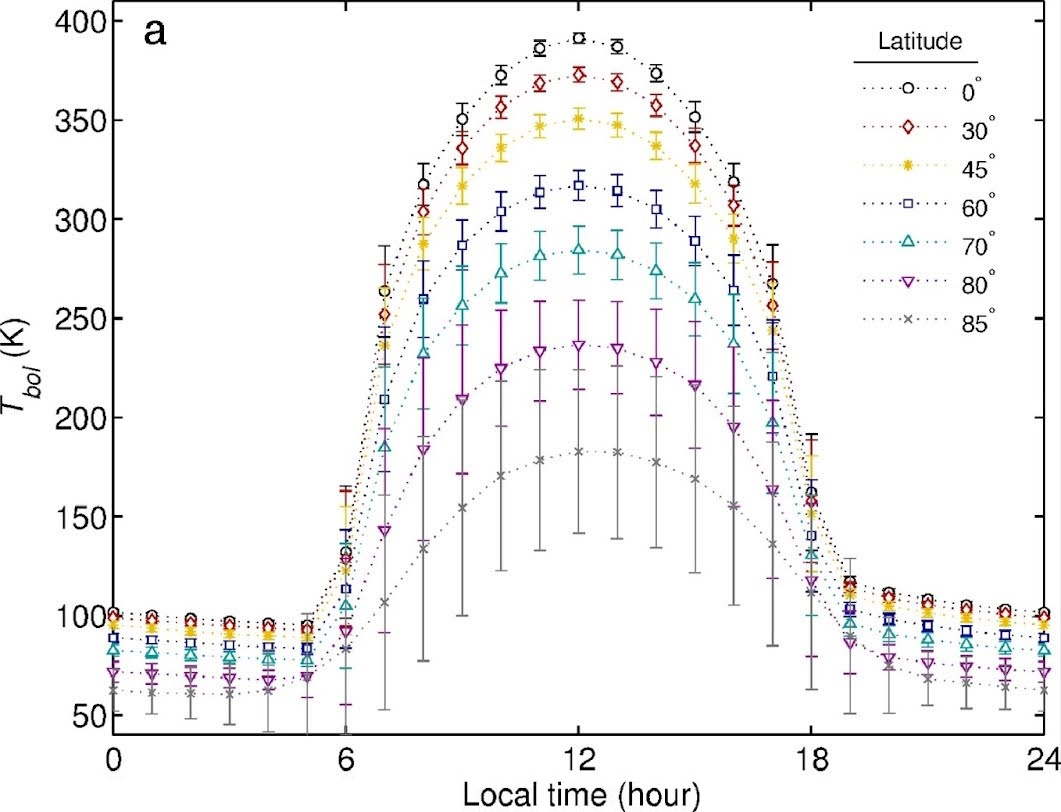}
    \caption{Latitudinal variation of temperature at the lunar surface based on the Diviner radiometer on the the Lunar Reconnaissance Orbiter (LRO) \cite{Williams2017Icar..283..300W}.}
    \label{fig-lunar_temps}
\end{figure}

Figure \ref{fig-lunar_temps} shows the temperatures over the course of a lunar diurnal cycle average over latitudinal bands, showing the large range between day and night temperatures, as well as the relative stability of temperatures during the nighttime.
Lunar surface temperatures have a wide range of daytime maximum temperatures and minimum nighttime temperatures, are unmitigated by a thick atmosphere, and are largely latitudinally dependent. Thermal and thermophysical properties have been used to reveal information about rock and regolith textures, near-surface conditions, and the variability of materials over local and regional scales. The \index{instruments!Diviner Lunar Radiometer Experiment}Diviner Lunar Radiometer Experiment \cite{Paige2010SSRv..150..125P} launched in 2009 on the Lunar Reconnaissance Orbiter (LRO) and has been obtaining continuous thermal measurements of the lunar surface ever since. These thermal measurements can be used to derive or infer thermophysical properties of the surface, such as the rock abundance, regolith density with depth (similar in nature as surface thermal inertia), or albedo of the surface.

Large temperature swings (\S \ref{sec-lunar-temperatures-day-night}) can be a challenge to electronics and optics that require thermal stability for operation and computation. High surface temperatures in many regions of the Moon could pose a threat to systems that require lower operating temperatures. The high thermal insulative properties of lunar regolith could be a resource or a detriment to a mission;  Lunokhod 2's\index{missions!Lunokhod 2} highly successful rover mission was ended when radiators became coated in dust \citep{wiki:Lunokhod_2}. Components buried in the regolith below a thermal skin depth (\S \ref{sec-subsurface-temps}) could be protected from large temperature swings, but even a few centimeters of lunar regolith covering an instrument could result in very rapid overheating.

Active and passive cooling systems can be considered for different configurations as a site investigation will be key to evaluating specific thermal conditions encountered by an observatory. One conclusion to that site investigation may be to take advantage of colder regions of the Moon, like permanently shadowed regions (PSRs, \S \ref{sec-permanently-shadowed-regions}) or high-latitude areas with less direct sunlight. These areas may provide unique advantages for specific types of observing, for example, in infrared wavelength ranges, where detectors and optics must be kept cool; using naturally colder regions of the Moon reduces the need for actively cooled instrumentation that may be required on Earth.

\subsection{Day versus night}\label{sec-lunar-temperatures-day-night}


Equatorial daytime temperatures can reach $\sim$400 K and drop to $\sim$100 K during the lunar night; polar temperatures may only reach $\sim$170 K maximum daytime temperatures in the sunlit areas, dropping as low as $\sim$30-40 K within shadowed craters during the night \cite{Vaniman1991lsug.book...27V}. The interferometric instrumentation will have to be designed to withstand large amounts of expansion, contraction, and fatigue stresses, depending on the location of the site.

\subsection{Subsurface temperatures}\label{sec-subsurface-temps}\index{Lunar Environment!temperatures, subsurface}

\begin{figure}
    \centering
    \includegraphics[width=0.95\linewidth]{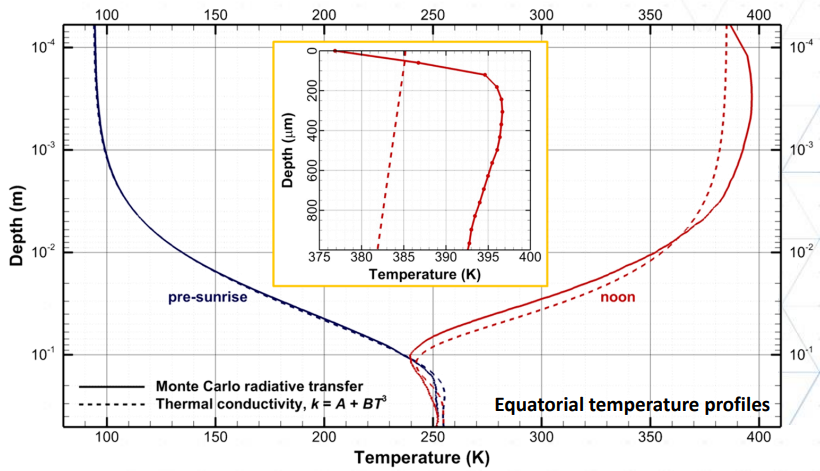}
    \caption{Lunar subsurface temperature at equatorial dawn (pre-sunrise) and noon, based on two different approaches to modeling radiative heat transfer. This illustrates the difference in computed subsurface temperature profiles when radiative heat transfer is modeled using a Monte Carlo approach, as opposed to being approximated by a temperature-dependent thermal conductivity.   \cite{Prem2019LPI....50.2425P}.}
    \label{fig-lunar_subsurface_temps}
\end{figure}

Apollo cores revealed a loosely stratified upper regolith layer with a fine layer of highly reworked dust component at the very top, and a mix of gravel and cobble-sized mix fractions below; the size distribution skews toward larger sizes with depth. Lateral and vertical mixing due to repeated impacts to the surface (known as impact gardening\index{impact gardening}) results in a highly space weathered upper layer of highly insulating regolith, below which mixed rocks and regolith are shielded from micrometeorite bombardment. The highly insulative outermost layer of regolith causes the temperature to drop hundreds of kelvin from maximum daylight temperatures within the first 30~cm. Below 50~cm, the temperatures become nearly isothermal with increasing depth \citep[Figure \ref{fig-lunar_subsurface_temps}; ][]{Langseth1977NASSP.370..283L,Prem2019LPI....50.2425P,Nagihara2023PSJ.....4..166N}.

This thermally insulated subsurface may be a valuable resource as a heat sink, but in areas where temperatures are optimal for observations or operations, heat generated by the instrumentation may be an asset more than a liability. Instrument components that are sensitive to large temperature swings may be buried under a 10~cm layer of regolith for thermal insulation. Regolith can also protect from solar and cosmic radiation, and can offer a solution to bury a radiogenic power source away from radiation-sensitive instruments. Future CLPS\index{CLPS} missions will  provide additional data on the lunar subsurface environment.

\section{Lighting}

\subsection{Permanently shadowed regions}\label{sec-permanently-shadowed-regions}\index{Lunar Environment!permanently shadowed regions}

\begin{figure}
    \centering
    \includegraphics[width=0.95\linewidth]{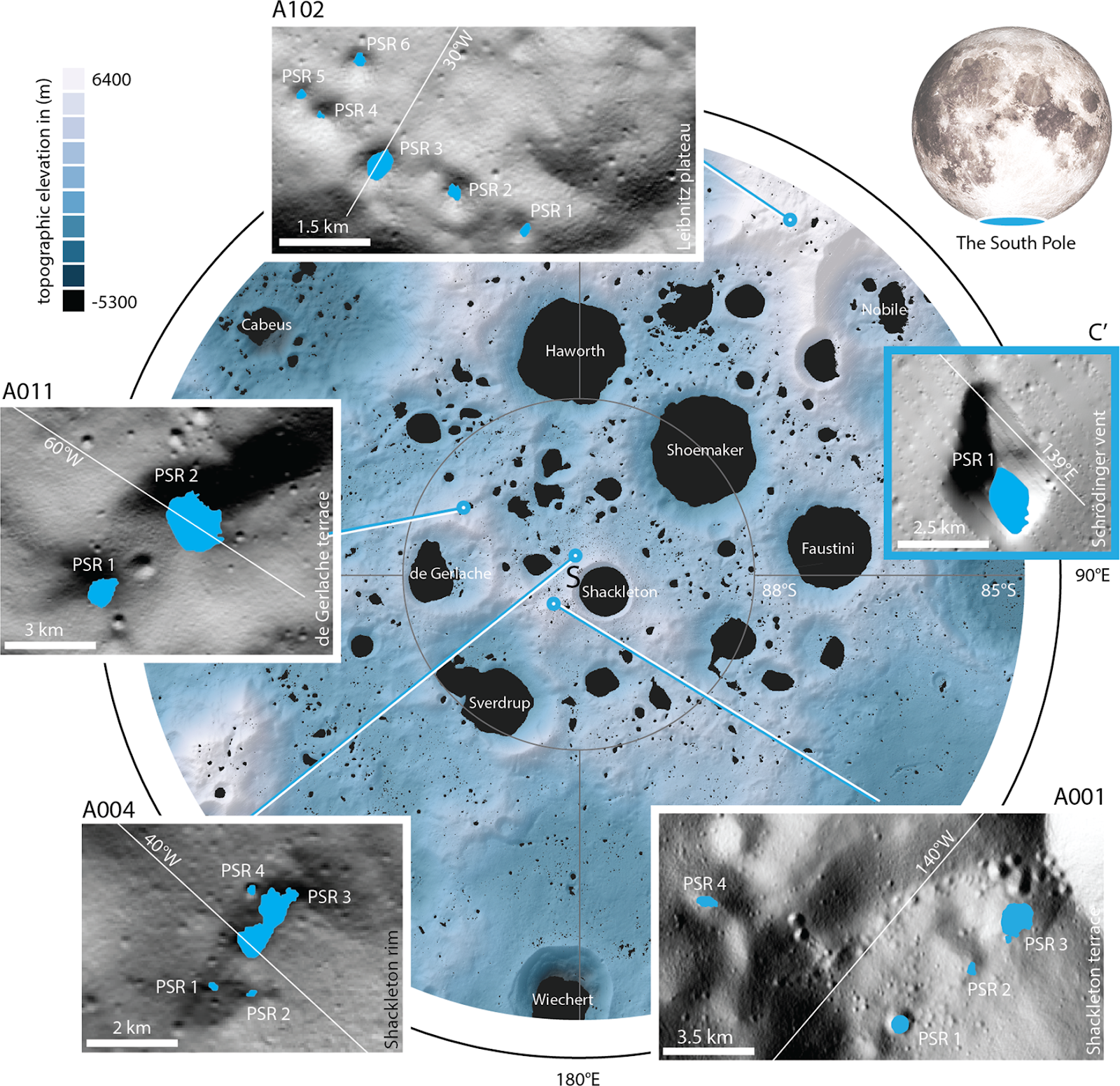}
    \caption{A topographic map of permanently shadowed regions on the lunar South Pole \cite{Prem2019LPI....50.2425P}.}
    \label{fig-lunar_PSRs}
\end{figure}

Permanently shadowed regions (PSRs) are constrained to the lunar North and South Poles (e.g., Figure \ref{fig-lunar_PSRs}) \cite{Mazarico2011Icar..211.1066M,Bickel2021NatCo..12.5607B}. As their name indicates, these areas receive no direct sunlight, and therefore have smaller temperature variations than most lunar regions. Many PSRs vary in temperature between 25 K--70 K. The regolith in PSRs is expected to be fine-grained and may have unique electrostatic and thermal properties due to cold trapping of volatiles. Additionally, because these features do not experience extremely high daytime temperatures as at higher latitudes, the regolith is less likely to experience compaction through processes like sintering, resulting in a less "fluffy" regolith texture in PSRs. As previously indicated, the lunar North and South Poles have greater changes in slope and terrain hazards due to topography, so caution should be exercised if placing an interferometer in a PSR due to elevation changes. Potentially elevated levels of settled dust particulates, combined with the unique (and not yet directly characterized) electrostatic environment, may result in challenges with dust that should also be considered.

\textit{Advantages}: Temperature swings are less extreme  compared to the equatorial regions. 
Lunar South Pole PSRs will be in relatively close proximity to expected Artemis\index{Artemis} surface access; crew members could be available if human servicing is needed and part of a facility's CONOPS. Such regions are also beneficial for mid-infrared/far-infrared (MIR/FIR) instrument cooling requirements.

\textit{Disadvantages}: PSRs are associated with micro-cold traps and therefore have a high probability of accumulating water ice \cite{Li2018PNAS..115.8907L}. Water and frost accumulation may affect optics, reducing scientific return. 
PSRs are located in regions of lower elevation in relation to their surrounding area; due to the light limitations, the geologic context within these shadows is currently being investigated by the ShadowCam imager aboard the Korea Pathfinder Lunar Orbiter (Danuri)\index{missions!Danuri}\index{missions!Korea Pathfinder Lunar Orbiter}\index{instruments!ShadowCam} \citep{Mahanti2023JASS...40..131M}. The south polar region will have high activity from NASA Artemis missions, causing an increase in levitated dust particulate and disturbed regolith from landing and launch.

\subsection{Persistently illuminated regions}\label{sec-persistently-lit}\index{Lunar Environment!persistently illuminated regions}

Certain areas at the poles, such as Mons Malapert, provide full sunlight for $\sim$90\% of the lunar year and partial sunlight for an additional $\sim$5\% of the lunar year \cite{Sharpe2003AdSpR..31.2467S}. Such locations would be suitable for the deployment of solar arrays to generate the power needed to support night survival.  Technologies under development for Artemis, such as power towers (\S \ref{sec-power-towers}) and power beaming (\S \ref{sec-power-beaming}), when deployed at the poles, could substantially relax the challenge for survive-the-night power requirements by making solar power\index{solar power} available for a large majority of the lunar day while sidestepping nuclear technologies.






\section{Ancillary lunar science returns}\label{sec-ancillary-lunar-science}




\begin{wrapfigure}{r}{7.25cm}
\vspace{-0.25cm}
\begin{minipage}[h]{1\linewidth}
\begin{tcolorbox}[colback=gray!5,colframe=green!40!black,title=Fringe fortune]
Lunar geophysical insights come serendipitously from interferometer engineering data.
\end{tcolorbox}
\end{minipage}
\end{wrapfigure}

Interesting elements of lunar science will be present  in the engineering data from any lunar optical interferometer.  Specifically, during check out and operations of any facility, an interferometer baseline solution will be determined and subsequently monitored at an extremely precise level throughout operations---typically to nanometers.  Such monitoring constitutes a highly sensitive seismic sensor; anecdotal evidence suggests the Palomar Testbed Interferometer\index{missions!Palomar Testbed Interferometer} collected unique seismic data during the magnitude 7.1 Hector Mine, California earthquake in 1999, as it was fringe tracking on stars during the event.

Additionally, there appear to be unmodeled physical effects in lunar libration \index{lunar libration}\citep{Williams2015JGRE..120..689W}---the "nodding" of the face of the Moon over a month as it orbits the Earth
.  Experiments such as the In-situ Lunar Orientation Measurement (ILOM)\index{instruments!In-situ Lunar Orientation Measurement} \citep{Sasaki2011epsc.conf.1135S,Hanada2012SCPMA..55..723H} have been proposed to explicitly monitor star paths across the sky at milli-arcsecond scales during the lunar month.  Such observations are done as a matter of routine operations by the tracking functions of an interferometer, and potentially constrain the size, density, and state of the core and lower mantle of the Moon \citep{Petrova2008AdSpR..42.1398P}.

%

\chapterimage{AeSI_Figure_3-1.jpg} 
\chapterspaceabove{1.75cm} 
\chapterspacebelow{8.75cm} 

\chapter{Interferometry Architectures on the Moon}

\section{General considerations of surface-based interferometry architecture}







Observational optical interferometry has been operational on-sky from terrestrial ground-based facilities for more than a century \citep{Michelson1921PNAS....7..143M}, and general considerations for the basic principles of modern operational interferometric facilities have been well understood for over three decades \citep{Shao1992ARA&A..30..457S}.  Expectations for the appeal of the technique from the lunar surface have been in place for a similar length of time as the current generation of instruments \citep{burns1988faom.work.....B} \citep[and see \S 3.6 of ][]{Shao1992ARA&A..30..457S}.

\subsection{Beam collection}\index{beam collection}
\begin{wrapfigure}{r}{8.75cm}
\vspace{-0.25cm}
\begin{minipage}[h]{1\linewidth}
\begin{tcolorbox}[colback=gray!5,colframe=green!40!black,title=Telescopes for lunar interferometry]
Anchored on a stable surface: very similar to their terrestrial counterparts.
\end{tcolorbox}
\end{minipage}
\end{wrapfigure}
The process of beam collection for an interferometer is essentially the same as for a standalone telescope, with a few additional considerations worth mentioning.  Typically, an altitude-azimuth telescope mount is used, though because many interferometric architectures on the Moon may have a very small field of regard, some additional consideration to observe at zenith or meridian crossing needs to be taken.  Another option often employed in interferometric designs for beam collectors is the use of a siderostat that feeds a fixed telescope \citep{Colavita1999ApJ...510..505C}.  While traditional Earth-based telescopes are housed in domes or enclosures to protect them from weather, this is likely not required in a lunar interferometric architecture.  Some type of covering for the optics for dust protection may be needed, especially during dust levitation periods during the lunar day-night transitions (\S \ref{sec-lunar-regolith-and-dust}).

Optics themselves will need to be made of high-quality materials with a low coefficient of thermal expansion (CTE) (e.g., Zerodur, ultra-low expansion glass/ULE, Cer-Vit) and have well-polished surfaces (to reduce scattered light) to survive the lunar temperature variations (\S \ref{sec-lunar-temperatures}) and produce high-quality, unaberrated beams for later interference.  Special attention should be paid to stiffness and precision-tracking capabilities of the mounts because assistive technologies such as adaptive optics correction and end-to-end metrology are unlikely to be implemented in the initial deployment of lunar interferometers.

Techniques for determining pointing models for individual telescopes are well understood \citep{Meeks2004SPIE.5497..140M} and can be rapidly ($<20$ minutes) established by pointing at, and solving for, a small number ($N<20$) of astrometric fields.
Redetermination of such pointing models on a lunar daily basis may be required due to the significant day-night temperature transitions.
Also, depending on the mechanism used to place the telescope on the lunar surface, and the consistency and stability of the regolith (and associated absence of bedrock), additional considerations associated with leveling, pointing and recovery from errors may be required.  This issue becomes more critical in terms of repeatability if the telescope is expected to be moved regularly to change the interferometric baselines.

Telescopes will likely have much smaller apertures (by 10--50$\times$) than terrestrial facilities due to the lack of atmosphere on the Moon, which allows for much longer coherent integration times, unconstrained by atmospheric seeing.
It is important to remember that the Moon’s lower gravity (1/6th that of the Earth) will change the natural vibrational frequencies of equipment on the lunar surface, which may be important for certain interferometric applications requiring long-term stability.

Depending upon the science requirements, telescopes may be fixed at a single location (like CHARA)\index{missions!CHARA} or movable (like VLTI)\index{missions!VLTI}.  Telescope relocation can solve certain problems of pathlength equalization (see \S \ref{sec-beam-relay}), but could also introduce new problems like the agitation of dust.

\subsection{Beam relay}\index{beam relay}\label{sec-beam-relay}
\begin{wrapfigure}{r}{8.75cm}
\vspace{-0.25cm}
\begin{minipage}[h]{1\linewidth}
\begin{tcolorbox}[colback=gray!5,colframe=green!40!black,title=Light relaying: free space versus fiber]
For free space beam relay, nighttime operations are best; fiber beam transport avoids scattered light during daytime operations.
\end{tcolorbox}
\end{minipage}
\end{wrapfigure}
The obvious technique for transporting light from the telescope to the beam combiner is free-space propagation.  This approach works over a wide range of distances and wavelengths: there is no atmosphere to cause turbulence, dispersion, or transmission losses; all that is needed is a clear line of sight from the telescope to the beam combination hub and a steerable mirror at each end of the path.

The main concern is scattered sunlight, which limits the sensitivity of daytime observations.  
The small field of view of an interferometer makes it possible to prevent scattered light from directly reaching the detector.  In the simplest case, two scatterings are needed. First, light must scatter off either the regolith or the telescope structure to reach the first mirror in the hub.  That scattered light must scatter off that mirror to reach the detector.  An order of magnitude estimate of the amount of scattered sunlight that reaches the detector is
\begin{equation}
    F_s = F_0\eta_s\Omega_B \left( \frac{\pi d^2 }{ 4} \right)\eta_M\omega_F
\end{equation}
where $F_0$ is the solar flux of $3.5 \times 10^{21}$ photons/sec (based on 1.4~$W/m^2$ at 0.5~$\mu$m), $\eta_s$=0.15 is the albedo of the lunar surface, $\Omega_B$ = 0.008~sr for solid angle of a notional 0.1~m diameter by 1~m long baffle, $d$=0.075~m mirror diameter, $\eta_M$ = $10^{-4}$ mirror scattering, and $\omega_F = 3.5 \times10^{-10}$~sr for the solid angle of a 2.4~$\lambda / d$ field stop.  With this example, 64 scattered photons per second per aperture reach the detector.


Scattered light means free space beam relay will have to be highly baffled in the case of daytime operations, or limited to nighttime operations with less (if any) baffling, or require daytime operations with fiber-based beam relay (\S \ref{sec-fiber-relay}).
Nighttime observations ratchet up the after-dark power requirements beyond just "survive the night" (\S \ref{sec-survive-the-night}), which in turn increases the need for a battery or other nighttime power sources.

A possible advantage of free-space propagation is the possibility of compensating differences in external optical pathlength by adjusting the placement of the telescopes for a given target pointing (see the discussion in \S \ref{sec-pupil-remapping} on pupil remapping).  

\subsubsection{Fiber versus free space}\label{sec-fiber-relay}

\begin{wrapfigure}{R}{0.66\textwidth}
    \centering
    \includegraphics[width=0.64\textwidth]{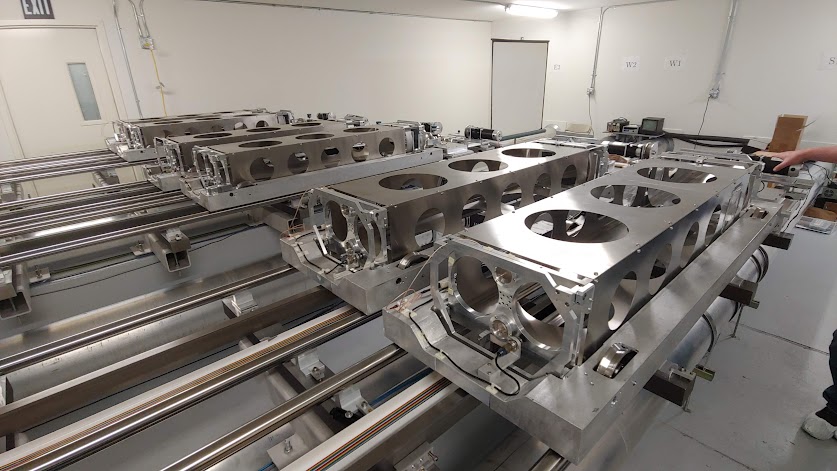}
    \caption{\label{fig-chara-delay-carts}Delay line carts at the CHARA interferometer.  (Image credit: Gerard van Belle)\index{missions!CHARA}} 
\end{wrapfigure}

An alternative, optical fiber\index{optical fibers} propagation, solves the scattered light problem.  Fibers are also potentially more tolerant to site selection since they eliminate the need for a clear line of sight between the stations and the combiner.  It is also possible the use of fibers can simplify maintaining alignment.  If delay compensation is at the telescope before injecting light into the fiber, the fibers can be connected directly to an integrated optics combiner, and the only alignment ever needed on the lunar surface will be for telescope pointing.

Optical fiber transport requires a physical connection between the telescopes and combiner.  
Such a cable could also carry power and communications; this may actually simplify system design by allowing the elimination of subsystems at the outboard telescopes, such as power/batteries, command and data-handling computers, and radio communications.

While they have been demonstrated to work for astronomical optical interferometry \citep{Perrin2006Sci...311..194P,Shivitz2014SPIE.9145E..0JS,Magri2025MNRAS.536..266M}, fibers carry with them some penalties.  Much of the industry technology development for fibers has focused on telecom wavelengths, meaning little is available shortwards of 500~nm or longwards of 1700~nm.
Maintaining coherence in the fiber could be a challenge; the most commonly available fibers are known to have chromatic dispersion \citep{Allured2021ApOpt..60.6371A} that must be accommodated by having near-exact matched fiber lengths.  
Still, dispersion can vary between fibers due to thermal environment mismatches; it has not been demonstrated that dispersion can be matched to the required level.
Fibers are effective Cherenkov detectors that could set a sensitivity limit, at least at shorter wavelengths.
Finally, there are possible concerns with survivability.  There are potential issues with thermal cycling, and fibers used in ground-based applications going dark from radiation \citep{Marshall1996ITNS...43..645M}; both of these issues are potentially remediated by burying the fiber, though this carries with it the technical challenge of trenching and re-covering.  Space-rated fibers exist, at least for low Earth orbit (LEO) \citep{Ciminelli2017SPIE10564E..1EC}, but these fibers have not been tested for interferometric observations; one of the goals of the STarlight Acquisition and Reflection toward Interferometry (STARI) \index{missions!STARI} \citep{Monnier2024SPIE13092E..2YM} technology demonstrator mission is to demonstrate this technology.

\subsection{Beam delay}\index{delay lines}

\begin{wrapfigure}{r}{7.25cm}
\vspace{-0.25cm}
\begin{minipage}[h]{1\linewidth}
\begin{tcolorbox}[colback=gray!5,colframe=green!40!black,title=Controlling pathlength]
Optical path control at nanometer scales (or less) is a well-proven technology.
\end{tcolorbox}
\end{minipage}
\end{wrapfigure}

For terrestrial optical interferometry, variable beam delay at the nanometer scale over 100+ meter ranges is accomplished by sending the light into and then out of a mechanically movable stage with retro-reflective optics; the extreme dynamic range of such systems is a hallmark of modern terrestrial facilities.  This technology has been well established for many decades \citep{Shao1977JOSA...67...81S,Colavita1991SPIE.1542..205C} and adopted for use at all modern facilities \citep[Figure \ref{fig-chara-delay-carts}; ][]{Colavita2000SPIE.4006..310C,Germain2002JOSAA..19...81G,Hogenhuis2003SPIE.4838.1148H,Launhardt2007ecf..book..265L,Fisher2010SPIE.7734E..49F}.  While multi-stage nested servo loops are commonly used for current long-throw ($>10$m) operational facilities, simpler single-stage off-the-shelf linear actuator solutions are commercially available for shorter throws ($<10$m).
Spaceflight-quality delay line articles have been produced for the unflown Space Interferometry Mission (SIM) \index{missions!SIM}\citep{Edberg2007SPIE.6693E..0DE}; these examples are likely over-engineered for a lunar application, given their momentum compensation aspects for a free-flying spacecraft and picometer-class tracking tolerances.

While beam delay takes place at a central beam laboratory for all current terrestrial facilities after beam relay, there is no fundamental reason that prevents beam delay from taking place at the telescope stations before relaying to a central lab.  Such an architecture could greatly simplify the complexity of the central beam combiner for a lunar facility.

\subsection{Beam recombination}\index{beam combiners}

Beam recombination of target light collected by multiple separate apertures is a well-developed area of technology for terrestrial optical interferometers.  Multi-beam combiners for up to six telescopes such as Michigan Young STar Imager at CHARA (MYSTIC)\index{Instruments!MYSTIC} \citep{Monnier2018SPIE10701E..22M} and Visible Imaging System for Interferometric Observations at NPOI (VISION) \citep{Garcia2016PASP..128e5004G}\index{Instruments!VISION} exist, as well as fiber- or integrated-optic-based combiners like Jouvence of FLUOR (JouFLU) \citep{Scott2014SPIE.9146E..1AS}\index{Instruments!JouFLU} and GRAVITY \citep{GRAVITY2017A&A...602A..94G}. Precision astrometric architectures are seen in GRAVITY\index{Instruments!GRAVITY}, ARrangement for Micro-Arcsecond Differential Astrometry (ARMADA) \citep{Gardner2022SPIE12183E..0ZG}\index{Instruments!ARMADA}, the NPOI astrometric combiner \citep{Benson2010SPIE.7734E..3KB}\index{missions!NPOI}, and even for spaceflight in the SIM astrometric architecture \citep{Laskin2006SPIE.6268E..23L}.

\section{Lunar advantages}

\subsection{Coherence volume}\label{sec-coherence-volume}

The concept of coherence volume\index{optical concepts!coherence volume} has proven useful in the context of long baseline optical interferometry: it quantifies the volume of space within which the enclosed photons are all sufficiently coherent to generate clean interference fringes. For terrestrial interferometry, calculation of the coherence volume is relatively straightforward because it is dominated by atmospheric seeing, which is well understood at least in terms of its statistical behavior. Under such conditions, the volume enclosing coherent photons is given by the size of a single phase-coherent patch in the incoming wavefront (parameterized by the Fried parameter\index{optical concepts!Fried parameter} $R_0$) and the coherence time (usually called $\tau_0$). Then, we write the coherence volume $\sim R_0^2 \tau_0 c$  where $c$ is the speed of light. Although these parameters vary by locale and with observing wavelength, in visible light at a competitive modern observatory on a good night we might expect $R_0$ to be of order $\sim$10~cm and $\tau_0$ of order $\sim$10~msec, so we have a coherence volume of $3 \times 10^4$ cubic meters. These length and time parameters scale approximately as lambda to the 6/5 power for the turbulent atmosphere, so the coherence volume scales strongly (with an 18/5 power) with wavelength. Large variations in seeing are driven by local weather conditions, although on the other hand, there is only moderate variation in median values for these parameters between observatories. An exception to this may be exotic sites on the Earth’s surface, such as the high Antarctic plateau, which can exhibit an order of magnitude larger coherence volume \cite{coudeduforesto2008poii.conf..543C, Saunders2009PASP..121..976S} (though still ultimately limited).

\begin{wrapfigure}{r}{8.75cm}
\vspace{-0.25cm}
\begin{minipage}[h]{1\linewidth}
\begin{tcolorbox}[colback=gray!5,colframe=green!40!black,title={No atmosphere, no problem}]
The lack of atmospheric turbulence means small lunar telescopes outperform even the largest terrestrial telescopes.
\end{tcolorbox}
\end{minipage}
\end{wrapfigure}
Another parameter of relevance to interferometry is the isoplanatic angle\index{optical concepts!isoplanatic angle} describing the scale over which the phase remains well correlated. Typical terrestrial values range from about 10 arcseconds in the visible up to an arcminute in the near infrared. This enters as a critical design driver for the performance of optical interferometers, specifically when observations over any nonzero field of view are required.  In the case of astrometry, where accurate registration of the separation between two or more objects is required over some angular separation on-sky.  For "dual star referencing\index{dual star referencing}," enabling the observation of very faint science targets, the instrument can obtain its cophasing signal from a bright reference source nearby on the sky---assuming the bright and faint targets share common disturbances from object to detector.  The origin for this parameter, principally, is the different columns of terrestrial air through which the two beams travel; as the angle between them increases, the degree to which atmospheric path disturbances are shared decreases.

While values for these parameters can be readily assigned when designing instrumentation limited by the atmosphere, it is far more difficult to do so for a lunar deployment---ironically, because they are so much more favorable on the Moon.  The reason for this can be found in the primary motivation for lunar interferometry in the first place. In principle, all three parameters---length scale, time scale, and isoplanatic angle---are infinite. Or, at least, unlike the case for observing through turbulence, there is no natural limit imposed by external conditions so that limits that do exist all arise from the stability of the instrument itself.

Any lunar equivalent of the isoplanatic angle would seem to be essentially unlimited. Practical limits to simultaneously observing two or more different field angles are only constrained by the ability to furnish a reasonable optical system to bring starlight arriving from diverse directions into the interferometer, together with sufficient throw in the delay line to account for the geometrical path lengths required. We therefore conclude the lunar "isoplanatic angle" essentially covers the entire visible sky, provided the optical infrastructure exists.			

Similarly, the concept of the coherence length $R_0$ \index{optical concepts!coherence length} seems less useful in the lunar context, being limited only by the ability to build and steer a large phase-coherent collector. Demonstrated cophasing stability outcomes for large cold mirrors in space are readily available from a number of missions; one example representative of deployment in a cold stable lunar environment (such as a shadowed crater floor) might be found from the experience of NASA's James Webb Space Telescope (JWST). Here, the long-term drifts in wavefront plateau at about 15~nm over longer timescales (50-100 hours), and about 2-4~nm for timescales under one hour. A lunar telescope outside of a cold trap, exposed to varying thermal loads over day/night or other disturbances from its surroundings, is unlikely to statically perform at such levels.  For the less demanding case of interferometric imaging where diffraction-limited performance requires only tens of nanometer control, this is not problematic; for astrometry or very high-contrast imaging, pathlength monitoring and/or control may be required.

Coherence time $\tau_0$\index{optical concepts!coherence time}, while again being theoretically unlimited, can perhaps be translated into the lunar context more successfully than the other quantities. Again, limits are imposed by the stability of optical hardware and, potentially, with input from the seismic stability of the lunar environment. However, as discussed in earlier sections, the Moon is an exceptionally quiet environment. With the potential exception of rare disturbance events, noise seems to lie below thresholds of detection for even dedicated seismic instruments. It therefore seems reasonable to conclude that geologically forced vibration will also be well below that already present in our instruments due to actuated or motorized components and electronics.

Such mechanically imposed pathlength changes along optical trains are to be expected, and will span a noise spectrum with power rising to lower frequencies. There are certain to be large terms due to imperfect knowledge of the interferometer baseline, as well as imperfections in gears and moving components. For sufficiently bright stars or with a bright phase reference present, both short-term fluctuations and long-term drifts in the phase center can be readily tracked and compensated with the delay line\index{delay lines}. The value of the coherence time applies more to a timescale over which no such tracking is possible.

From the ground, the strategy of envelope tracking may be adopted so that even where it is not possible to track the phase, the coherent fringe pattern can be kept within range, albeit at the expense of signal to noise. Such a strategy could be readily translated to the Moon provided a sufficiently good baseline solution, and would effectively extend integration times indefinitely.

On the other hand, observations requiring full phase coherent integration will encounter limits. Determining these limits is difficult to calculate without knowledge of the opto-electronic performance of the hardware. Certainly, it seems reasonable that $\tau_0$   lies in the realm of seconds, perhaps minutes to hours with well-designed apparatus.  \textbf{Thus, lunar-based interferometers will be significantly more sensitive than their terrestrial counterparts.}

In summary, the values of the quantities that bound the coherence volume\index{optical concepts!coherence volume} are somewhat awkward to translate to lunar conditions, largely because the excellence of the site makes their intrinsic values tend to infinity, and while some limitations will certainly arise from instabilities in the optics, these occupy a dramatically different realm than ground-based constraints.



\subsection{Astrometry from the lunar surface}\label{sec-astrometry-from-lunar-surface}

\begin{wrapfigure}{R}{8.75cm}
    \vspace{-0.25cm}
    \begin{minipage}[h]{1\linewidth}
        \begin{tcolorbox}[colback=gray!5,colframe=green!40!black,title={Astrometry with long baselines}]
        Very long (100~m+) baselines on the lunar surface mean relaxed mechanical tolerances for astrometric measurements relative to shorter orbital facilities.
        \end{tcolorbox}
    \end{minipage}
\end{wrapfigure}

A high-precision astrometric interferometer\index{Astrometry} on the lunar surface could serve as a powerful tool for identifying Earth-like exoplanets for the Habitable Worlds Observatory (\S \ref{sec-masses-for-HWO}).  As shown in Figure \ref{fig-astrometric_exoearths}, the astrometric signatures of such planets will typically have amplitudes of $\sim 0.2 \mu$as orbiting nearby stars.  In this section, we describe an astrometric interferometer point design capable of achieving $0.1 \mu$as relative astrometry with access to the full sky. With its single fixed baseline, the interferometer measures motion along a single direction; like the extreme precision radial velocity technique (EPRV), it measures a projected component of the stellar reflex motion. The combination of EPRV and astrometry measures all orbital components, including the planet mass, modulo symmetry about the interferometer baseline.  The full complement of orbital components could also be resolved with a second, orthogonal baseline.

The highly successful VLTI GRAVITY instrument \cite{GRAVITY2017A&A...602A..94G}\index{Instruments!GRAVITY} routinely performs relative astrometry to 10 $\mu$as, an incredible achievement from a ground-based system.  Its large apertures have the sensitivity to observe faint reference stars within 1 arcsecond of the target stars.  GRAVITY does not monitor the baseline length and orientation; with a reference star so close, the baseline instability, which may be on the order of microns, has a negligible effect on astrometry.  Internal pathlengths from the collecting apertures to the beam combiners are measured with a metrology system capable of nanometer accuracy \cite{Gravity_metrology_lippa}.

The lunar interferometer described here is inspired by GRAVITY but differs in several important ways.  First, the astrometric requirement, 0.1 $\mu$as, is 100$\times$ finer than GRAVITY's.  For a similar baseline, the metrology precision is proportionally tighter, requiring no worse than 50~pm pathlength precision distributed over the reference and target star measurements.  On the other hand, the lunar environment is much more stable, with expected tilts of just a few nanoradian (see \S \ref{sec-seismology}), corresponding to motions of optics of a few nm if placed a meter above the surface. This relaxes the complexity compared to GRAVITY, where a large fraction of the system is dedicated to compensation for turbulence and vibrations.

\begin{figure}
    \centering
    \includegraphics[width=1.0\linewidth]{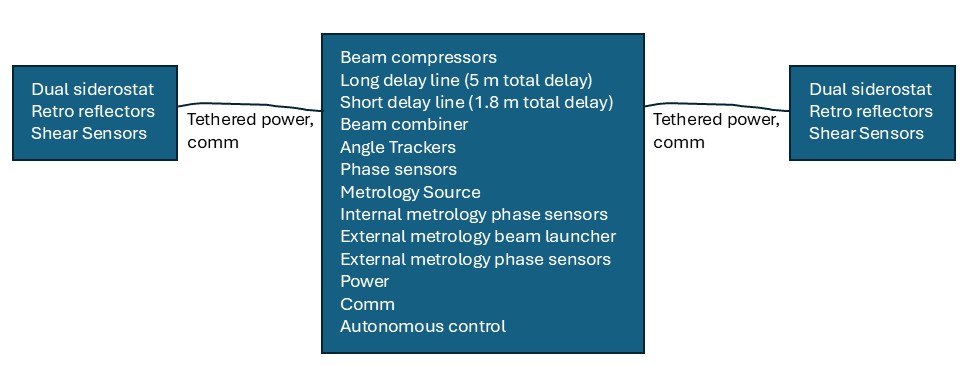}
    \caption{Astrometric interferometer block diagram. } 
    \label{fig:block_diagram}
\end{figure}

\begin{figure}
    \centering
    \includegraphics[width=0.75\linewidth]{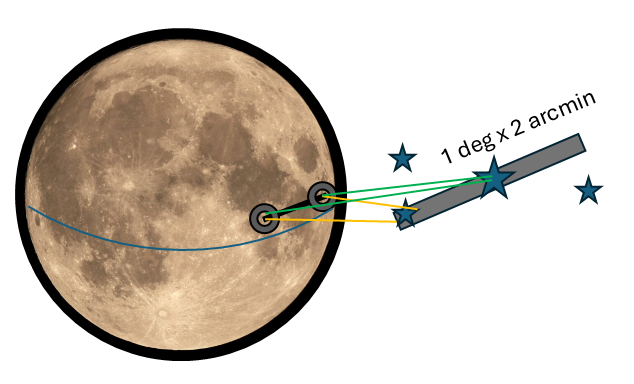}
    \caption{Astrometric interferometer baseline (not to scale). } 
    \label{fig:baseline_and_field}
\end{figure}

The point design has the attributes summarized in Table \ref{tab:astrometric_design}.  Figure \ref{fig:block_diagram} shows a block diagram. The instrumentation is fully contained in the central station except for the siderostats and a set of shear sensors. Figure \ref{fig:baseline_and_field} shows the preferred baseline orientation, east-west and near the Moon's equator.  This allows access to the whole sky once per month and with a minimum delay line length.

\begin{table}
\centering
\begin{tabular}{c|c|p{2.5in}}
   \textbf{Parameter}  &  \textbf{Value} & \textbf{Comments}\\ \hline\hline
    Baseline & 100 m    & Balances metrology precision, delay line length, and ease of setup. \\ \hline
    Orientation, Position & E-W, near equator & full sky access with minimum delay line length \\ \hline
    Bandpass & 450-850~nm & Throughput, fringe width, sensitivity \\ \hline
    Aperture Diameter  & 30~cm & Minimum size sensitive to V=12 reference stars with a few hours integration. \\ \hline
    \begin{tabular}{@{}c@{}}Beam Combiner  \\ Position\end{tabular} & \begin{tabular}{@{}c@{}}midway between  \\ siderostats\end{tabular}  & Symmetric diffraction. Alternative is to co-locate with one siderostat.\\ \hline
    Propagation & Free Space & Requirements of 50~pm path measurement are problematic for fibers. Beam is compressed 3:1 at the central station.\\ \hline
    Ref Star Magnitude & V $\leq$ 12 & On average, one reference star every 1$^\circ$ x 2 arcmin. \\ \hline
    \begin{tabular}{@{}c@{}}Baseline Length \\ Measurement\end{tabular}& $\sigma$ = 2~nm & Set by $\pm$ 0.5$^\circ$ field parallel to baseline. \\ \hline
    Baseline 3-D Stability & $\sigma$ = 200 mn & Set by $\pm$ 1.5 arcmin field orthogonal to baseline; only length is monitored. \\ \hline
    \begin{tabular}{@{}c@{}}Main Delay \\ Line Length\end{tabular} & 5 m & Function of lunar rotation rate and integration time on reference star \\ \hline
    \begin{tabular}{@{}c@{}}Differential Delay \\ Line Length\end{tabular}  & 1.9 m & Set by $\pm$ 0.5$^\circ$ field parallel to baseline \\ \hline \hline

\end{tabular}
\caption{Astrometric interferometer characteristics}
\label{tab:astrometric_design}
\end{table}

\begin{figure}
    \centering
    \includegraphics[width=0.95\linewidth]{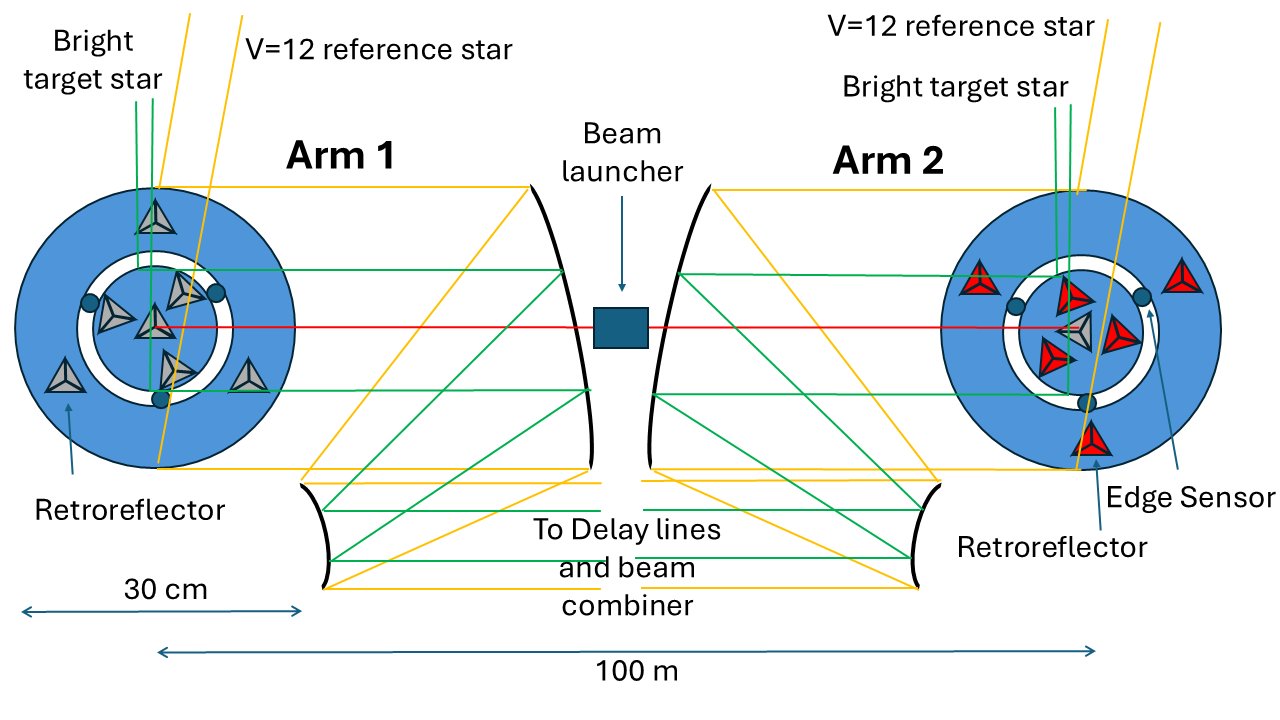}
    \caption{Astrometric interferometer baseline (not to scale). } 
    \label{fig:astrom_baseline_aps}
\end{figure}

\begin{figure}
    \centering
    \includegraphics[width=0.95\linewidth]{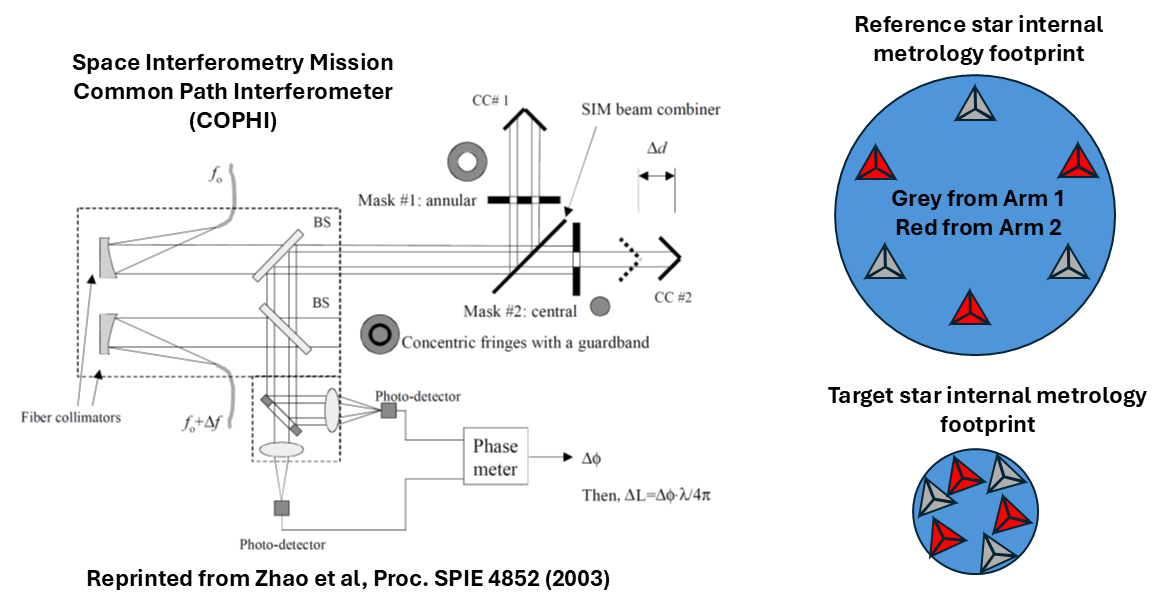}
    \caption{Astrometric metrology system based on the SIM COPHI architecture. The pathlength through each arm, for each subaperture, is measured using four sets of retroreflectors. } 
    \label{fig:astrom_metrology}
\end{figure}

\begin{figure}
    \centering
    \includegraphics[width=0.95\linewidth]{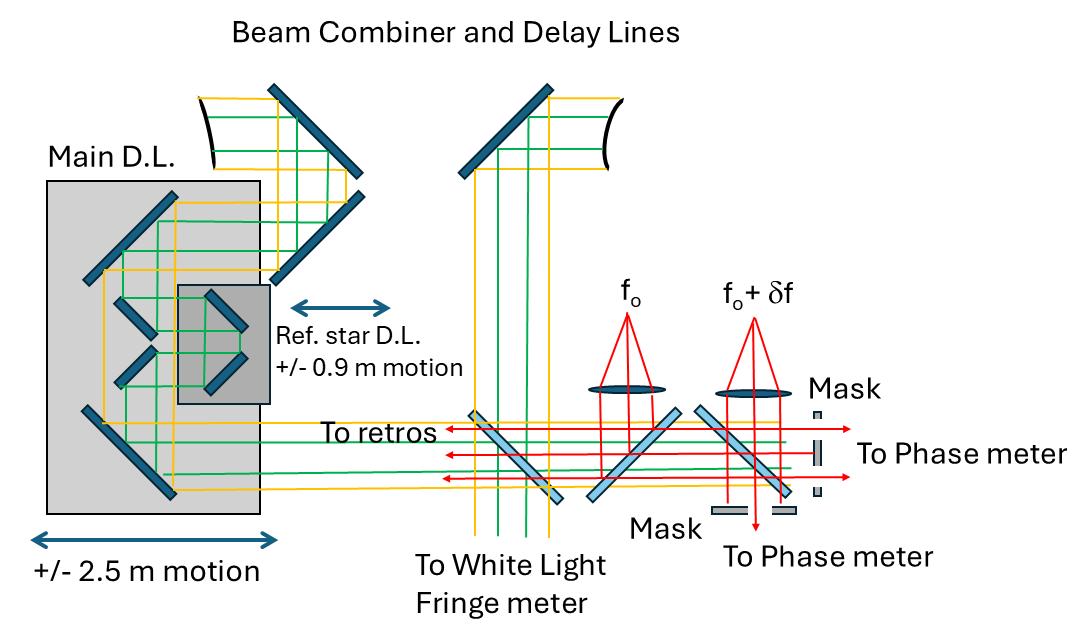}
    \caption{Beam combiner and delay line.  The delay line consists of a long motion to allow for hours of integration on a reference star, and a short motion to account for the angle between the reference and target star. A set of masks separates metrology and starlight.} 
    \label{fig:astrom_bc_dl}
\end{figure}

The concept of operations is to simultaneously observe the reference and target star using concentric siderostats (Figure \ref{fig:astrom_baseline_aps}).  This is driven by the sensitivity of the astrometric measurement to baseline length changes; a 2~nm change in baseline length introduces a 0.05 $\mu$as relative astrometry error across the 0.5$^\circ$ separation between the reference and target star. Meanwhile, the internal metrology that traces the starlight through the interferometer simultaneously measures the optical path to the inner and outer siderostats, retroreflecting from corner cubes mounted on the optical surfaces (Figure \ref{fig:astrom_metrology}).  A set of masks in the beam combiner separates the return beams from each arm before mixing with a reference beam (Figure \ref{fig:astrom_bc_dl}). The metrology scheme is based on the common-path heterodyne interferometer (COPHI) architecture \cite{Zhao_COPHI_2003} developed for SIM\index{missions!SIM}. The measurement requirement on this metrology is $\sim$ 20~pm per arm such that the overall differential delay measurement is below 50~pm.  Any number of metrology modulation/electronics approaches can meet this requirement, including Laser Interferometer Space Antenna (LISA)\index{missions!LISA}-style metrology \citep{Weise2017SPIE10567E..0QW}, Gravity Recovery and Climate Experiment (GRACE-C) \citep{Abich2019PhRvL.123c1101A}, and systems studied for large aperture ultra-stable telescopes \citep{Nissen_metrology_2017}.

The laser metrology scheme leaves one degree of freedom unmeasured: the relative piston of the inner and outer siderostats.  To account for these motions, precision edge sensors are placed between the inner and outer siderostats.  The precision and repeatability requirement on these sensors is 2~nm, to ensure no more than 0.05~$\mu$as astrometric error as the siderostats track the target and reference stars.

The duration of each observation requires $\sim$ four hours for a $m_V=12$ reference star. During this time, the main delay line moves 4 m to track the white-light fringe\index{delay lines}. Meanwhile, a second delay line, riding on the main one, adds up to $\pm$0.9 m delay depending on the position of the reference star.  The two sets of white light fringes, the four internal metrology paths (reference and target siderostats in both arms), and the external baseline metrology are all recorded so that the relative position of the target and reference are measured to 0.1 $\mu$as with each monthly target observation.  This approach can observe up to 120 HWO\index{missions!Habitable Worlds Observatory} target stars on a monthly cadence.

The technology to perform these measurements exists on Earth. All sensor and stability requirements have been demonstrated in the laboratory or in the field.

\subsection{Lunar stations versus formation flying}\label{sec-formation-flying}
\begin{wrapfigure}{R}{8.75cm}
    \vspace{-0.25cm}
    \begin{minipage}[h]{1\linewidth}
        \begin{tcolorbox}[colback=gray!5,colframe=green!40!black,title={Gravity: it's a feature, not a bug}]
        When you put something on the lunar surface, it stays put---no complicated formation-flying infrastructure.
        \end{tcolorbox}
    \end{minipage}
\end{wrapfigure}

\begin{wrapfigure}{R}{0.55\textwidth}
    \centering
    \includegraphics[width=0.53\textwidth]{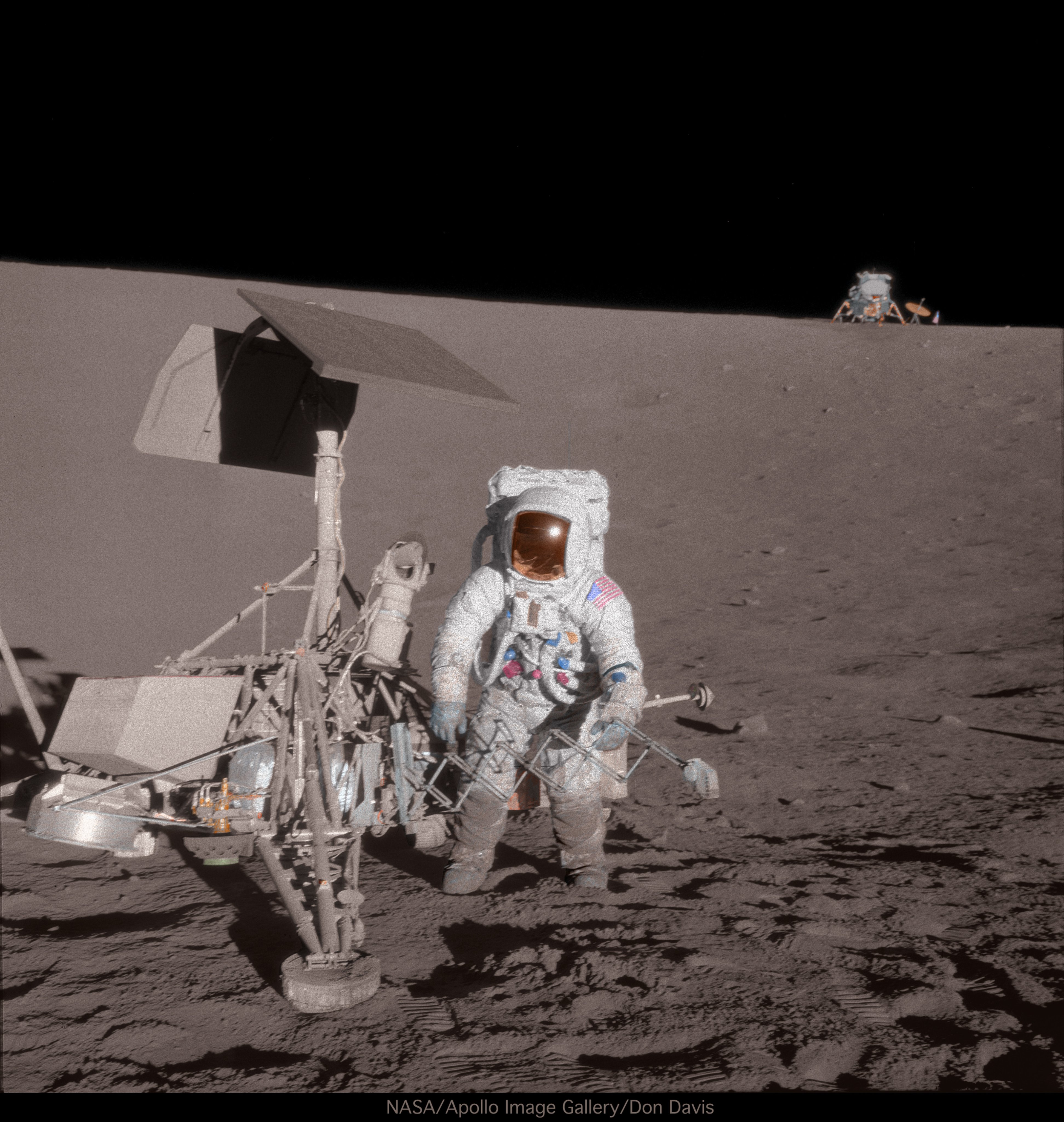}
    \caption{\label{fig-A12S3}Fixed stations for free on the Moon: The Surveyor 3 lander (foreground) and Apollo 12 descent stage (background) have remained at a constant distance of 180 m for the past 50 years, without being explicitly designed for such station-keeping. (Image credit: NASA/Don Davis)}
\end{wrapfigure}

An interferometer fixed to the surface of the Moon offers certain advantages over orbital counterparts, either free-floating in space (discussed in this section) or structurally connected (\S \ref{sec-orbital-structural}), albeit with the challenges of the lunar environment (\S \ref{sec-lunar-disadvantages}).  While small lunar telescopes can be relocated as needed, the option to easily fix them in place allows for their relative positions to be more easily determined for imaging or even more challenging astrometry programs (Figure \ref{fig-A12S3}).  The requirement of sub-nanometer baseline knowledge for sub-microarcsecond astrometric searches of exo-Earths may not be possible for a free-flying array\index{free-fliers} of telescopes.  In contrast, the metrology systems first designed for the structurally connected SIM \index{missions!SIM} could be directly applied to a lunar array.  This is identified as a key difference that motivates an astrometric science mission later in \S \ref{sec-astrometry-from-lunar-surface}.

To achieve non-astrometric science goals, we would likely use movable small telescopes, and here the lunar surface also has some advantages over the formation-flying architecture.  Rover-based telescopes can be mobile and the natural vacuum of space allows for the required reconfigurable long-delay lines to be vastly simplified compared to ground-based counterparts.  The curvature of the lunar surface does mean the horizon distance is about 1.8~km for the 1.75~m height of a typical astronaut (about half the distance of the terrestrial horizon). This does complicate deployment for ultra-long baselines greater than 5~km, but most of the demanding science cases considered in this report do not require such baseline lengths.

The lifetime of a lunar interferometer is potentially much longer than for a free-flying mission. A free flyer expends limited fuel whenever changing targets, potentially limiting mission lifetime and taking significant time to reconfigure.  On the lunar surface, the interferometer could rapidly change targets by adjusting delay lines through electromechanical means, requiring only electrical power and no need for propulsion expendables.  These operational advantages would allow a lunar interferometer to serve as a general-purpose observatory with many concurrent observing programs.

A comprehensive review of free-flier missions to date \citep[see Table 2 of][]{Monnier2019BAAS...51g.153M} illustrates the challenges of getting such constellations flown in the first place, and operating them at the required levels of precision needed for optical interferometry.  Efforts are being made to advance the technology on this front; the STARI\index{missions!STARI} mission \citep{Monnier2024SPIE13092E..2YM} is a NASA-funded orbital technology demonstration mission that will attempt to address some of these concerns.


\subsection{Lunar stations versus structurally connected orbital facilities}\label{sec-orbital-structural}
A large number of orbital space interferometer concepts have been considered, many of which incorporate a structurally connected architecture\index{structurally connected orbital interferometer}, rather than formation flying.  This includes the SIM astrometric mission \citep{Unwin2008PASP..120...38U}\index{missions!SIM} and the Space Interferometer for Cosmic Evolution (SPICE)\index{missions!SPICE} far-infrared mission \citep{Leisawitz2023AAS...24116004L,Urry2023HEAD...2010378U}, which is based on the Space Infrared Interferometric Telescope (SPIRIT) design \citep{Leisawitz2007AdSpR..40..689L}; for a more complete listing, see Table 1 of \citep{Monnier2019BAAS...51g.153M}.
Novel solutions like the Optimast concept\index{missions!Optimast} have examined in-situ manufacturing of structural components via additive manufacturing \citep{vanBelle2022SPIE12183E..1DV}.  This approach has the dual advantages that a significantly larger structure can be part of the facility, and that structure does not need to be engineered to survive launch or fit within the confines of a payload shroud.

A key advantage of structurally connected architectures is that the entire instrument can be re-pointed along the vector towards the target of interest, with two benefits.  First, the amount of differential path delay from the input apertures can be kept to a minimum, meaning active delay lines need a relatively short range.  Second, the relative placement of the input apertures as mapped onto the target may remain constant if desired, meaning that the overall input pupil is well constrained. In the SPICE concept, the input apertures are movable to set baseline length, and the structure rotates to cover baseline angles.  For beam combiners that employ Fizeau recombination techniques (i.e., image plane recombination), this can be an advantage (see \citep{Faucherre1990SPIE.1237..206F} for a discussion of Michelson versus Fizeau techniques).  Depending on the chosen orbit, a space-based interferometer may have a largely unobstructed view of the sky and therefore large mission field of regard.

Structurally connected systems avoid collision concerns of formation-flying systems, and potentially have longer lifetimes since input elements do not need consumables to maneuver relative to each other.  However, as with formation flying, stabilization of the input apertures relative to each other, and relative to the recombination element, can be difficult.  In structural systems, boom oscillations can exhibit significant excursions, complicating array phasing.  And in both structurally connected and formation-flying systems, stabilization of a free-floating baseline relative to the sky for phasing is a non-trivial complication.  This is a key attraction for lunar optical interferometry: station-keeping of input elements relative to each other is significantly simplified.

\section{Lunar disadvantages}\label{sec-lunar-disadvantages}

In comparison to an orbital interferometer, a lunar interferometer has certain advantages as noted previously (\S \ref{sec-formation-flying}, \S \ref{sec-orbital-structural}).  However, there are also disadvantages to be confronted when considering a lunar-based optical interferometry architecture.

\subsection{Dust and seismic}\label{sec-dust_seismic_sky_thermal}

\textit{Dust environment.}  As discussed in \S \ref{sec-lunar-regolith-and-dust}, the lunar dust is a significant consideration for any optical system on the Moon's surface\index{Lunar Environment!dust}.  This could be particularly problematic for architectures that intend to repeatedly relocate the input apertures.  For fixed station interferometry, the most significant dust stir-up is a one-time concern during deployment.
Overall, years of operations of the LUT\index{Instruments!LUT}  aboard the Chang'e 3 lander (Figure \ref{fig-change3})---even in the presence of surface disturbances from the mission's Yutu rover---demonstrate that lunar dust is a problem that can be satisfactorily mitigated.

\textit{Seismic noise.}  As with dust, seismic noise\index{Lunar Environment!seismic noise} is a concern to be addressed but is not a fundamental barrier to operation of an optical interferometer on the lunar surface (\S \ref{sec-seismology}).  The mid-90's ESA study on lunar interferometry \citep{Bely1996kbsi.book.....B} assessed this---incorrectly---as a significant problem; this was later challenged \citep{Mendell1998sp98.conf..451M}.  The overall background level is not problematic for imaging interferometry; for interferometric astrometry (\S \ref{sec-astrometry-from-lunar-surface}), laser metrology monitoring systems are probably necessary, though the tolerances of astrometry would demand this in virtually any setting---and the effectively absolute reference frame of the lunar surface can potentially simplify those systems.  Localized events (e.g., nearby, significant micrometeorite impacts) will drive the need for periodic re-calibration of interferometric baselines, though on an infrequent ($>$weeks) basis; probably such re-calibrations will be  necessitated by the lunar day-night cycle.

\subsection{Sky coverage}\label{sec-sky_coverage}

Nominally, any space-based facility (\S \ref{sec-formation-flying}, \ref{sec-orbital-structural}) will have a full $4\pi$ steradians of sky coverage\index{Lunar Environment!sky coverage} over the course of a year.  A surface-based facility will have some measure less than that, depending on latitude and allowable zenith angle for observing.  If we consider for the sake of argument that the latter is $\sim 45^o$, then an equatorially located interferometer will see roughly $2.8 \pi$~sr of sky coverage (70\% of the whole sky), whereas a polar facility would only see $0.58 \pi$~sr  (15\% of the whole sky).  At $60^o$ this increases to 87\% and 25\% sky coverage, respectively.  Increasing zenith angle has additional complications for baseline foreshortening (\S \ref{sec-pupil-remapping}).

\subsection{Thermal challenges}\label{sec-thermal_challenges}

\begin{wrapfigure}[15]{r}{0.5\textwidth}
 \vspace{-20pt}
\centering
    \includegraphics[width=0.5\textwidth]{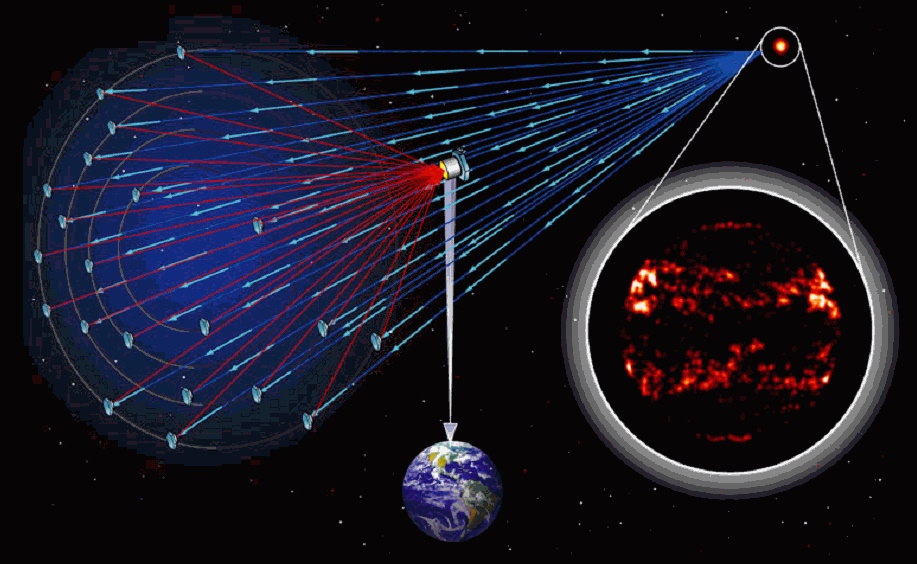}
    \caption{The Stellar Imager mission concept \citep{Carpenter2004AAS...204.0810C}.  Arrangement of the apertures in a paraboloid surface means each aperture's pathlength to the central combiner is nearly equal.  (Image credit: NASA)}\label{fig-stellar_imager}
\end{wrapfigure}

Of the topics in this section, the thermal challenges\index{Lunar Environment!thermal challenges} from the lunar day-night cycle are probably the most problematic.  This is due both to the extremes---between $\sim$50~K and 400~K---and the excursion amplitude, which is up to 275~K at the equator (Figure \ref{fig-lunar_temps}).  Extreme temperature swings can complicate optical alignments, battery operations, and electronics survivability.

\subsection{Pupil remapping, delay lines}\label{sec-pupil-remapping}

Surface-based interferometry, either terrestrial or lunar, has the complication that the entrance pupil\index{optical concepts!entrance pupil}---the arrangement of individual apertures relative to each other, as they constitute the whole array---will change over time as the object moves across the sky as the Earth or Moon rotates.  A key advantage of an orbital interferometer---formation flying (\S \ref{sec-formation-flying}) or structurally connected (\S \ref{sec-orbital-structural})---is that the pupil can be kept constant (e.g., by arranging the apertures on a paraboloid surface whose normal vector points back at the target, or at fixed locations on a structure whose rotation axis points to the target).  As noted in the previous sections, this advantage comes at various costs: stability/relative placement of the entire array is non-trivial to achieve, and there may be fuel costs for station re-positioning.  The "Vision Mission" Stellar Imager concept \citep{Carpenter2004AAS...204.0810C} had a swarm of individual apertures that would reposition themselves before relaying light to a central combiner (Figure \ref{fig-stellar_imager}); this repositioning meant the entire array of apertures could be arranged on the same paraboloid appropriate for the target being observed.  A second advantage in space is the pathlengths are followed from the target to the central combiner; each individual aperture is nominally equal, meaning beam recombination can be accomplished with a minimal range of pathlength control.

\begin{wrapfigure}[]{r}{0.65\textwidth}
\centering
   \includegraphics[width=0.65\textwidth]{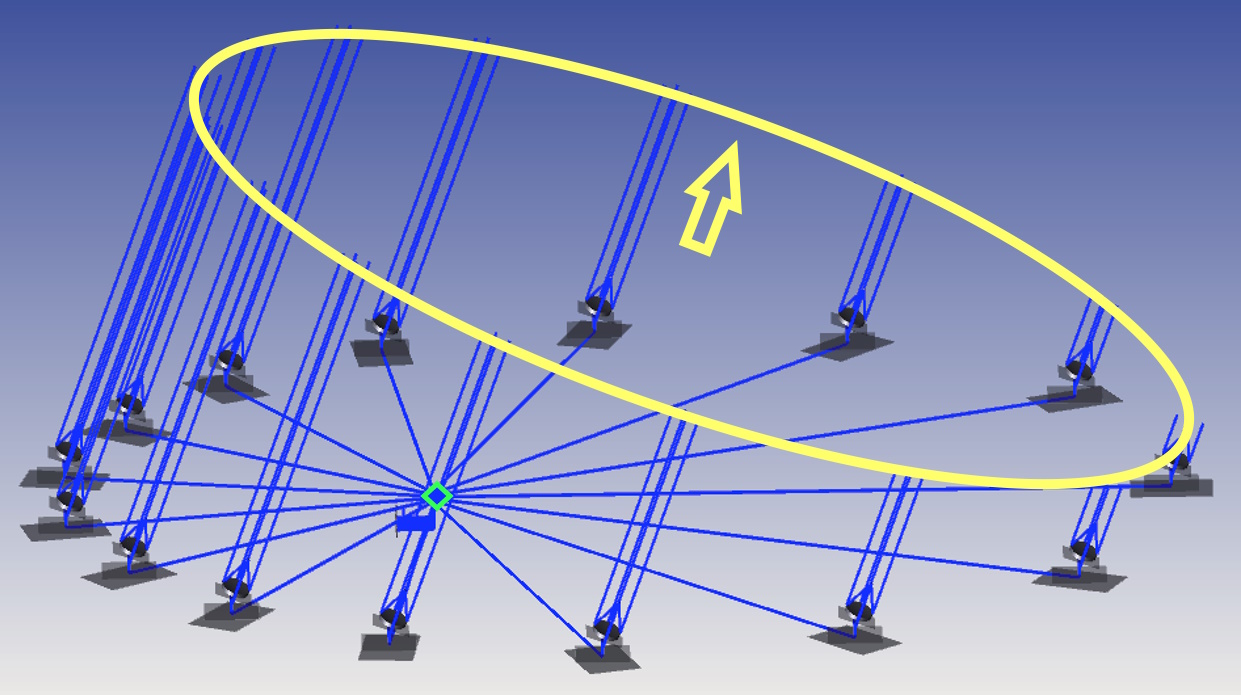}
    \caption{A surface-based array of telescopes.  The direction of the target (yellow arrow) specifies a plane (yellow circle) indicative of the plane wave arriving from that target.  The central combiner (blue diamond) is where each aperture's pathlength needs to be equal.}\label{fig-pupil-remapping} 
\end{wrapfigure}
In the case of an array of telescopes on a given surface, the array of telescopes is (typically) on a plane with a normal vector pointing at the zenith, from which the target in question has a pointing vector with some offset  (Figure \ref{fig-pupil-remapping})\index{optical concepts!pupil remapping}.  If the telescopes can be moved relative to the central beam combiner, then the mapping of the individual aperture's projected circular footprint back towards the target in question can be accomplished, which means the pathlength followed by the beam of each aperture is again roughly equal, and minimal delay compensation range is required.  The Artemis-enabled Stellar Imager (AeSI)\index{Artemis!Artemis-enabled Stellar Imager} proposal's adaptation of the Stellar Imager concept \citep{Carpenter2025arXiv250302105C} incorporates this approach.

Alternatively, if the telescope positions are fixed, pathlength equalization will require substantial amounts of available optical path delay (OPD)\index{optical concepts!optical path delay}\index{delay lines} range to accomplish a substantial amount of all-sky pointing---typically on order of the amount of telescope-to-telescope separation.  Such a large amount of OPD generally leads to a large optical system itself.  For example, the CHARA\index{missions!CHARA} delay lines shown in Figure \ref{fig-chara-delay-carts} have $\sim$200~m of OPD, and required a $\sim$100~m building.  This can be mitigated in a number of ways and with a few factors.  First, the enclosures required on Earth because of weather can be omitted on the Moon.  Second, using multiple optical passes in and out of a delay line multiplies the OPD for a given linear range of motion.  Finally, a compromise can be made by having limited but optimal sky coverage with a shorter delay line. In the case of the MoonLITE concept\index{missions!MoonLITE} \citep{vanBelle2024SPIE13092E..2NV}, a short ($<3$m) delay line supports a pair of apertures on a 100~m east-west line, and that combination of baseline orientation and the Moon's rotation means targets at all declinations will eventually rotate through the meridian strip on-sky accessible to the system.

Early lunar interferometers might make do with limited delay line throw and fixed telescopes, with more capable ensuing facilities having either movable telescopes or longer-range delay lines.  For the latter, in-situ manufacturing of rails (\S \ref{sec-ISRU}) could be useful for both repeatable alignment and reduction of dust disturbance as delay lines vary in pathlength.  One key complication of terrestrial systems---the need for an enclosure and/or extensive vacuum systems to take out "lab seeing"---is eliminated because of the lunar vacuum.


\section{Operations: day versus night }


Operation of a lunar interferometric array during daytime or nighttime has a significant trade space to consider (Table \ref{tab-day_vs_night}).
There is significant temperature variability throughout the lunar day, with a range of up to 300~K (\S \ref{sec-lunar-temperatures}). Temperature variations can impact the performance of the optical systems over the lunar day. This will have to be mitigated through a combination of thermal control systems and accurate modeling and control of the optical systems.  

Conversely, the lunar night provides
more stable (though frigid) temperatures for observing; this can be useful in the infrared due to lower background noise. The low absolute temperature is generally good for detectors, but potentially harmful to electronics. Electronic components, and especially batteries\index{batteries}, will need to be kept within their operational temperature range.

%
\begin{table}
\footnotesize
\centering
\begin{tabular}{lll}
 &
  \cellcolor[HTML]{CBCEFB}Daytime Operations &
  \cellcolor[HTML]{CBCEFB}Nighttime Operations \\ \cline{2-3}
\hline \multicolumn{1}{l|}{\cellcolor[HTML]{CBCEFB}} &
  \cellcolor[HTML]{FD6864}\begin{tabular}[c]{@{}l@{}}Free space: Scattered light a\\ significant concern\end{tabular} &
  \cellcolor[HTML]{67FD9A} \\
\multicolumn{1}{l|}{\multirow{-3}{*}{\cellcolor[HTML]{CBCEFB}\begin{tabular}[c]{@{}l@{}}Beam\\ Relay\end{tabular}}} &
  \cellcolor[HTML]{FFFFC7}\begin{tabular}[c]{@{}l@{}}Fiber: thermo-mechanical concerns,\\ radiation yellowing\end{tabular} &
  \multirow{-3}{*}{\cellcolor[HTML]{67FD9A}\begin{tabular}[c]{@{}l@{}}Free space: no\\ scattered light\end{tabular}} \\
\hline
\multicolumn{1}{l|}{\cellcolor[HTML]{CBCEFB}Power} &
  \cellcolor[HTML]{67FD9A}\begin{tabular}[c]{@{}l@{}}Solar power\\ \\ Battery-based "survive the\\ night" still needed\end{tabular} &
  \cellcolor[HTML]{FD6864}\begin{tabular}[c]{@{}l@{}}Battery-based: mass,\\ thermal concerns\end{tabular} \\
\rowcolor[HTML]{FFFFC7}
\hline
\multicolumn{1}{l|}{\cellcolor[HTML]{CBCEFB}Thermal} &
  Greater temperature excursions &
  \begin{tabular}[c]{@{}l@{}}Frigid but stable\\ \\ Better for MIR observing\end{tabular}
\end{tabular}%
\caption{Considerations for daytime versus nighttime operations for beam relay, power, and thermal.  Red indicates problematic areas that will affect mission budget and/or performance; yellow is for areas of concern to watch; green is for areas with known and/or readily accessible solutions.}
\label{tab-day_vs_night}
\end{table}

Power is a major consideration, especially for observing overnight. For missions that aim to observe during the day, it should be much cheaper to maintain enough power for the entire system. If observing is to be done during the night, mass and cost budgets will have to accommodate larger power and storage systems. Section \ref{sec-survive-the-night} elaborates on this topic, and it should be a critical consideration for CONOPS.


\chapterimage{intuitive_machines_launch.jpg} 
\chapterspaceabove{8.0cm} 
\chapterspacebelow{8.25cm} 

\part{The Missions}

\chapter{Supporting Technologies}\label{sec-lunar-supporting-technologies}

\section{Surface access}\label{sec-surface-access}

Since 2010, attempted lunar landings have resumed after a substantial hiatus following the US and United Soviet Socialist Republic/Russian efforts of the 1960s and 1970s (Table \ref{tab-lunar_landers}).  This includes several new national entrants, such as China, Japan, and India, as well as private companies that blur the line between commercial enterprise and national identity, such as SpaceIL (Israel), ispace (Japan), Astrobotic Technology, Firefly Aerospace, and Intuitive Machines (US). NASA-led efforts include small-to-large CLPS landers, as well as Human Landing System (HLS) vehicles in both crewed and cargo variants.

\subsection{CLPS landers}\label{sec-CLPS-landers}

\begin{wrapfigure}{R}{8.75cm}
    \vspace{-0.25cm}
    \begin{minipage}[h]{1\linewidth}
        \begin{tcolorbox}[colback=gray!5,colframe=green!40!black,title={Regular surface access}]
        Multiple CLPS landers are being sent to the Moon every year.
        \end{tcolorbox}
    \end{minipage}
\end{wrapfigure}

NASA’s \href{https://science.nasa.gov/lunar-science/clps-deliveries/}{Commercial Lunar Payload Services (CLPS)} initiative\index{CLPS} allows for the acquisition of lunar delivery services from US companies for payloads that NASA has developed to accomplish its goals at the Moon. A considerable variety of payloads have either been or are being developed at NASA centers, academia, or at industry partners for delivery to the Moon. Once payloads have been manifested, task orders for lunar delivery are awarded via a competitive process among a pool of companies that can target landing sites distributed across the surface of the Moon, including the south polar region and the farside.

Pathways for delivering elements of a lunar interferometer system on a CLPS lander involve the following steps:
\begin{enumerate}
    \item
Securing instrument development funding through a variety of different programs (e.g., \href{https://nspires.nasaprs.com/external/solicitations/summary.do?solId=%7B8C8C2628-3976-F04F-DC88-70706B738C94%7D&path=&method=init}{Pioneers}\index{Mission Opportunities!Pioneers}, \href{https://nspires.nasaprs.com/external/solicitations/summary.do?solId=%7bAD1DEAD1-7060-2C93-8CD1-780AF8FC9D54%7d&path=&method=init}{Payloads and Research Investigations on the Surface of the Moon/PRISM}, \href{https://nspires.nasaprs.com/external/solicitations/summary.do?solId=%7b383AAAC1-8CEF-F372-7DA6-4D6A3279A820%7d&path=&method=init}{SALSA}, \href{https://nspires.nasaprs.com/external/solicitations/summary!init.do?solId={9987D88F-0A12-5203-FC25-423773FAF134}&path=open}{Space Technology Mission Directorate/STMD tipping point}). Some of these programs come with "guaranteed" inclusion on a CLPS lander while others don’t necessarily include CLPS delivery.
\item
Manifesting on a CLPS lander occurs through the CLPS Manifest Selection Board (CMSB)\index{CLPS!CLPS Manifest Selection Board}. This can occur in various ways---if the payload requires multiple landers, then manifesting could occur via multiple task orders awarded to the same or different companies. A single task order could also be awarded to one company for multiple landers.
\item
Once the task order is awarded, then payload requirements are agreed to and the mission proceeds towards launch and landing.
\item During flight, CLPS task orders include mission operations, where the lander vendor provides support and coordination for the NASA payload to conduct its surface operations.
\end{enumerate}

The simplest lunar interferometry missions are a good match for the CLPS program, as the accommodability requirements are well within the capabilities of the CLPS vendors. A single lander lunar interferometer requires a small rover for deployment of a collector over manageable distances and then requires nighttime survival of the payload, which may be supported by the CLPS lander. All of these features can be accommodated within the range of capabilities provided by the CLPS program.

More complex lunar interferometry missions would likely require larger landers, or multiple landers, with mobility, and nighttime survival and would likely be more challenging to accomplish within the CLPS program. Since CLPS landers are rarely committed to a single payload---these more complex missions may be better accomplished using a different surface access program, or modifications to the CLPS program.

The CLPS program aims to send eight landers to the Moon over four years. As of the time of this report, there are 13 active CLPS providers.  CLPS enables rapid, low-cost access to the lunar surface; the current \$2.6B contract with NASA ends in 2029 \citep{nasaNASAAnnouncesCLPS}. Discussions on the future of CLPS are ongoing.

\subsection{HLS Artemis deployed payloads}\label{sec-HLS-Artemis}\index{Artemis}

\begin{wrapfigure}{R}{7.75cm}
    \vspace{-0.25cm}
    \begin{minipage}[h]{1\linewidth}
        \begin{tcolorbox}[colback=gray!5,colframe=green!40!black,title={Astronaut-enabled science}]
        The ADI program follows on from the "suitcase science" of the Apollo era.
        \end{tcolorbox}
    \end{minipage}
\end{wrapfigure}
The Artemis Deployed Instruments  (ADI) program \index{Artemis!Artemis Deployed Instruments}(\href{https://nspires.nasaprs.com/external/solicitations/summary!init.do?solId={76053627-8933-47CD-B627-C5DCDD474076}&path=open}{A3DI}, \href{https://nspires.nasaprs.com/external/solicitations/summary.do?solId=%7bA2E6F92B-D1CD-85F6-C583-C2ACC955B340%7d&path=&method=init}{A4DI}) intends to provide a means for science instruments to be deployed on the surface of the Moon during the upcoming crewed Artemis landings. Payloads selected through this program element would be part of an individual Artemis mission payload manifest and would be deployed on the surface during the crewed activities on the lunar surface. The Artemis landers will target sites in the south polar region of the Moon, within 6$^o$ of latitude from the lunar South Pole.  Proposed deployed instruments can address science objectives related to "observing the universe and the local space environment from a unique location."

The mass limit for ADI payloads is 60~kg, which will constrain potential interferometer payloads to small instruments.
This limit is determined by NASA guidelines on maximum payload size that can be safely handled by astronauts on the lunar surface during extravehicular activities (EVAs). 
Current ADI guidelines indicate the entire instrument should be within this mass envelope, but a future instrument could be consistent with updated handling guidelines that allow multiple components that are each below the 60~kg mass limit.
Certain deployment elements could require unspooling optical fiber and power cables that could be handled well by astronauts. In addition, interferometry payloads would benefit from specific positioning on the surface to ensure line of sight and proper distance between elements.

\subsection{HLS cargo}\label{sec-HLS-cargo}
\begin{wrapfigure}{R}{8.75cm}
    \vspace{-0.25cm}
    \begin{minipage}[h]{1\linewidth}
        \begin{tcolorbox}[colback=gray!5,colframe=green!40!black,title={Large payloads with HLS}]
        Multiple tons of cargo can be delivered to the lunar surface with a single HLS lander.
        \end{tcolorbox}
    \end{minipage}
\end{wrapfigure}

The large cargo capacity of HLS\index{Human Landing System} landers will enable transportation of huge amounts of material to a point where it may be challenging to fill up the capability. These landers include the SpaceX Lunar Starship, and the Blue Origin Blue Moon lander.  While NASA contracts currently exist for delivery of single large items, such as a habitat and pressurized rover, additional space may be available on the HLS cargo flights or additional demonstration flights of HLS landers.

Lunar interferometer instrument elements would be well suited for delivery in this manner, especially if robotic deployment via a large rover was available. Lunar interferometers may have flexibility specific to the deployment site and may not require long transport. Furthermore, deployment requirements may be limited to delivery of elements to specific locations without necessitating other more complex manipulations.
Currently, it is unclear how payloads could be delivered via HLS flights; contract modifications and non-NASA funding may be required.






\begin{table}[]
\centering
\resizebox{\textwidth}{!}{%
\begin{tabular}{lllll}
\textbf{Provider} &
  \textbf{Lander} &
  \textbf{\begin{tabular}[c]{@{}l@{}}Payload\\ Capacity\\(appx.)\end{tabular}} &
  \textbf{\begin{tabular}[c]{@{}l@{}}Landing /\\Status\end{tabular}} &
  \textbf{Notes} \\
\hline
\textit{Launched}\\
\hline
China             & \href{https://nssdc.gsfc.nasa.gov/nmc/spacecraft/display.action?id=2013-070A}{Chang’e 3} & & 2013-12-14$^a$ & Landed Mare Imbrium\\
China             & \href{https://nssdc.gsfc.nasa.gov/nmc/spacecraft/display.action?id=2018-103A}{Chang’e 4} & & 2019-01-03$^a$ & Landed Von Karman crater on the farside\\
SpaceIL (Israel)  & \href{https://nssdc.gsfc.nasa.gov/nmc/spacecraft/display.action?id=2019-009B}{Beresheet}              & 5~kg            & 2019-04-04$^c$                 & Lunar X Prize participant\\
India             & \href{https://nssdc.gsfc.nasa.gov/nmc/spacecraft/display.action?id=2019-042A}{Chandrayaan 2}        & 26–27~kg        & 2019-09-07$^c$       & Lunar South Pole region-focused               \\
China             & \href{https://nssdc.gsfc.nasa.gov/nmc/spacecraft/display.action?id=2020-087A}{Chang’e 5} & & 2020-12-01$^a$ & \begin{tabular}[c]{@{}l@{}}Landed Mons R\"{u}mker region of Oceanus\\ Procellarum; 1.73kg sample return\end{tabular}\\
India             & \href{https://nssdc.gsfc.nasa.gov/nmc/spacecraft/display.action?id=2023-098A}{Chandrayaan 3}        & 26–27~kg        & 2023-09-23$^a$      & Lunar South Pole region-focused               \\
ispace            & \href{https://nssdc.gsfc.nasa.gov/nmc/spacecraft/display.action?id=2022-168A}{Series 1}               & 30~kg           & 2023-04-26$^c$                 & \begin{tabular}[c]{@{}l@{}}Japan-based; small payload\\delivery to Mare Frigoris \end{tabular}      \\
Russia            & \href{https://nssdc.gsfc.nasa.gov/nmc/spacecraft/display.action?id=2023-118A}{Luna 25}                & 30~kg           & 2023-10-19$^c$ & Intended for Boguslawsky crater       \\
Astrobotic        & \href{https://nssdc.gsfc.nasa.gov/nmc/spacecraft/display.action?id=PEREGRN-1}{Peregrine 1}  & 90 kg           & 2024-01-18$^c$                 & CLPS\index{CLPS}; equatorial regions, Sinus Viscositatis       \\
Japan              & \href{https://nssdc.gsfc.nasa.gov/nmc/spacecraft/display.action?id=2023-137D}{SLIM}                   & 30~kg           & 2024-01-20$^b$                 & \begin{tabular}[c]{@{}l@{}}Precision landing technology demo,\\ landed Shioli crater\end{tabular}   \\
Intuitive Machines& \href{https://nssdc.gsfc.nasa.gov/nmc/spacecraft/display.action?id=IM-1-NOVA}{Nova-C / IM-1}          & 130~kg          & 2024-02-22$^b$                 & CLPS; Landed at Malapert A \\
China             & \href{https://nssdc.gsfc.nasa.gov/nmc/spacecraft/display.action?id=CHANG-E-6}{Chang’e 6} & & 2024-06-01$^a$ & \begin{tabular}[c]{@{}l@{}}Landed farside South Pole--Aitken\\  basin; 1.94kg sample return\end{tabular}\\
Firefly Aerospace & \href{https://nssdc.gsfc.nasa.gov/nmc/spacecraft/display.action?id=BLUEGHOST}{Blue Ghost M1}          & 150~kg          & 2025-03-02$^a$              & CLPS; Landed at Mare Crisium  \\

ispace              & \href{https://ispace-inc.com/m2}{Series 1 RESILIENCE}               & 30 kg      & 2025-01-15$^e$              &
\begin{tabular}[c]{@{}l@{}}HAKUTO-R Mission 2\\  May-June 2025 landing at Mare Frigoris \end{tabular}\\
Intuitive Machines& \href{https://www.intuitivemachines.com/im-2}{Nova-C / IM-2}          & 130 kg          & 2025-03-07$^b$              & CLPS; PRIME-1 mission     \\
\hline

\textit{Planned}\\
\hline
Blue Origin/Sierra& \href{https://www.blueorigin.com/blue-moon/mark-1}{Blue Moon MK1}          & 3,000~kg   & 2025              & Cargo variant      \\
Astrobotic        & \href{https://nssdc.gsfc.nasa.gov/nmc/spacecraft/display.action?id=VIPER}{Griffin}                & 500~kg          & 2025                     & \begin{tabular}[c]{@{}l@{}}CLPS; NASA VIPER (canceled), replaced\\by Astrobotic FLIP rover, to the lunar South Pole\end{tabular}\\
Intuitive Machines& Nova-C / IM-3          & 130~kg          & 2026                     & CLPS; PRISM mission    \\
                  & Nova-C / IM-4          & 130~kg          & 2027                     & CLPS     \\
                  & Nova-D                 & 500–1,000~kg    & Proposed                 & Larger lander for future missions       \\
Firefly Aerospace & \href{https://fireflyspace.com/missions/blue-ghost-mission-2/}{Blue Ghost M2}          & 150 kg          & 2026                     & CLPS; farside      \\
                  & Blue Ghost M3          & 150~kg          & 2028                     & CLPS; Gruithuisen domes      \\
Draper            & \href{https://nssdc.gsfc.nasa.gov/nmc/spacecraft/display.action?id=DRAPER}{SERIES-2}               & 300–500~kg      & 2026                     & CLPS; Targeting Farside Schr\"{o}dinger basin     \\
ispace & \href{https://spacenews.com/ispace-revises-design-of-lunar-lander-for-nasa-clps-mission/}{APEX 1.0} & 300~kg & 2026 & CLPS; US-based\\
SpaceX            & \href{https://www.spacex.com/humanspaceflight/moon/}{Lunar Starship}         & \textgreater{}100 tons & \begin{tabular}[c]{@{}l@{}}2026 demo\\(uncrewed)\end{tabular}       & Artemis III/IV HLS provider                 \\
Blue Origin/Sierra& Blue Moon MK2          & 20~tons  & \begin{tabular}[c]{@{}l@{}}2027 demo\\(uncrewed)\end{tabular}       & Artemis V HLS provider in 2029      \\
China             & \href{https://nssdc.gsfc.nasa.gov/nmc/spacecraft/display.action?id=CHANG-E-7}{Chang’e 7}             & 30–100~kg       & 2026  & Farside missions, sample return, polar  \\
Russia            &  Luna 26/27          & 30~kg           & Planned & Returning to lunar exploration          \\
ESA               & EL3                    & 1.5~tons & 2028                     & Supporting Artemis and science missions \\
\hline
\end{tabular}%
}
\footnotesize{$^a$ successful; $^b$ partial success; $^c$ failed; $^d$ launched and en route; $^e$ manifest for launch}\\
\caption{Lunar landers recently flown and/or currently under development, as of February 1, 2025.}
\label{tab-lunar_landers}
\end{table}

\subsection{Mobility}\label{sec-mobility}


\begin{wrapfigure}{R}{8.75cm}
    \vspace{-0.25cm}
    \begin{minipage}[h]{1\linewidth}
        \begin{tcolorbox}[colback=gray!5,colframe=green!40!black,title={Getting around}]
        A large number of rovers, of all sizes, are under development for lunar mobility.
        \end{tcolorbox}
    \end{minipage}
\end{wrapfigure}

Early landers are carrying micro rovers such as the ispace RESILIENCE mission carrying the TENACIOUS rover \citep{ispace_tenacious}; this rover is 26~cm tall, 31.5~cm wide, 54~cm long, and will weigh approximately 5~kg.
The NASA Volatiles Investigating Polar Exploration Rover\index{missions!VIPER} (VIPER), a larger (430~kg) rover originally intended for a late 2025 launch aboard the Astrobotic Griffin\index{missions!Astrobotic Griffin} lander, has been canceled/delayed \citep{theguardian_viper}, but the Astrolab Flexible Logistics and Exploration (FLEX) Lunar Innovation Platform\index{missions!Astrolab FLIP} (FLIP) rover is now in its place \citep{astrobotic-FLIP} on the Astrobotic Griffin lander slated to fly in either case.  The FLIP rover is designed to carry payloads up to 30~kg, and will demonstrate a capability that can be utilized in future missions.  It will also demonstrate technologies for Astrolab's much larger FLEX rover\index{missions!Astrolab FLEX}, which is designed to carry payloads up to 1,600~kg in mass and 3~m$^3$ in volume.

For crewed missions, NASA's \href{https://www.nasa.gov/suits-and-rovers/lunar-terrain-vehicle/}{Lunar Terrain Vehicle} (LTV) competition recently selected three finalists: Astrolab \href{https://www.astrolab.space/flex-rover/}{FLEX} rover, Intuitive Machines \href{https://www.intuitivemachines.com/post/intuitive-machines-unveils-moon-racer-ltv}{Moon Racer}, and Lunar Outpost \href{https://www.lunaroutpost.com/ltv}{LTV}.
More ambitiously, JAXA's pressurized \href{https://global.toyota/en/mobility/technology/lunarcruiser/}{Lunar Cruiser} is being developed by Toyota as part of the Artemis\index{Artemis} program, which will enable 2-4 astronauts to make sorties across the lunar surface lasting a month at a time.

\section{Power}\label{sec-survive-the-night}

Two interconnected issues for any lunar facility are the generation of power and the ability to either operate, or at least survive, during the deep cold of the lunar night.  The lunar cycle is nominally 14 Earth days long in sunlight, then 14 Earth days of night; near-pole locations with persistent lighting can dramatically alter that balance (\S \ref{sec-persistently-lit}) in favor of system architectures that benefit from more sunlight.  In many cases, nighttime operations are not a part of CONOPS, but sufficient "survive-the-night" capability is built such that nighttime does not fatally damage a system and sunrise can re-activate a hibernating system.

\subsection{Power generation and heating}

Power for operations and heating on the Moon likely relies either on solar\index{solar power} or nuclear power generation (\S \ref{sec-nuclear-power}).  Solar power has been the primary source for the robotic missions to date.
The latter approach are typically either radioisotope thermoelectric generators (RTGs)\index{radioisotope thermoelectric generators}, which can generate hundreds of watts of power and significant additional heat, or radioisotope heater units (RHUs)\index{radioisotope heater units}, which are small units with grams of radioactive material producing on the order of 1~watt of heat for nighttime support. Solar approaches need to deal with long lunar nights, necessitating significant battery mass\index{batteries}, while nuclear approaches face additional up-front costs and logistical challenges due to the regulation of nuclear materials.

Of these two approaches, solar and RHUs have been employed with the recent missions to the Moon; no lunar missions of the "modern" era (after 1972) have employed RTGs.  The Apollo missions included RTGs with their Apollo Lunar Surface Experiments Package (ALSEP) packages \citep{Lewis2008LPI....39.1356L}.

\subsection{Power storage}\label{sec-power-storage}\index{batteries}

\begin{wrapfigure}{R}{8.75cm}
    \vspace{-0.25cm}
    \begin{minipage}[h]{1\linewidth}
        \begin{tcolorbox}[colback=gray!5,colframe=green!40!black,title={Power through the darkness}]
        Power storage for operations---or simply survival---is needed for, and complicated by, the long and cold lunar night.
        \end{tcolorbox}
    \end{minipage}
\end{wrapfigure}
Details of power storage will depend on instrument CONOPS---day versus night operations (\S \ref{tab-day_vs_night}). Power storage demand depends on the complexity of the interferometer, which may require precise instruments with optical components, cooling systems, and data acquisition hardware that consume significant power. Additionally, power storage considerations for data processing and transmission need to be considered.

Extended lunar nights (14 Earth days or more depending on topographic shielding of the horizon) impose further challenges for power storage capabilities. Relying on solar-powered systems means power storage must sustain the entire operation during lunar nights. Depending on whether the interferometer will be operating during the night or solely staying alive will greatly affect the power storage requirements.

Power storage systems must have high energy density to sustain long operational periods. Advanced lithium-ion (Li-ion) batteries, solid-state batteries, or other high-capacity systems are candidates. In general, the case against rechargeable batteries is that these units have lower energy density, therefore, more volume or weight would be required to provide equivalent power storage capacity. Non-rechargeable primary batteries can have specific energy capacity greater than rechargeable batteries by a factor of a few, so they may be desired for missions operating only over one or two nights.

Using most modern Li-ion rechargeable battery chemistries, a rule of a thumb for the battery mass required for overnight energy needs is roughly 5~kg of battery per watt of average overnight power requirements. Therefore, even overnight survival likely requires tens of kilograms of battery mass, while more significant operations may require hundreds of kilograms of battery mass.  Batteries that are in use in proximity to human operations will have additional requirements for meeting safety protocols related to human spaceflight.


\section{Communications}\label{sec-communications}


For lunar nearside operations, a direct line of sight with Earth potentially enables high data rate communications.  The complexity of a given interferometry facility will dictate the need for both local and remote data exchange, in terms of volume, rates, and allowable latencies.  Current terrestrial facilities operate at frame rates in the kilohertz range, which in turn levies requirements upon station-to-station communications.  This demand is typically met at those facilities by either standard gigabit ethernet links, or even serial links like RS-422.  However, much of those frame rates are derived from terrestrial atmospheric fluctuation rates, so it is possible that the quieter lunar environment will relax those requirements.  Additionally, much of the raw collected data is collected, processed, and averaged down before utilization, leading to a sizable reduction in data exchanged between stations or relayed back for scientific utilization.  A simple experiment like MoonLITE (\S \ref{fig:MoonLITE}) is expected to produce only a few tens of megabytes of non-realtime science observation data per lunar day, which is well within the expected CLPS data allocation for a hosted experiment.

Missions to the lunar farside require orbital relay infrastructure; for example, the Chang'e-4 and -6 landers\index{missions!Chang'e-4, -6} set down and operated on the farside, and used the Queqiao-1\index{missions!Queqiao-1, -2} and -2 ("Magpie Bridge") communications relay satellites, respectively, to contact Earth \citep{Reuters_queqiao2,nasaNASANSSDCA-queqiao2}.  Relay spacecraft enable farside operations, but will be a limiting factor in data relay bandwidth, latency, and quantity.  Optical interferometry---unlike radio interferometry---does not gain a significant noise reduction advantage in being located on the farside.  

\section{Parallel investments}

\begin{wrapfigure}{R}{8.75cm}
    \vspace{-0.25cm}
    \begin{minipage}[h]{1\linewidth}
        \begin{tcolorbox}[colback=gray!5,colframe=green!40!black,title={A broad swath of developments}]
        A large number of supporting technologies are being developed for sustaining prolonged-to-permanent activity on the lunar surface
        \end{tcolorbox}
    \end{minipage}
\end{wrapfigure}
Industry investments in lunar infrastructure enhance the prospects of optical interferometry on the Moon.
The \href{https://lsic.jhuapl.edu/}{Lunar Surface Innovation Consortium} (LSIC) at Johns Hopkins University Applied Physics Laboratory is tracking multiple areas of development, including in-situ resource utilization (ISRU), surface power, excavation and construction, and crosscutting capabilities such as dust mitigation and surviving extreme environments.  Overall, it is reasonable to expect the rapidly emerging lunar industrial base will provide capabilities useful for a wide variety of lunar astrophysics facilities, including interferometers.


\subsection{Solar power: polar power towers}\label{sec-power-towers}\index{solar power}

Commercial companies are providing technical solutions suitable for deployment on Mons Malapert. If a lunar interferometer were deployed in a PSR at the base of the mountain, the kilometers of vertical distance from the peak of Mons Malapert to the base of the mountain would require power beaming and line of sight with the interferometer’s receiver and delivery hub (\S \ref{sec-power-beaming}). It is also conceivable that tall towers could be deployed on crater rims near the interferometer, rather than being deployed on Mons Malapert.

\begin{wrapfigure}{r}{6.75cm}
    \centering
    \includegraphics[width=0.95\linewidth]{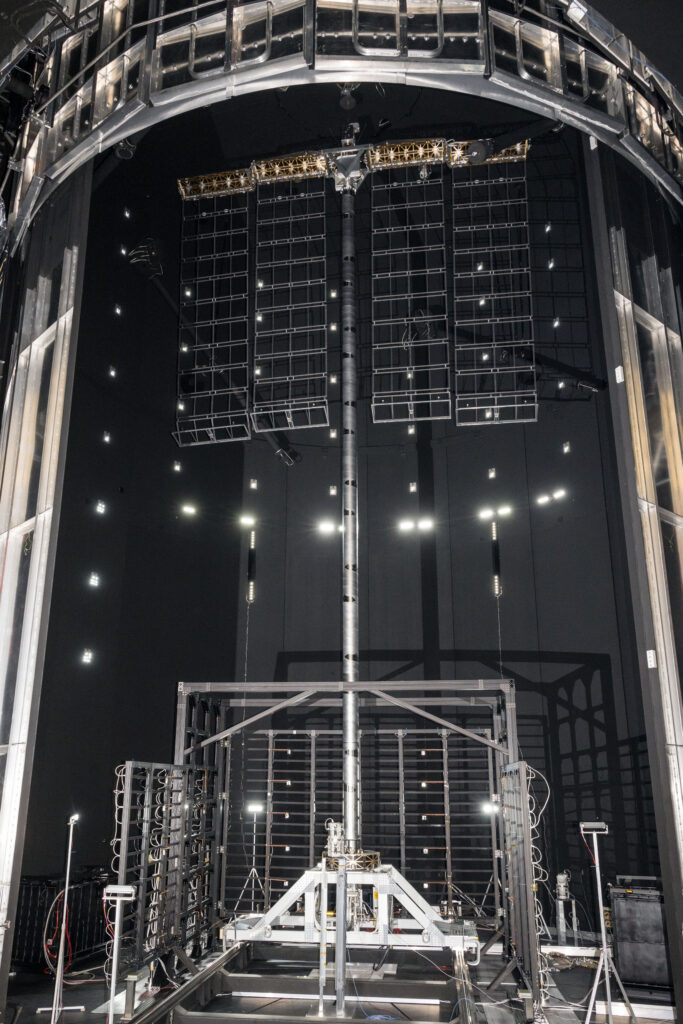}
    \caption{Thermal Vacuum Testing of LAMPS at NASA Johnson Space Center. (Photo credit: NASA/David DeHoyos)}
    \label{fig-LAMPS}
\end{wrapfigure}

Due to its constant view of Earth and long periods of sunlight, Mons Malapert is considered a prime location for lunar infrastructure (i.e., telescopes, lunar bases, communication systems, etc.). However, the peak of Mons Malapert stands kilometers from the base of the mountain where short- to medium-term exploration and resource exploitation are likely to take place (i.e., in PSRs). Moreover, access to Mons Malapert's peak from the base would be extremely difficult for astronauts and rovers since local slopes are often >20$^{\circ}$.

To avoid navigating operational complexities associated with Mons Malapert massif, commercial companies are developing tall towers to artificially create the illumination conditions and DTE communications with Earth experienced at the mountain top. For example, Honeybee Robotics has developed the \href{https://www.honeybeerobotics.com/news-events/honeybee-robotics-deploys-lamps-at-nasa-johnson-space-center/}{Lunar Array Mast and Power System} (LAMPS) (Figure \ref{fig-LAMPS}), a 20 m robotically deployable solar array that rotates and tracks sunlight and generates 10 kW of power. LAMPS is restowable and utilizes a Deployable Interlocking Actuated Band for Linear Operations (DIABLO) for its retracting mast. Dust-tolerant connectors are utilized for surface charging. LAMPS has successfully been deployed and retracted in NASA Johnson Space Center Chamber A.

\begin{figure}
    \centering
    \includegraphics[width=0.8\linewidth]{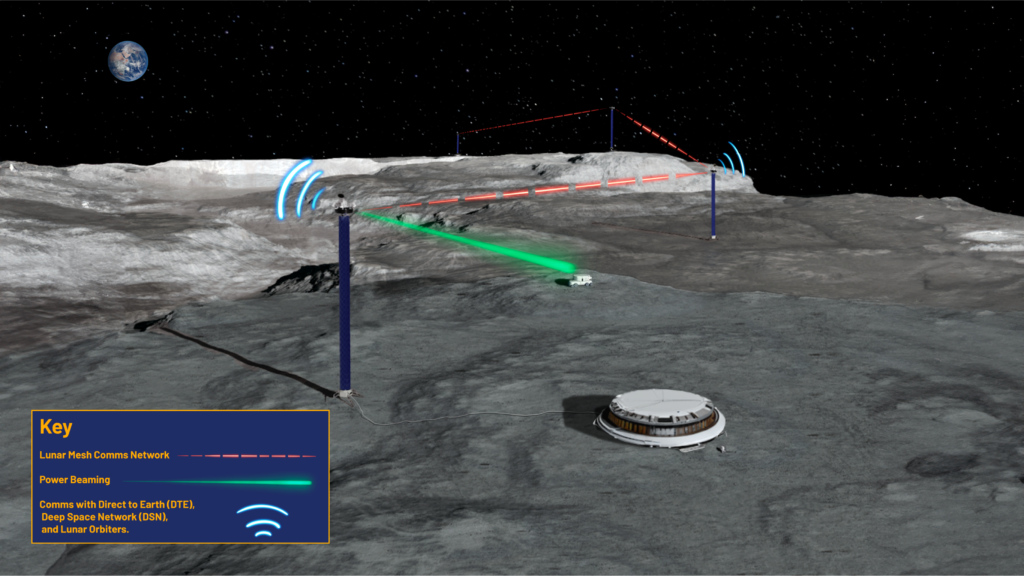}
    \caption{Schematic depicting LUNARSABER services to various assets on the Lunar Surface. (Image credit: Honeybee Robotics/Grayson Glazer)}
    \label{fig-lunarsaber}
\end{figure}

While LAMPS would be suitable for a small-class mission, systems in the Explorer or Probe/Flagship Class may require towers capable of providing more than just power. \href{https://www.honeybeerobotics.com/news-events/honeybee-robotics-to-develop-lunarsaber-for-darpas-luna-10-program/}{Lunar Utility with Navigation, Advanced Remote Sensing, and Autonomous Bearing} (LUNARSABER; Figure \ref{fig-lunarsaber}), for example, is a deployable structure concept that integrates solar power, power storage and transfer, communications, mesh network, PNT (position, navigation, and timing), and surveillance into a single infrastructure. LUNARSABER's mast uses Honeybee's DIABLO to deploy up to 100 m in height, is capable of producing 100 kW of power, and has a payload mast of 1,000~kg at its masthead. It is conceivable that a power-beaming system could be deployed at the top of LUNARSABER's mast. If a 100 m tower were deployed, the square area of land where power could be distributed increases dramatically. Systems could then conceivably be deployed in darkness and near continuous power from LUNARSABER wirelessly. Moreover, a network of LUNARSABERs could provide continuous communications from an optical interferometer located at the poles to a base at the mid-latitudes.

\subsection{ISRU manufacturing}\label{sec-ISRU}

In-situ resource utilization (ISRU) on the Moon has progressed from theoretical studies to funded projects, laboratory demonstrations, and upcoming field trials. These technologies are crucial for reducing reliance on Earth-based supply chains and enabling long-term lunar operations. These government and commercially supported initiatives are critical to reducing the costs and logistical complexity of lunar infrastructure development. They are particularly important for enabling large-scale projects, such as lunar-based observatories, by providing construction materials, energy systems, and mobility solutions using local resources. There are a number of key ongoing ISRU developments and demonstrations:\\

\textit{Resource Identification and Mapping.}
Missions like NASA’s Lunar Reconnaissance Orbiter\index{missions!Lunar Reconnaissance Orbiter} \citep{nasaLunarReconnaissance} and India's Chandrayaan 1\index{missions!Chandrayaan-1} \citep{chandrayaan1} have mapped polar water ice and volatiles.
NASA’s 
missions, picking up on the efforts and intentions of the unfortunately unsuccessful Lunar Trailblazer\index{missions!Lunar Trailblazer}, can enhance our understanding of the form, abundance, and distribution of water on the Moon and the lunar water cycle. Future missions to support the characterization of water on the lunar surface and subsurface can provide detailed constraints on this critical energy feedstock \citep{lunartrailblazer}.\\

\textit{Regolith-Based Construction.}
ESA and NASA have demonstrated 3-D printing technologies using simulated regolith to build structures, paving, and radiation shielding \citep{dyoungESAsInnovative,techbriefsRegolithPolymerPrinting}.
ICON, under NASA funding, is adapting this technology for lunar infrastructure construction, including roads and platforms for observatories \citep{iconbuildICONDevelop}.

\textit{Metal and Oxygen Extraction.}
Blue Origin’s Blue Alchemist\index{instruments!Blue Alchemist} program \citep{blueoriginBlueAlchemist}, funded by NASA, has demonstrated molten regolith electrolysis (MRE) to extract oxygen and metals, including aluminum and silicon, for construction and solar cell manufacturing. ESA’s Protolab\index{instruments!Protolab} achieved a 96\% oxygen yield from simulated regolith, with metals suitable for building components as byproducts \citep{ESAoxygenmetal}. A variety of lunar regolith simulants have been evaluated for oxygen yield and will likely be a growing research area by many space agencies in the coming years \citep{shi2022,strangeaddo,Lomax2025AcAau.234..287L}. Other metals, like titanium and iron, could be extracted from the lunar regolith as by-products of primary oxygen extraction, especially in cases where titanium-rich deposits (e.g., titanium basalts and "black spot" pyroclastic deposits) have been shown to yield more oxygen than titanium-poor regions (e.g., highlands).

\textit{Energy Systems.}
Blue Origin has demonstrated ISRU-derived solar cells from lunar regolith, supporting scalable and sustainable energy production. NASA’s Kilopower fission nuclear reactors were intended to provide continuous power for ISRU and observatory operations, particularly during the two-week lunar night \citep{nasaKilopowerNASA}, along with the follow-on Fission Surface Power project \citep{nasaFissionSurface}.

\textit{Mobility Solutions.}
The Lunar Surface Innovation Consortium (LSIC) has identified regolith-based roads and rail systems as enabling technologies for surface mobility (LSIC Vision) \citep{LSICvision}. NASA is funding early-stage concepts for regolith-based paving to reduce dust and improve transportation efficiency. Regolith construction systems mentioned previously could be a critical component of this infrastructure development.

\vspace{5pt}

ISRU capabilities provide essential support for constructing and operating advanced astronomical observatories on the Moon, such as optical interferometers, which require precise alignment, mobility, and energy.

\textit{Railways and Interferometer Delay Lines.}
Large-scale optical interferometers on the Moon require precise and stable delay lines to account for differences in light path lengths between telescope apertures.
ISRU can enable the construction of regolith-based or metal railway systems for mobile delay line units, allowing precise, vibration-free movement.

\textit{Aperture Mobility and Precision Alignment.}
Optical interferometers often rely on widely spaced apertures for high-resolution imaging. Mobility systems (e.g., rovers or rail-based platforms) enable the movement and reconfiguration of these apertures. ISRU-derived materials (e.g., metals from MRE or basalt fibers from regolith) can be used to construct lightweight, durable tracks or mobility systems for telescopes.
NASA-funded regolith compaction and paving technologies (NASA's Surface Innovation Initiative) ensure stable bases for moving apertures, minimizing vibrations and alignment drift.

\textit{Construction of Observatory Infrastructure.}
ISRU-derived structural materials (e.g., aluminum, silicon) can be used to fabricate the supports, frames, and components of interferometers. Regolith-based construction can provide shielding for sensitive instruments, protecting them from radiation and thermal fluctuations.


\vspace{5pt}

%
%



\subsection{Power beaming}\label{sec-power-beaming}

\begin{wrapfigure}{right}{8cm}
\includegraphics[width=0.52\textwidth, angle=0]{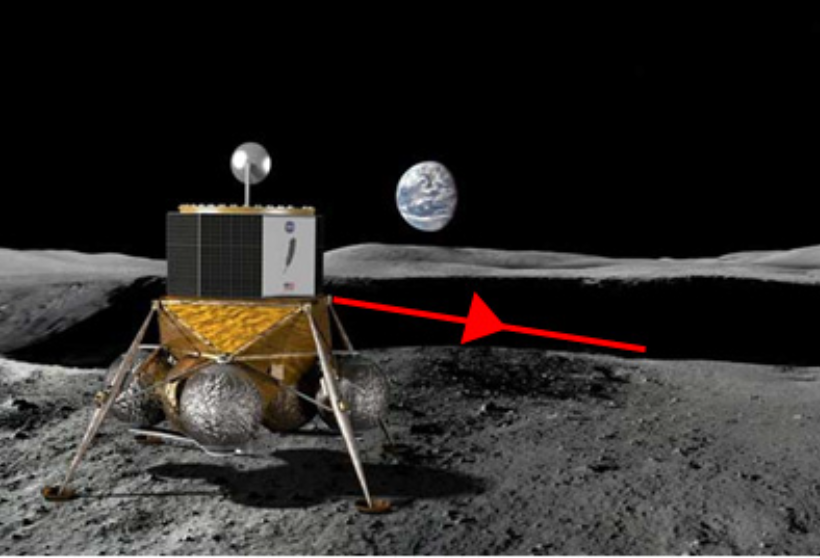}
\caption{Power collection station beaming energy into a permanently shadowed region. (Image credit: Blue Origin)}  
\label{fig-power_beaming}
\end{wrapfigure}

While a number of types of energy storage (\S \ref{sec-power-storage}) may be available to enable survival and operation of elements of an interferometry mission through periods of lunar night, there may be instances where operation in extended periods (or regions) of darkness may be desired that make energy storage solutions untenable.  One solution that has the potential to enable prolonged operations in dark areas is power beaming.  This approach allows the siting of a power generation unit, such as a large solar array in an area where sunlight is abundant.  A portion of the power generated by the solar array can then be converted into a transmissible form, usually as laser or microwave energy, which can then be received by a suitable collector at a remote user location.  This method becomes especially effective for siting systems in permanently shadowed regions (PSRs) on the Moon near the poles (Figure \ref{fig-lunar_PSRs}), where peaks that experience near permanent solar illumination are adjacent to craters containing PSRs.  Provided there is a line-of-site view between the two, near-continuous power can be assured.

A notional system for power beaming is described in \cite{Grandidier2021BAAS...53d.302G} and illustrated in Figure \ref{fig-power_beaming}.  This reference describes a laser power beaming system of a size that might be applicable to a medium-sized interferometry mission concept.  At the emitter location, a solar photovoltaic array would generate about 3~kW of electrical power.  The array would power a 1~kW 1064~nm wavelength fiber laser using a 20~cm optical element to deliver that power to the receiver at a distance of about 15~km. The receiver comprises an array of semiconductor cells tuned to the laser frequency, enabling a high efficiency (goal of 65\%) conversion back to electricity that can be used to directly power the remote facility or periodically charge batteries\index{batteries} for extended operations.  The receiving array size would be about 0.3~m$^2$.

For such a system, the mass would be divided between the emitter, assumed to be located on a lander in the sunlit area, and the receiving system at the interferometer site.  A system providing $\sim$650~watts as described would require a total mass of about 70~kg, of which 69~kg would be on the emitter side (comprising the solar array, laser, optical and tracking elements, thermal control, and power electronics) and only about 1~kg at the receiver.  For multiple elements in an interferometer array, a receiver package could be incorporated in each element with the emitter periodically providing power to charge batteries in each element in sequence.

While the technology readiness level (TRL) of an end-to-end power beaming system of the type described is currently relatively low, the elements exist and are being actively tested and developed as discussed in the referenced white paper.  With a suitable application identified, it is expected that a workable system could be achieved in the near future.

Beaming from orbital stations has also been proposed \citep{nasaPowerBeaming,spacenews_volta}.
In this area, it is worthwhile to note that LEO solutions are being pursued by startups like Star Catcher \citep{starcatcher}. There are already economic motivations for developing this technology commercially, which greatly increases the likelihood it would be available for implementation in a lunar context.


\subsection{Nuclear power}\label{sec-nuclear-power}

\begin{wrapfigure}{right}{8cm}
\includegraphics[width=0.52\textwidth, angle=0]{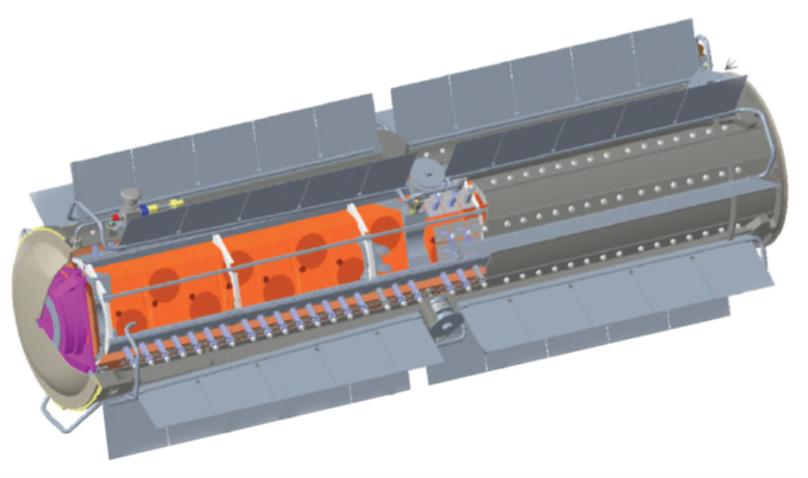}
\caption{Example radioisotope thermoelectric generator, the NASA Next Generation Mod 1 RTG. (Image Credit: Aerojet Rocketdyne)}
\label{fig-RTG}
\end{wrapfigure}

While beamed power as described in the preceding section can enable a certain class of missions, it remains a new technology and may have limited applicability at sites away from the poles.  For missions at lower latitudes that may need to survive and operate over the full duration of lunar nights, a more direct solution may be found using nuclear power systems, which can provide continuous electrical power regardless of the availability of sunlight.

Current flight-proven power systems include radioisotope thermoelectric generators (RTGs) and radioisotope heater units (RHUs).  RTGs have been used for electrical power and heat on space missions for six decades, including on the lunar surface during the Apollo program.  Currently, only one type of RTG is being produced: the Multi-Mission RTG (MMRTG) \citep{MMRTG}, which is currently powering the Curiosity and Perseverance rovers on Mars and is baselined for the upcoming Dragonfly mission to Titan.  The MMRTG is a 45~kg unit that produces about 115~W of electrical power at the beginning of life (decreasing at a rate of $\sim$3.5\% per year), as well as about 1800~watts of waste heat that can be used for other purposes.  The NASA Radioisotope Power Systems (RPS) program \citep{nasaRadioisotopePower} is also currently developing another RTG type, currently designated the Next Generation Mod 1 RTG (Figure \ref{fig-RTG}).  This version, set to be available in 2030, should produce about 245~watts of electricity at the beginning of life (decreasing at a rate of $\sim$1.6\% per year) at a mass of about 56~kg.  It also provides up to 3755~watts of waste heat.

\begin{wrapfigure}{right}{8cm}
\includegraphics[width=0.5\textwidth, angle=0]{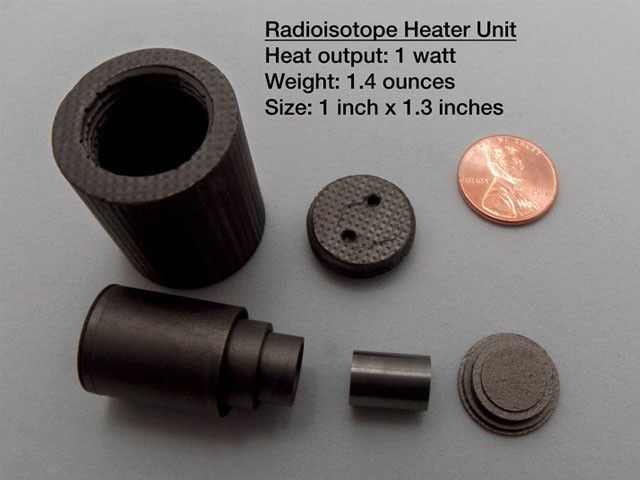}
\caption{Example radioisotope heater units. (Image credit: NASA/US Department of Energy)}
\label{fig-RHU}
\end{wrapfigure}

RHUs (Figure \ref{fig-RHU}) provide an option for small heat sources where needed.  They can replace electrical heaters, allowing reduced power requirements for overnight survival.  Each RHU is about the size of a C-cell battery (a cylinder about 3.2~cm tall and about 2.6~cm diameter) and has a mass of about 40~g.  RHUs contain a small amount of Pu-238 producing about 1~watt of heat output at fueling, decaying at a rate of slightly less than 1\% per year.  While available to projects at a relatively low cost, it should be noted that the use of RHUs could incur additional expense for launch approval activities associated with the ground processing and flight of radioactive material.

In addition to these existing and planned radioisotope power systems, work is continuing on advanced options that may be available to future missions.  Stirling radioisotope generators that operate using a dynamic system with greatly increased efficiency are in development.  Such a system using Pu-238 fuel could provide power levels on par or higher than the Next Gen Mod 1 RTG using roughly one-third the amount of fuel.  Further, work is underway in the UK and a number of other countries to develop both RTGs and Stirling-based RPS using Am-241 as an alternative fuel \citep{Ambrosi2019SSRv..215...55A}.  This isotope of americium has a longer half-life than plutonium, resulting in somewhat less power per unit mass, but the longer half-life will result in a slower decrease in unit power and the isotope itself may be significantly easier to acquire than the plutonium alternative.

Efforts are also being invested in developing fission reactor systems for lunar applications.  This includes NASA's Fission Surface Power project \citep{nasaFissionSurface}, which builds on the previous Kilopower project \citep{nasaKilopowerNASA}, as well as the Rolls-Royce Space Micro-Reactor \citep{rollsroyceRollsRoyceUnveils} in the UK.  These proposed power systems are substantially more massive and expensive than RTGs and RHUs, but also would have much greater power outputs; the principal motivation currently for developments on this front is supporting long-term human activity on the surface.

%
%
%

\chapterimage{AS16-117-18848HR_crop.jpg} 
\chapterspaceabove{7.0cm} 
\chapterspacebelow{7.25cm} 

\chapter{Mission Opportunities}\index{Mission Opportunities}\label{sec-mission-opportunties}

\section{Introduction}

Of all the potential pathways to getting optical interferometry off the surface of the Earth, not all are equally viable. The opportunity landscape is a discriminator in this regard and NASA’s technology maturation and flight programs are particularly influential. The Artemis\index{Artemis} program has broad international support and is destined to develop infrastructure on the lunar surface that will enable or enhance science missions. NASA’s Commercial Lunar Payload Services (CLPS)\index{CLPS} initiative is working with industry to deliver science and technology to the lunar surface. Science missions are sponsored by NASA’s Science Mission Directorate (SMD) and address community priorities espoused in the decadal surveys and reflected in NASA strategic goals. NASA’s Space Technology Mission Directorate offers funding opportunities ranging from the NASA Innovative Advanced Concepts (NIAC)\index{Mission Opportunities!NIAC} program to explore visionary concepts to flight opportunities that fund agency partners to develop and demonstrate innovative technologies. NIAC currently supports studies of UV/optical and radio interferometers on the surface of the Moon.

Within SMD, the Astrophysics Division offers opportunities to propose science missions ranging from Pioneers (\$20M)\index{Mission Opportunities!Pioneers} to Probes (\$1B) and develops grander flagship-class missions, such as the James Webb Space Telescope and the Nancy Grace Roman Space Telescope in response to National Academies decadal priorities and recommendations. Suborbital-class research investigations, including those motivated to demonstrate new technologies, can be proposed under the Research Opportunities in Space and Earth Science (ROSES) Astrophysics Research and Analysis (APRA) program up to a cost of \$10M. The Planetary Science Division (PSD) sponsors lunar technology maturation and research programs, including the ROSES Development and Advancement of Lunar Instrumentation (DALI) and Artemis IV Deployed Instruments (A4DI) programs.

\section{NASA Astrophysics}\label{sec-nasa-astrophysics-mission-classes}

The NASA Astrophysics Division (APD) conducts a wide range of missions for exploring the universe. Table \ref{tab-AO-table} lists the standard APD mission proposal opportunities. In the following sections, the salient features of those completed mission proposal opportunities of various scopes are discussed.

\begin{table}[]
\centering
\resizebox{\textwidth}{!}{%
\begin{tabular}{llll}
\textbf{AO/NRA}   & \textbf{Cost Cap}          & \textbf{Next Opportunity}             & \begin{tabular}[c]{@{}l@{}}\textbf{Launch Vehicle and}\\ \textbf{Lander}\end{tabular}         \\ \hline
Pioneers          & \$20M                      & $\sim$Annual NRA  (ROSES)      & F9/Vulcan, CLPS-small       \\
PRISM             & \$25-50M                   & $\sim$Annual NRA  (ROSES)      & F9/Vulcan,   CLPS-sm/med    \\
SMEX              & \$150M              & Spring 2025 AO,   2029 AO?            & F9/Vulcan,   CLPS-medium    \\
MIDEX             & \$300M           & 2027 AO?                              & F9/Vulcan,   CLPS-med/large \\
Probe &
  $\sim$\$1B  &
  \begin{tabular}[c]{@{}l@{}}2033 AO? (after FIR / X-ray  Probe),\\ White paper for Astro2030\end{tabular} &
  \begin{tabular}[c]{@{}l@{}}HLS   (+astronaut deployed) /\\ CLPS-large\end{tabular} \\
Large   Strategic & \textgreater{}\$2B (\$3-10B) & After HWO, white paper for Astro2030 & HLS (+astronaut deployed)
\end{tabular}%
}
\caption{NASA Announcements of Opportunity (AOs)/Research Announcement (NRAs).}
\label{tab-AO-table}
\end{table}

\subsection{Small Missions: Pioneers, APRA}\index{Mission Opportunities!APRA}\index{Mission Opportunities!Pioneers}\label{sec-small-missions-APRA}

NASA's \href{https://science.nasa.gov/astrophysics/programs/astrophysics-pioneers/}{Pioneers Program} has supported low-cost ($\leq$\$20M) science missions; historically, this has included accepting proposals for payloads delivered to the lunar surface via CLPS landers, as well as CubeSat, SmallSat, balloon, and modest International Space Station-attached payloads. NASA issues a call for Pioneers proposals at a nominally annual cadence through the ROSES omnibus solicitation. According to the ROSES 2024 solicitation, "[t]echnology development and maturation within the proposed project is allowed, but the primary review criterion for selection is the merit of the proposed science investigation." For lunar missions, NASA's 2023 solicitation covered the cost of launch and integration outside the PI’s \$20M cost cap.  The 2024 Pioneers solicitation was canceled, and the 2025 solicitation has deleted all language for lunar surface payloads.

NASA's \href{https://science.nasa.gov/astrophysics/programs/astrophysics-division-technology/}{Astrophysics Research and Analysis (APRA)} program offers "suborbital" class opportunities with a cost cap of $\leq$\$10M; the "suborbital" moniker is reflective of scope and risk rather than a boundary condition of the flight nature of the proposals.  Cubesats such as the \href{https://heasarc.gsfc.nasa.gov/docs/halosat/}{HaloSat} observatory \citep{LaRocca2020JATIS...6a4003L} have successfully flown under the APRA program.

A strawman small mission concept for a lunar optical interferometer is presented in \S \ref{sec-strawman-small}.

\subsection{Explorers: small and medium}\index{Medium-class Explorers (MIDEX)}\index{Small Explorer missions (SMEX)}\index{Mission Opportunities!MIDEX}\index{Mission Opportunities!SMEX}

NASA’s \href{https://explorers.gsfc.nasa.gov/}{Explorers Program} provides frequent flight opportunities for world-class scientific investigations from space utilizing innovative, streamlined and efficient management approaches. NASA issues Announcements of Opportunity (AO) for \href{https://explorers.gsfc.nasa.gov/smex.html}{Small Explorer missions (SMEX)} and \href{https://explorers.gsfc.nasa.gov/midex.html}{Medium-class Explorers (MIDEX)}, each at approximately four-year intervals. SMEX missions are cost-capped at $\sim$\$150M, while MIDEX missions are capped at $\sim$\$300M, and NASA separately pays for launch costs. Contributions valued at up to one-third of the PI-managed mission cost are allowed. Past AOs have accepted proposals for missions that operate in lunar orbits but not on the lunar surface.

NASA has historically also solicited for smaller \href{https://explorers.gsfc.nasa.gov/unex_mo_intern.html}{Missions of Opportunity (MoOs)} during the Explorer calls for proposals, which include both add-ons to existing missions as well as stand-alone missions.  The \href{https://nspires.nasaprs.com/external/solicitations/summary!init.do?solId={3C237E13-1C54-67EE-14DB-85141543DEAC}&path=open}{current 2025 draft call for SMEX mission proposals} does not solicit for MoOs; it is unclear when such investigations might again be invited to apply to upcoming Explorers calls.

\subsection{Explorers: Probe class}

NASA responded to an Astro2020 Decadal Survey recommendation by offering a \href{https://explorers.larc.nasa.gov/2023APPROBE/}{Probe-class Explorer AO} with an expected launch rate of one per decade. Probes have a PI-managed cost cap of \$1B. Like SMEX and MIDEX missions, contributions are allowed up to one-third of the PI-managed cost cap, and launch costs are covered separately. Probe AOs call for concepts in priority areas identified by decadal surveys. Thus, the first Probe AO called for far-infrared or x-ray astrophysics mission concepts. With demonstrable community interest, later calls could include lunar-hosted payloads.

\section{NASA Planetary Science}\label{sec-nasa-PSD}


\begin{wrapfigure}{R}{8.75cm}
    \vspace{-0.25cm}
    \begin{minipage}[h]{1\linewidth}
        \begin{tcolorbox}[colback=gray!5,colframe=green!40!black,title={Rapid reconnaissance}]
        Sensitive, high-resolution imaging could provide a "flyby without the flyby" capability for objects across the solar system.
        \end{tcolorbox}
    \end{minipage}
\end{wrapfigure}

Beyond NASA Astrophysics, high angular resolution optical systems would be of great value to NASA’s Planetary Science Division (PSD).  The potential for rapid, high-cadence, in-depth investigation of scores of small solar system bodies (\S \ref{sec-ultra-high-res-imaging-science}) has the potential for significant impact in planetary science.  What previously was accomplished by small ($\sim$10 cm) telescopes aboard spacecraft---with long wait times---could be carried out by an interferometric imager, minus the wait time.

PSD offers small opportunities via yearly ROSES calls. Specifically applicable to lunar interferometry concepts are the PRISM (Payloads and Research Investigations on the Surface of the Moon), PRISM Stand Alone Landing Site-Agnostic (SALSA), and evolving Artemis\index{Artemis} science instrument calls. 
These opportunities require instrumentation at the TRL 6 level and higher at the time of proposal submission to be able to be flight-worthy at the end of the award period, typically 3-5 years. These opportunities generally have strict mass, power, and volume constraints and are in the \$10-\$50M cost range per instrument payload.

PRISM and SALSA payloads follow a route to the Moon, with landing and deployment handled by CLPS landers (\S \ref{sec-CLPS-landers}). Some PRISM calls identify a specific destination. It is highly recommended that lunar surface payloads be site-agnostic to facilitate matching to multiple CLPS landers.\index{CLPS}

PSD offers additional opportunities, usually every two years, for technology development of planetary science payloads over 2-3 years through the Planetary Instrument Concepts for the Advancement of Solar System Observations (\href{https://www1.grc.nasa.gov/space/pesto/investment-areas/picasso/}{PICASSO}; TRL 1-3), Development and Advancement of Lunar Instrumentation (\href{https://www1.grc.nasa.gov/space/pesto/investment-areas/dali/}{DALI}; TRL 4-5), and Maturation of Instruments for Solar System Exploration (\href{https://www1.grc.nasa.gov/space/pesto/investment-areas/matisse/}{MatISSE}; TRL 4-6) programs in ROSES. Technology maturation projects are funded at $\sim$\$1M/year for a typical three-year duration. A lunar optical interferometer instrument with planetary science goals (e.g., outer planets, small bodies, other lunar studies)  could be matured via these programs. Technology roadmap details for NASA’s Planetary Science, including needs and focus areas, are maintained through NASA’s Planetary Exploration Science Technology Office (\href{https://www1.grc.nasa.gov/space/pesto/}{PESTO}).

Finally, PSD offers a range of robotic mission classes ranging from small, focused investigations (mid-size: Discovery, large: New Frontiers) to larger strategic flagships. The science cases for these missions must be tied to recommendations of the Planetary Science Decadal Survey (2023-2032) \citep{NASEM2023_PSD}.

\section{Artemis}\label{sec-artemis}\index{Artemis}

\begin{wrapfigure}{R}{8.25cm}
    \vspace{-0.25cm}
    \begin{minipage}[h]{1\linewidth}
        \begin{tcolorbox}[colback=gray!5,colframe=green!40!black,title={Returning humans to the Moon}]
        Human activities on the lunar surface will include science uniquely enabled by the Moon.
        \end{tcolorbox}
    \end{minipage}
\end{wrapfigure}
NASA plans to release periodic calls for independent science investigations supporting the long-term human exploration Artemis program. Each call will typically be sent first in draft form followed by a formal release. 

Artemis III is the first planned crewed landing with a primary destination at the lunar South Pole. At the time of this report, there are nine landing sites under consideration. Calls for Artemis III science investigations (ROSES A3DI) were solicited in 2023\footnote{NASA ROSES \href{https://science.nasa.gov/researchers/solicitations/roses-2023/amendment-26-artemis-iii-deployed-instruments-final-text-and-due-dates/}{F.12 Artemis III Deployed Instruments}}. Three payloads were selected to investigate the lunar environment, analyze properties of the lunar regolith, and characterize a lunar habitat environment for crop growth experiments.

In November 2024, there was a call for instruments to be deployed on the surface of the Moon during Artemis IV (ROSES A4DI), the second crewed landing at the lunar South Pole.  This call references objectives in the \href{https://www.nasa.gov/wp-content/uploads/2022/09/m2m-objectives-exec-summary.pdf}{2022 Moon-to-Mars Objectives}, the
Artemis III Science Team Report \citep{Weber2021LPI....52.1261W}, and the \href{https://science.nasa.gov/wp-content/uploads/2023/11/implementationplan-draft.pdf}{2023 SMD Lunar Science Strategy}. The M2M Astrophysics Objective (PPS-01-L) is presently limited in scope to astrophysical observations from the radio-quiet farside.  However, the more recent Artemis III Science Team Report, which includes additional community recommendations, identifies  "Ultra-High resolution optical imaging of astronomical objects" as a medium-priority investigation supporting "Artemis Science Goal 5a. Astrophysical and Basic Physics Investigations using the Moon" \citep{NASA_A3SD_imaging}.

Future Artemis missions have augmented capabilities such as a lunar terrain vehicle and cryogenic sample collection.  It is anticipated that subsequent science payload calls will be commensurate with the identified objectives per mission.
Additionally, it is expected there will be opportunities, via community advisory groups to inform the Moon-to-Mars \href{https://www.nasa.gov/moontomarsarchitecture-architecturedefinitiondocuments/}{Architecture Definition Document} (ADD), which is planned to be updated yearly.  Building and maintaining astronomical instruments from the lunar surface can be aligned with many objectives related to crew and autonomous/semi-autonomous operations there.

Within NASA’s SMD, the Exploration Science and Strategy office (\href{https://science.nasa.gov/lunar-science/}{ESSIO}) leads NASA’s objectives to achieve Moon-to-Mars (M2M) science objectives. These objectives are derived from community-driven decadal surveys. This office runs community workshops and conducts studies using the National Academy of Sciences.

\section{Other federal partners}\label{sec-other-federal-partners}


\subsection{Department of Defense (DoD)}

The US Space Systems Command has observed that the lack of space situational awareness (SSA) in the cislunar domain presents increased risk for spacecraft operating in this domain \citep{siew_cislunar_2022}. The lack of adequate cislunar SSA is compounded by the complex three-body dynamics governing orbits in this regime. This may provide motivation for the DoD to collaborate with lunar surface interferometry missions, and for that reason, the DoD should be included in stakeholder analyses.  Additionally, the DARPA 10-Year Lunar Architecture (LunA-10) capability study \citep{darpaFrameworkOptimized} is designed to explore and develop a framework for a sustainable commercial lunar economy. It was launched in 2023 to assess the technological, logistical, and operational needs for long-term activities on the Moon.

\subsection{Department of Energy (DoE)}

The Department of Energy (DoE) plays a significant role in the field of astronomy, primarily by funding research related to fundamental particles/astroparticle physics, directly connected to particle physics research areas that fall under DoE's purview.  The Dark Energy Camera \citep[DECAM;][]{Flaugher2015AJ....150..150F} is a good example of DoE investments in astrophysics that are directly related to its areas of research.  Experiments that shed insight into the nature of the universe on cosmological scales (e.g., \S \ref{sec-additional-single-baseline-cases}) would be of interest here.

\section{International partners}\label{sec-international-partners}

The global astronomical community has relied on national and international collaborations to accomplish large programs for many decades, with great success. The Astro2020 Decadal Survey \cite{NASEM_Decadal_2021pdaa.book.....N} states "new large telescopes and missions...with international partners...are an essential base upon which the survey’s scientific vision is built" and "sustaining broad observational capabilities is crucially dependent on international partnerships and missions."  International partnerships leverage expertise and resources, build resilience, and increase science return. Many examples of successful collaboration exist with ground- and space-based observatories. The lunar exploration window is newly reopened through Artemis with many international signatories to the \href{https://www.nasa.gov/artemis-accords/}{Artemis Accords}.
The Artemis\index{Artemis} Accords "promote international cooperation...for the benefit of all humanity."  As such, it is important to recognize that the potential for global partnership on lunar missions, including lunar-based astronomy and interferometry, exists within the Artemis signatories (NASA, ESA, JAXA, CSA, United Arab Emirates/UAE, etc.) as well as outside of them.

While lunar interferometry is new, many of the building blocks are well established within the international astronomy community. One key milestone towards space-based interferometry (including lunar) is the advancement of terrestrial interferometry at small to large scales (e.g., VLTI\index{missions!VLTI}, CHARA\index{missions!CHARA}, VLA, ALMA, EHT). Another milestone towards space interferometry is the advancement of space astronomy capabilities, for which there exist a plethora of examples of international collaboration, both NASA-led as well as led by international partners/agencies with key NASA contributions (e.g., Hubble, Herschel, Euclid, JWST, JUICE, XRISM).  While "astronomy continues to become more global and interconnected," an emphasis on "the highest-priority sustaining activity [in space] is a multi-messenger program"  further fortifies the utility of international partnerships both broadly across terrestrial and space astronomy, and in lunar astronomy in particular \cite{NASEM_Decadal_2021pdaa.book.....N}.


%
%
%

\chapterimage{openart-array-of-telescopes_crop.png} 
\chapterspaceabove{7.5cm} 
\chapterspacebelow{8.25cm} 

\chapter{Strawman Mission Concepts}\label{sec-strawmen}

\section{Introduction}

During the course of the KISS Workshop, our working group examined a number of notional mission architectures.  These were intended to be representative---but not constraining---of the possibilities offered by the various mission classes noted for NASA Astrophysics (\S \ref{sec-nasa-astrophysics-mission-classes}); additional possibilities within Astrophysics, elsewhere within NASA (\S \ref{sec-nasa-PSD}), and with other partners both domestic (\S \ref{sec-other-federal-partners}) and abroad (\S \ref{sec-international-partners}), are all within the broad swath of opportunities offered by interferometry from the lunar surface.

\section{Small class}\label{sec-strawman-small}


The small-class strawman mission concept is the well-developed MoonLITE (Lunar Interferometry Express)\index{missions!MoonLITE} proposal \citep{vanBelle2024SPIE13092E..2NV}.  MoonLITE is a NASA Astrophysics Pioneers\index{Mission Opportunities!Pioneers} submission that proposes to deliver a Michelson interferometer experiment within the parameters of that Call for Proposals. The broad parameters of the ROSES-2023 Pioneers call for proposal allowed proposers to request a portion of a CLPS host to the lunar surface, along with support from a rover provided as part of the CLPS service.  The payload mass had to be below 50 kg, with no more than 15~kg of that onboard the rover; 200 W of daytime power was to be provided, along with communications service and other support infrastructure from the lander.  The CLPS support included $<$20W of survive-the-night power, and operations for a minimum of two lunar days was the threshold CLPS service (with six being the goal).\index{CLPS}

\begin{figure}
    \centering
    \includegraphics[width=0.98\linewidth]{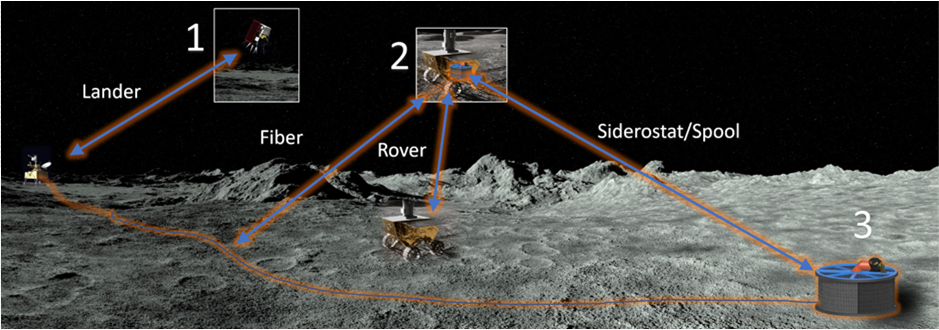}
    \caption{The MoonLITE experiment.  (1) A CLPS-provided lander and rover land with MoonLITE.  (2) The rover deploys the outboard station 100 m away from the lander.  (3) The outboard telescope relays light back to the lander via fiber.  At the lander, that light is interfered with light from the inboard station, providing milli-arcsecond-sized measurements of faint objects. (Image credit: Redwire Space)}
    \label{fig:MoonLITE}
\end{figure}

Within these parameters, the MoonLITE concept was built up to deliver and operate a very simple---yet capable---two-element Michelson interferometer.  Two small 50~mm (2-inch) telescopes will be delivered, with separate inboard and outboard stations.  The lander's rover will be pre-loaded with the outboard station before mission launch; deployment will consist of the rover taking the outboard station 100 m away from the lander along an east-west line, and setting down that station. With a single step, the experiment will be ready for operations (Figure \ref{fig:MoonLITE}).  An umbilical line with a fiber optic and power/communications wiring will be unspooled during the deployment.  The fiber-optic line will be a polarization-maintaining, single-mode fiber, and will relay the outboard telescope light back to the lander, where a beam combiner is the other major element of the experiment.  The inboard telescope will have an identical optical path via fiber, for pathlength equalization and matched longitudinal dispersion of the relayed target light, with the main difference being that telescope's fiber will be simply coiled up.  The combiner has a short-throw ($<$3 m of optical path difference) mechanical delay line, which means for a 100 m east-west baseline and the Moon's slow rotation rate, objects at any declination crossing the meridian line of the sky could be observed for many hours.  The lack of an appreciable lunar atmosphere means that interference fringes could be observed for many tens if not hundreds of seconds, in contrast to the $\sim1$ms terrestrial atmospheric coherence time.  These substantially increased coherence times mean that, even with extremely modest apertures, MoonLITE's sensitivity would easily exceed that of even the largest terrestrial optical interferometers by many orders of magnitude.

MoonLITE's single interferometric baseline would be a powerful tool for the single-baseline cases noted above (\S \ref{sec-single-baseline-science}), including direct size measurements of brown dwarfs (\S \ref{sec-brown-dwarf-sizes}).
The MoonLITE proposal remains under development by its proposal team. It was submitted for the ROSES-2023 call, and not selected; the ROSES-2024 Pioneers call was canceled; and the ROSES-2025 Pioneers call had all CLPS language removed.

\section{Explorer class}\label{sec-strawman-explorer-class}

A KISS subgroup focused on a strawman medium-scale mission.  A mission of this scope would be enabled by a dedicated medium-to-large-scale CLPS lander and selection as a SMEX or MIDEX mission.  \index{Mission Opportunities!MIDEX}\index{Mission Opportunities!SMEX} The primary goal of a high-precision lunar-based astrometric interferometer (LAI) is the discovery of exo-Earth targets for the Habitable Worlds Observatory.  Knowledge of exoplanet orbits and mass will greatly enhance the exoplanet spectral characterization yield of HWO for both coronagraph-based and starshade-based direct imaging approaches \cite{morgan_spie_2022,morgan_spie_2023}.

The required astrometric precision is $\sim$0.1~$\mu$as. A sweet spot for the interferometer baseline length is $\sim$100~m. This is short enough to fit in numerous sites, but with a metrology error budget of 50~pm (= 0.1 $\mu$as $\times$ 100~m), it is within reach of extant laser metrology.  As described in \S \ref{sec-astrometry-from-lunar-surface}, the combination of visual magnitude m$_V$=12 reference stars and 30~cm apertures leads to observations that are several hours long, requiring delay lines $\sim$ 5~m long.  The delay lines can be folded so that the physical motion is a meter or less.  Thus, there are no unwieldy or extraordinarily challenging components in the system.

LAI can be compared to two high-precision astrometric instruments, the now-operational VLTI GRAVITY \citep{Gillessen2010SPIE.7734E..0YG,GRAVITY2017A&A...602A..94G}, and the Space Interferometry Mission (SIM) which was 
canceled in Phase C \citep{Unwin2008PASP..120...38U}.  Compared to the GRAVITY instrument, the lunar astrometric interferometer requires fewer control systems since it is not battling atmospheric turbulence. However, it does require higher-precision metrology and the ability to precisely measure (to $\sim$2~nm) the baseline length, something that GRAVITY does not need with its 10~$\mu$as precision.  Compared to SIM, LAI performs only relative astrometry, and its baseline is naturally inertially stabilized by being attached to the Moon. Thus, LAI uses just one baseline (whereas SIM needed three to determine its orientation in space), and LAI has just two pairs of siderostats and two delay lines (one for science target, one for the reference star).    With its 100~m baseline, 14$\times$ longer than SIM, it has relaxed metrology requirements even with a 10$\times$ better astrometric precision.
In addition to these advantages, such a mission is expected to be significantly less expensive than the projected \$1.9Bn cost (2010 dollars) for SIM \citep[][p. 196]{2010nwnh.book......}.

\section{Probe/Flagship class}\label{sec-strawman-probe-to-flagship-class}

A mission of this scope would be enabled by the multi-ton downmass capability of the HLS landers (\S \ref{sec-HLS-Artemis}, \ref{sec-HLS-cargo}), and potentially also human-assisted deployment and/or maintenance.  Significant trade space elements for such concepts would be part of a likely lunar South Pole-focused location for HLS landers, such as sky coverage  (\S \ref{sec-dust_seismic_sky_thermal}) versus significant infrastructure (power, communications, etc.) available at a human base.

The KISS subgroup that focused on this scope of mission quickly zeroed in on the possibility of a UV/optical High Energy Astrophysics Imager (HEAI).\index{missions!High Energy Astrophysics Imager}  Such a facility would be able to image the inner region of nearby active galactic nuclei powered by supermassive black holes, down to one Schwarzschild radius ($R_S$) or less.  At a distance of 20 Mpc, such a $10^9 M_\odot$ central black hole would subtend an angle of roughly 1 microarcsecond.  Additional features of interest would also be resolvable: the photon ring (a general relativistic feature) should be apparent at 3/2 $R_S$, and the surrounding accretion disk would be strongly emitting in the ultraviolet at 10-100 $R_S$ (Figures \ref{fig:agn_examples}, \ref{fig-agn_models}).  Additional capabilities would be imaging supernova explosions--including asphericity--and resolved-disk imaging of degenerate matter objects like white dwarfs (\S \ref{sec-ultra-high-res-imaging-science}).

\begin{figure}[h]
    \centering
    \includegraphics[width=0.495\textwidth]{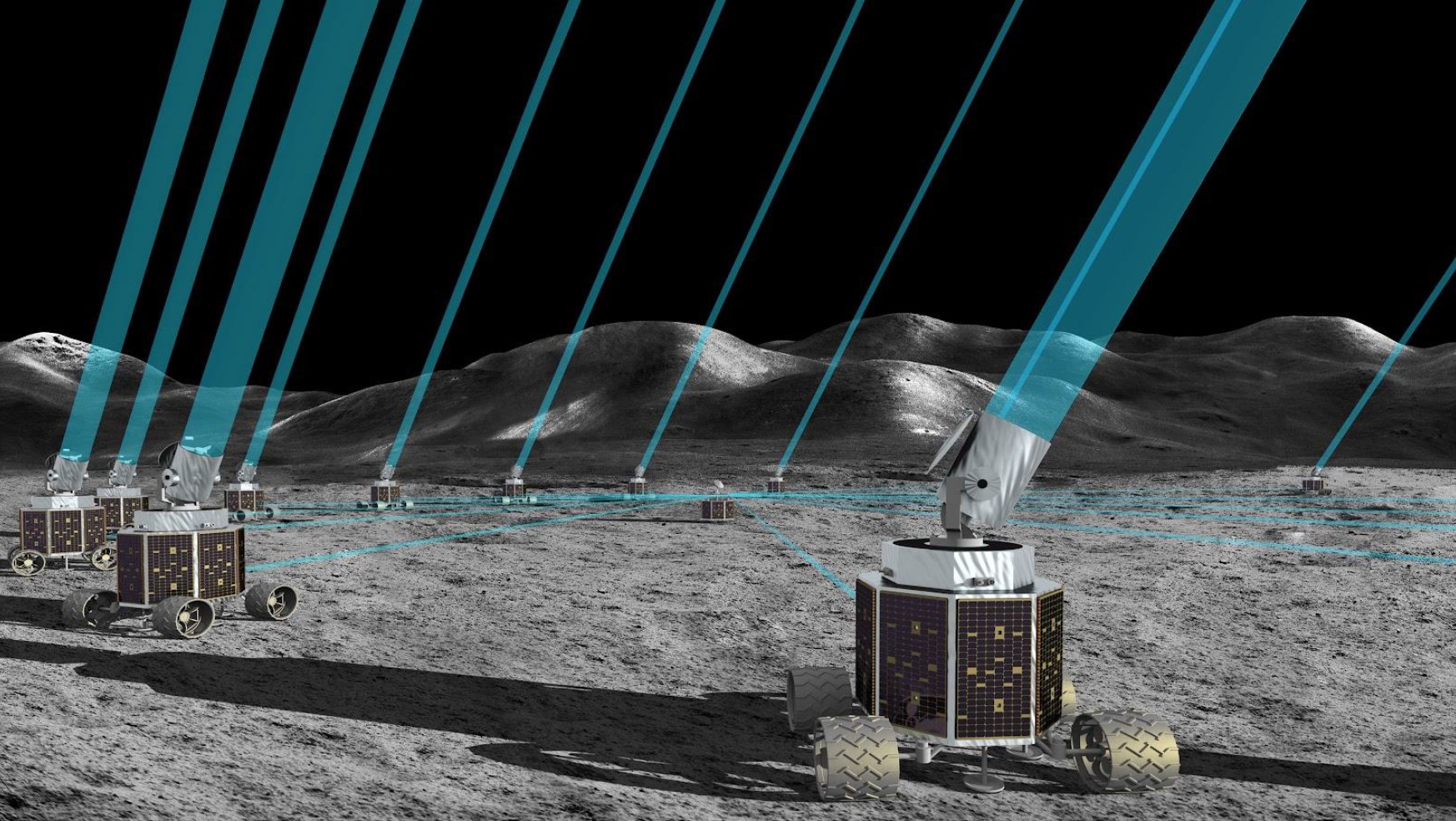}
    \includegraphics[width=0.495\textwidth]{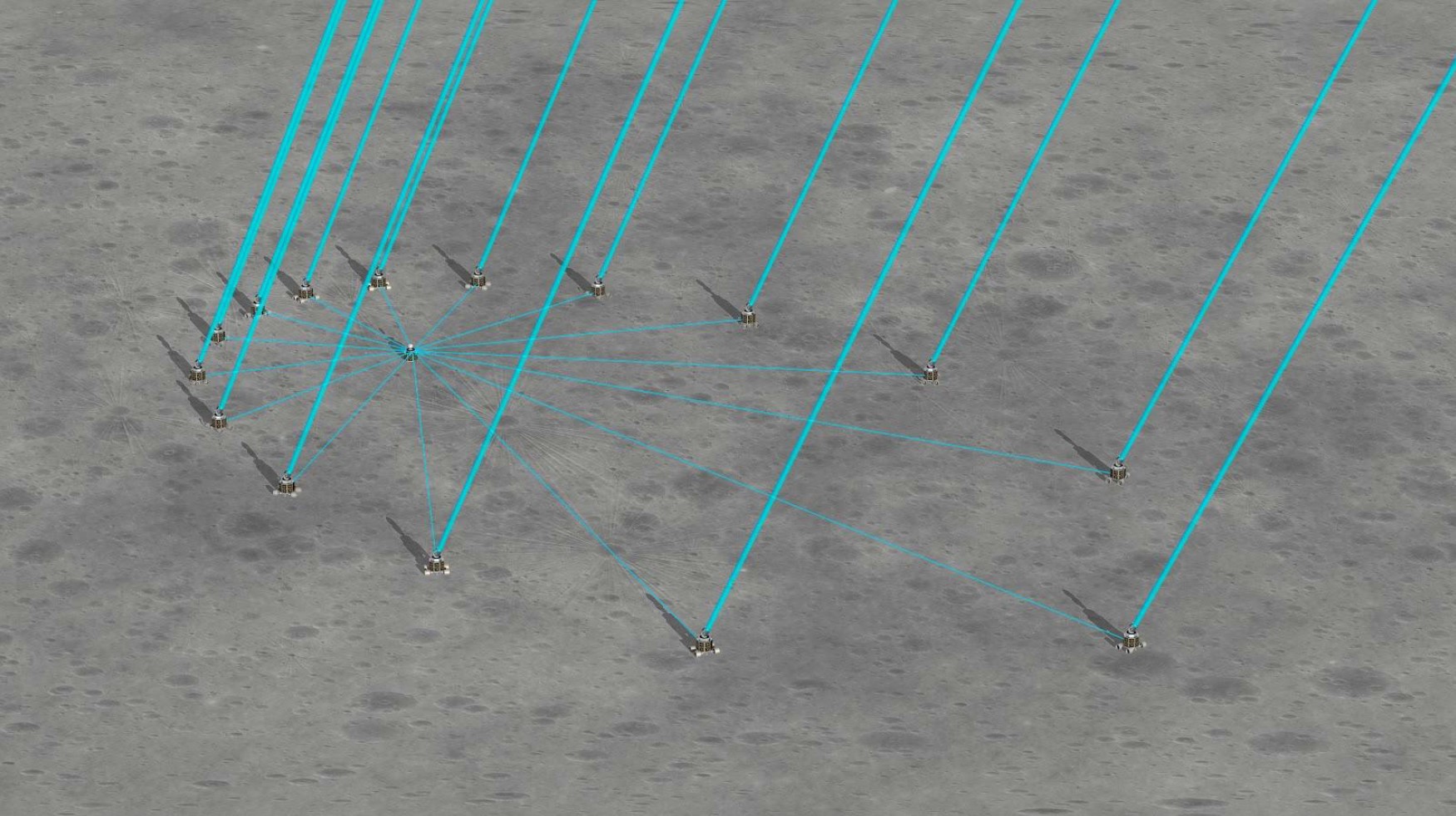}
    \caption{Artemis-enabled Stellar Imager (AeSI; from \citep{Carpenter2025arXiv250302105C}, Figures 3.1 and 3.3, image credit: Britt Griswold).  AeSI would have meter-class apertures on mobile platforms to allow for pupil reconfiguration and account for gross amounts of optical delay (left); the overall facility could start with a small number of apertures ($\sim$6) but then expand to 15-30 (right).}
    \label{fig-AeSI}
\end{figure}

The Artemis-enabled Stellar Imager (AeSI)\index{Artemis!Artemis-enabled Stellar Imager} is actually quite analogous to this concept, with the only significant difference being maximum baseline length: AeSI's science case requires a 1~km baseline, and the HEAI would need roughly twice that.  A NASA Innovative Advanced Concepts (NIAC)\index{Mission Opportunities!NIAC} study of AeSI has resulted in a comprehensive write-up on the concept \citep{Carpenter2025arXiv250302105C}, which has $\sim$15-to-30$\times$1-meter telescopes that can relocate relative to a central combiner (Figure \ref{fig-AeSI}).  An AeSI/HEAI facility would take advantage of the airless environment and operate down below the Lyman-$\alpha$ cutoff of 1216\AA.

\part{Conclusion}


%
%
%

\chapterimage{KISS_Lunar_Cover_20250326_crop.jpg} 
\chapterspaceabove{4.5cm} 
\chapterspacebelow{12.25cm} 

\chapter{Conclusion}

\section{Summary of big questions---and answers}\label{sec-Big-Questions-and-Answers}


Some notional "Big Questions" considered by this workshop included:

\textbf{What are key milestones on the way towards an interferometric lunar observatory?}  A top-level summary of these milestones---some of which are already complete as denoted with checked boxes---are as follows:
\begin{itemize}
    \item \makebox[0pt][l]{$\square$}\raisebox{.15ex}{\hspace{0.1em}$\checkmark$} Flight opportunities to the surface of the Moon, with a wide range of opportunities becoming available (\S \ref{sec-surface-access}).  Table \ref{tab-lunar_landers} outlines the significant number of flown and upcoming missions to the lunar surface.
    \item \makebox[0pt][l]{$\square$}\raisebox{.15ex}{\hspace{0.1em}$\checkmark$} Demonstration of mature terrestrial optical interferometry operations is noted in \S \ref{sec-mature-operations}.
    \item \makebox[0pt][l]{$\square$}\raisebox{.15ex}{\hspace{0.1em}$\checkmark$} Demonstration that the lunar seismic environment is not problematic for interferometry; this is reviewed in \S \ref{sec-seismology} and shown to be a non-issue for optical interferometry.
    \item \makebox[0pt][l]{$\square$}\raisebox{.15ex}{\hspace{0.1em}$\checkmark$} Demonstration that lunar dust is not an obstacle for optics and optomechanical systems. More than three years of operations of the LUT observatory aboard the Chang'e 3 lander confirms that, with a reasonable CONOPS, telescopes can operate from the lunar surface (\S \ref{sec-lunar-regolith-and-dust}).
    \item \makebox[0pt][l]{$\square$}\raisebox{.15ex}{\hspace{0.1em} } ~~Proposal opportunities for a lunar optical interferometer.  The 2023 NASA ROSES Pioneers\index{Mission Opportunities!Pioneers} call (\S \ref{sec-small-missions-APRA}) opened an initial door for astrophysics proposals, but only temporarily.  Future Pioneers calls and additional funding lines (\S \ref{sec-mission-opportunties}) could allow for competitive selection of missions that take advantage of this opportunity.
    \item \makebox[0pt][l]{$\square$}\raisebox{.15ex}{\hspace{0.1em} } ~~Development of supporting lunar infrastructure technologies.  This includes surface mobility (\S \ref{sec-mobility}), communications / data infrastructure (\S \ref{sec-communications}), and nighttime survival capability (\S \ref{sec-survive-the-night}).\index{CLPS}\index{Artemis}
    \item \makebox[0pt][l]{$\square$}\raisebox{.15ex}{\hspace{0.1em} } ~~Proposal development, selection \& flight.  At least one small proposal has already been submitted, MoonLITE  (\S \ref{sec-strawman-small}).
\end{itemize}

\textbf{What has changed in the last 5 to 10 years to make lunar surface interferometry a possibility?  What forthcoming developments will further enable this?}  A key development has been the NASA CLPS program (\S \ref{sec-CLPS-landers}), and the larger overarching Artemis opportunity (\S \ref{sec-artemis}).  Within the framework of CLPS robotic landers, small- to medium-class missions appear readily achievable.

Additionally, the demonstrated maturity of terrestrial optical interferometry techniques, and science results from terrestrial facilities (\S \ref{sec-mature-operations}), support the pursuit of lunar facilities that can extend the reach of this technique to the distant edges of the universe in the absence of a distorting, turbulent atmosphere. \\

\textbf{What can be done within the scope of each of the NASA Astrophysics funding lines---Pioneers, SMEX, MIDEX, Probe, Flagship?}
The suggested strawman missions (\S \ref{sec-strawmen}) are indicative of how science return scales at the various project scopes.  These range from simple, short-lived feasibility demonstration instruments to full-fledged general astrophysics observatories with unprecedented capabilities; each of the mission categories is expected to provide a significant scientific return commensurate with the level of investment.\\

\textbf{Are robotic or crewed missions best for the implementation of these ideas?}
Robotic missions to the lunar surface are already taking place, and are appropriate for the small-to-medium facility concepts.  The largest facility ideas (\S \ref{sec-strawman-probe-to-flagship-class}) probably need HLS-scale delivery, which in turn likely carry with them a crewed option (though not necessarily so, if dedicated HLS cargo missions could be used for robotic missions).  Small-scale experiments could also be part of the Artemis Deployed Instruments\index{Artemis!Artemis Deployed Instruments} program (\S \ref{sec-artemis}), which further blurs the line.  \\

\textbf{Are there implications that significantly impact the past Astrophysics Decadal Survey, or the next one?}  The 2020 Astrophysics Decadal Survey \citep{NASEM_Decadal_2021pdaa.book.....N} strongly advocated for searching for the biosignatures of $\sim$25 habitable zone planets; the spectra supporting such a search would come from the notional Habitable Worlds Observatory.  However, discovery and characterization of those planets in advance of HWO\index{missions!Habitable Worlds Observatory}---especially their masses---is currently a lingering unknown.  Lunar astrometry (\S \ref{sec-masses-for-HWO}, \ref{sec-astrometry-from-lunar-surface}) can fill in that gap.  Other capabilities---such as sensitive, sub-milliarcsecond-class imaging from a larger facility (\S \ref{sec-strawman-probe-to-flagship-class})---are sufficiently revolutionary as to leapfrog well past the anticipations of the Decadal Survey.\\

\textbf{What are the greatest challenges for---or misunderstandings about---astronomy from the lunar surface?}  The lunar seismic and dust environments are not obstacles, as incorrectly assessed in 1996 \citep{Bely1996kbsi.book.....B} and demonstrated (at least for dust) by the Chang'e-3 lander's LUT  \citep{Wang2015Ap&SS.360...10W}.  Improved understanding of the environment and handling of the challenges presented by that environment are already being characterized in detail by pathfinder experiments on the first generation of CLPS landers.

\textbf{How do the cost, risk profiles, and science return of interferometric lunar observatories compare to orbital facilities?}
The elimination of formation-flying infrastructure, or structural connection infrastructure, has the potential to simplify lunar facilities relative to orbital concepts.  As seen in the case of astrometry (\S \ref{sec-masses-for-HWO}, \ref{sec-astrometry-from-lunar-surface}), this can result in an expectation for increased capability at lower complexity and cost.

\section{Findings}



\textbf{Finding 1.}  The intersection of mature optical interferometry technology (\S \ref{sec-mature-operations}), and rapidly maturing lunar surface access and survival technology (\S \ref{sec-lunar-supporting-technologies}; Table \ref{tab-lunar_landers}), presents an opportunity to achieve optical imaging systems with angular resolutions orders of magnitude greater than currently possible with current space observatories, at sensitivity levels at orders of magnitude greater than terrestrial interferometric facilities (\S \ref{sec-coherence-volume}).  

\textbf{Finding 2.}  Existing NASA Astrophysics (\S \ref{sec-nasa-astrophysics-mission-classes}) and Planetary Science (\S \ref{sec-nasa-PSD}) mission funding lines could take advantage of this potential by allowing missions to be competitively proposed alongside orbital facilities.  These missions could be selected on their expected scientific return as weighed against their cost and risk profiles, as is the case for proposed orbital missions.

\textbf{Finding 3.}  
A near-term, small mission could conclusively demonstrate the feasibility and potential for astronomical interferometry from the lunar surface (\S \ref{sec-strawman-small}).  The Pioneers call in ROSES-2023 was a briefly opened opportunity for this; upcoming ROSES calls could reinstate this opportunity for a competitively selected small instrument aboard a CLPS lander.

\textbf{Finding 4.}  
A medium-class mission could further exploit the advantage of observing from the lunar surface for a more capable effort, including more advanced high-precision interferometric techniques such as astrometry or nulling.  One such strawman medium-class mission concept could leverage the lunar opportunity and provide uniquely enabling advanced exoplanet reconnaissance for HWO (\S \ref{sec-strawman-explorer-class}).
The upcoming SMEX and MIDEX calls from NASA Astrophysics could be appropriately scoped opportunities for competitive consideration of such a concept.

\textbf{Finding 5.}  
A major, Probe- to Flagship-class mission could provide unparalleled capabilities and resulting science return, with highly sensitive, milli- to micro-arcsecond imaging at wavelengths from the UV to the MIR.  The KISS workshop strawman concept for a facility of this scope (\S \ref{sec-strawman-probe-to-flagship-class}) illustrated the unprecedented exploratory value of such a facility, with a much-expanded examination of this possibility in the AeSI study report \citep{Carpenter2025arXiv250302105C}.  Such a facility could take advantage of the infrastructure of a burgeoning lunar economy expected to be developed in the coming decades (\S \ref{sec-lunar-supporting-technologies}).


\section{Engagement opportunities}

\subsection{International partnerships}

The potential for international cooperation could further enhance potential lunar interferometric facilities. The terrestrial optical international interferometry community is well established, highly collaborative \citep{Choquet2024sf2a.conf..187C}, and has collectively matured the technology over the past three decades.  International partners can provide training opportunities (e.g., European interferometry workshops \citep{Lykou2023eas..conf.2371L}) and other programs can be utilized for subsystem technology advancement by individual countries.  Flexibility is required to accommodate the alignment of funding cycles between international partnerships.  Given that interferometers are distributed systems, many opportunities exist for individual partner contributions---e.g., one partner could provide outboard telescopes, another could provide the backend beam combiner, etc. The exchange of students and early-career researchers internationally within collaborations will also greatly enhance leadership training and collaborative networking.

\subsection{Public outreach}

Communication is a critical component of any mission profile and the development or support of any science effort or new technology. Formal communications can inspire collaboration from the scientific community and enhance development through research projects, manuscript-based progress, and sharing at conferences. Informal communication can help motivate support from students and the public via news/press releases, social media, and public outreach. Having a common language and collaborative development surfaced through open-access reports and data provides community solidarity for sharing ideas and information and advancing goals more easily for products like proposals, manuscripts, and policy documents.

\section{A vision for transformative astrophysics from the Moon}

The lunar surface offers a unique and compelling platform for next-generation optical astronomy. With maturing access technologies and decades of progress in Earth-based optical interferometry, the Moon is now poised to host observatories capable of achieving imaging resolutions far beyond what is possible from Earth or orbit. This is not a distant dream---it is an emerging reality made possible by converging technological and programmatic developments.

\begin{itemize}
\item \textbf{Unprecedented Imaging}: Lunar-based interferometry can unlock sub-milliarcsecond resolution and sensitivity across UV to MIR wavelengths.

\item \textbf{Leverage Existing Programs}: NASA’s current funding mechanisms can support missions to test and deploy these capabilities.

\item \textbf{From Pathfinders to Flagships}: Small-scale demonstrations today can pave the way for medium-class science missions and ultimately flagship-class observatories delivering breakthrough science.

\end{itemize}

We stand at the intersection of technical readiness and lunar opportunity. By advancing lunar interferometry missions today, we can lay the foundation for a revolutionary new era of astrophysics. Seizing this moment now with small, achievable steps can build toward a future where the sharpest eyes in the universe watch from the Moon.

\stopcontents[part] 


\chapterimage{} 
\chapterspaceabove{2.5cm} 
\chapterspacebelow{2cm} 



\bibliographystyle{spiebib2b}
\bibliography{bibliography}
\addcontentsline{toc}{chapter}{\textcolor{ocre}{Bibliography}} 

%
%
%

%
%
%


\cleardoublepage 
\phantomsection
\addcontentsline{toc}{chapter}{\textcolor{ocre}{Index}} 
\printindex 


\chapterimage{AS16-117-18848HR_crop.jpg} 
\chapterspaceabove{6.75cm} 
\chapterspacebelow{7.25cm} 

\begin{appendices}

\renewcommand{\chaptername}{Appendix} 


\chapter{Appendices}

\section{Acronyms}
\begin{longtable}[c]{ll}
A3DI   & Artemis III Deployed Instruments \\
A4DI   & Artemis IV Deployed Instruments  \\
AAS    & American Astronomical Society \\
ADD    & architecture definition document \\
ADI    & Artemis Deployed Instruments  \\
AeSI   & Artemis-enabled Stellar Imager\\
AGN    & active galactic nucleus\\
ALMA   & Atacama Large Millimeter \\
       & and Submillimeter Array  \\
ALSEP  & Apollo Lunar Surface Experiments Package\\
AO     & Announcement of Opportunity \\
APRA   & Astrophysics Research and Analysis \\
arcmin & arcminutes \\
ARMADA & ARrangement for Micro-Arcsecond \\
       & Differential Astrometry\\
AU     & astronomical unit \\
B      & billion \\
Caltech& California Institute of Technology\\
CAN    & Cooperative Agreement Notice \\
CHARA  & Center for High Angular Resolution Astronomy \\
CLPS   & Commercial Lunar Payload Services   \\
cm     & centimeter \\
CMSB   & CLPS Manifest Selection Board \\
CONOPS & concept of operations               \\
COPAG  & Cosmic Origins Program Analysis Group \\
COPHI  & common-path heterodyne interferometers \\
CubeSat& class of nanosatellite \\
CSA    & Canadian Space Agency  \\
CTE    & coefficient of thermal expansion    \\
DALI   & Development and Advancement \\
       & of Lunar Instrumentation \\
DARPA  & Defense Advanced Research Projects Agency\\
DC     & direct current  \\
DECAM  & Dark Energy Camera\\
deg    & degrees  \\
DENIS  & Deep Near-Infrared Survey of the Southern Sky \\
DIABLO & Deployable Interlocking Actuated \\
       & Band for Linear Operations \\
DL     & delay line   \\
DoD    & Department of Defense   \\
DoE    & Department of Energy  \\
DR3    & data release 3  \\
DTE    & data terminal equipment  \\
EDS    & Electrodynamic Dust Shield  \\
EHT    & Event Horizon Telescope  \\
EM     & electromagnetic  \\
EPRV   & extreme precision radial velocity   \\
ESA    & European Space Agency \\
ESSIO  & Exploration Science and Strategy office \\
EVA    & extravehicular activities \\
EXOPAG & Exoplanet Exploration Program Analysis Group  \\
FAR    & Federal Acquisition Regulation\\
FBO    & Federal Business Opportunities \\
FGS    & fine guidance sensor \\
FIR    & far infrared     \\
FLEX   & Flexible Logistics and Exploration  \\
FLIP   & F(lex) Lunar Innovation Platform  \\
GIS    & Geographic Information System \\
GLINT  & Guided-Light Interferometric \\
       & Nulling Technology  \\
GRACE-C& Gravity Recovery and Climate Experiment \\
GRAVITY& capitalized, but not an acronym \\
GSFC   & Goddard Space Flight Center \\
GSU    & Georgia State University \\
HEAI   & High Energy Astrophysics Imager \\
HLS    & Human Landing System \\
HST    & Hubble Space Telescope  \\
HWO    & Habitable Worlds Observatory \\
Hz     & Hertz \\
IAU    & International Astronomical Union \\
ICON   & Name of NASA contractor (not an acronym)  \\
ILOM   & In-situ Lunar Orientation Measurement \\
IOTA   & Infrared-Optical Telescope Array  \\
IR     & infrared      \\
ISRU   & in-situ resource utilization \\
JAXA   & Japan Aerospace Exploration Agency\\
JouFLU & Jouvence of FLUOR \\
JPL    & Jet Propulsion Laboratory \\
JSC    & Johnson Space Center\\
JUICE  & Jupiter Icy Moons Explorer  \\
JWST   & James Webb Space Telescope \\
K      & kelvin  \\
KISS   & Keck Institute for Space Studies \\
kg     & kilogram  \\
kW     & kilowatt  \\
KISS   & Keck Institute for Space Studies \\
LAI    & lunar-based astrometric interferometer  \\
LAMPS  & Lunar Array Mast and Power System \\
LBTI   & Large Binocular Telescope Interferometer  \\
LEAG   & Lunar Exploration Analysis Group  \\
LEO    & low Earth orbit \\
LISA   & Laser Interferometer Space Antenna  \\
LIT    & Lunar Intelligent Tower \\
LMC    & Large Magellanic Cloud \\
LOLA   & Lunar Orbiter Laser Altimeter  \\
LRO    & Lunar Reconnaissance Orbiter \\
LROC   & Lunar Reconnaissance Orbiter Camera \\
LSIC   & Lunar Surface Innovation Consortium \\
LunA-10& 10-Year Lunar Architecture \\
LUNARSABER & Lunar Utility with Navigation, Advanced\\
           &  Remote Sensing, and Autonomous Bearing \\
LUT    & Lunar Ultraviolet Telescope \\
LRO    & Lunar Reconnaissance Orbiter \\
m      & meter \\
M      & million \\
M2M    & Moon-to-Mars \\
MatISSE& Maturation of Instruments for \\
       & Solar System Exploration \\
MIDEX  & Medium-Class Explorer\\
MIR    & mid infrared  \\
MIT    & Massachusetts Institute of Technology\\
MMRTG  & multi-mission radioisotope \\
       & thermoelectric generator  \\
MoO    & Missions of Opportunity \\
MoonLITE & Moon Lunar InTerferometry Express \\
Mpc    & megaparsec \\
MRE    & molten regolith electrolysis\\
MROI   & Magdalena Ridge Observatory Interferometer  \\
MYSTIC & Michigan Young STar Imager at CHARA\\
NAC    & Narrow-Angle Camera  \\
NASA   & National Aeronautics and Space Administration \\
NIAC   & NASA Innovative Advanced Concepts \\
NIR    & near infrared     \\
nm     & nanometer \\
NPOI   & Navy Precision Optical Interferometer \\
NRA    & NASA Research Announcement\\
NSPIRES & NASA Solicitation and Proposal Integrated\\
       & Review and Evaluation System \\
OMB    & Office of Management and Budget  \\
OPD    & optical path delay \\
OSTP   & Office of Science and Technology Policy    \\
pc     & parsec \\
PESTO  & Planetary Exploration Science Technology Office \\
PI     & Principal Investigator \\
PICASSO& Planetary Instrument Concepts for the\\
       & Advancement of Solar System Observations \\
pm     & picometer \\
PNT    & position, navigation and timing  \\
PRISM  & Payloads and Research Investigations \\
       & on the Surface of the Moon \\
PRV    & precision radial velocity \\
PSD    & Planetary Science Division \\
PSR    & permanently shadowed region \\
PTI    & Palomar Testbed Interferometer   \\
QSO    & quasi-stellar objects   \\
RESILIENCE & ispace lunar lander \\
       & (capitalized, not an acronym) \\
RHU    & radioisotope heater units \\
RPS    & radioisotope power systems \\
ROSES  & Research Opportunities \\
       & in Space and Earth Science \\
RTG    & radioisotope thermoelectric generator \\
RV     & radial velocity \\
SALSA  & Stand Alone Landing Site-Agnostic  \\
SED    & spectral energy distribution       \\
SIM    & Space Interferometry Mission       \\
SIMPLE & Substellar and IMaged PLanet  \\
       & Explorer Archive of Complex Objects  \\
SmallSat& small spacecraft \\
SMBH   & supermassive black hole \\
SMD    & Science Mission Directorate  \\
SMEX   & Small Explorer \\
SN     & supernova(e) \\
SOA    & state of the art \\
SPICE  & Space Interferometer for Cosmic Evolution \\
SPIE   & Society of Photographic \\
       & Instrumentation Engineers   \\
SPIRIT & Space Infrared Interferometric Telescope \\
SSA    & Space Situational Awareness   \\
STARI  & STarlight Acquisition and Reflection \\
       & toward Interferometry \\
STEM   & science, technology, engineering, \\
       & and mathematics \\
STMD   & Space Technology Mission Directorate  \\
TENACIOUS & ispace lunar rover \\
       & (capitalized, not an acronym) \\
TRL    & technology readiness level \\
ULE    & ultra-low expansion glass\\
UltracoolSHEET & capitalized, not an acronym\\
US     & United States  \\
UV     & ultraviolet \\
VIPER  & Volatiles Investigating Polar Exploration Rover \\
VISION & Visible Imaging System for Interferometric \\
       & Observations at NPOI \\
VLA    & Very Large Array \\
VLBI   & very long baseline interferometry  \\
VLT    & Very Large Telescope            \\
VLTI   & Very Large Telescope Interferometer \\
W      & Watt \\
XRISM  & X-Ray Imaging and Spectroscopy Mission \\
YORP   & Yarkovsky-O'Keefe-Radzievskii-\\
       & Paddack effect \\
YSO    & young stellar object\\

\caption{Acronyms and initialisms used in this document.}
\label{tab-acronyms}
\end{longtable}







\end{appendices}


\end{document}